\documentclass[aps,prb,amsmath, amssymb, english, twocolumn,reprint,floatfix, superscriptaddress]{revtex4-1}
\usepackage{multirow}
\usepackage{hhline}
\usepackage{bbm}
\usepackage{comment}
\usepackage{mathtools}

\usepackage{pgfplots}
\usepackage{microtype}
\usepackage{epstopdf}
\usepackage{epsfig} 
\usepackage{amsmath}

\usepackage{hyperref}
\hypersetup{
    colorlinks,
    citecolor=blue,
    filecolor=blue,
    urlcolor=blue}

\newcommand{\ket}[1]{\left|#1\right>}
\newcommand{\bra}[1]{\left<#1\right|}

\newcommand{\bR}{\mathbf{R}}

\newcommand{\cc}{c^{\ }}
\newcommand{\bq}{\mathbf{q}}
\newcommand{\bQ}{\mathbf{Q}}

\newcommand{\bk}{\mathbf{k}}

\newcommand{\cd}{c^\dag}

\graphicspath{{./Images/}}

\title{Broken Symmetry Phases}

\date{}
\pgfplotsset{compat=1.16}
\begin{document}
\title{The dynamical vertex approximation for many-electron systems \\ with spontaneously broken SU(2)-symmetry}
\author{Lorenzo Del Re} 
\affiliation{Department of Physics, Georgetown University, 37th and O Sts., NW, Washington, DC 20057, USA}
\affiliation{Erwin Schr\"{o}dinger International Institute for Mathematics and Physics, Boltzmanngasse 9 A-1090 Vienna, Austria}
\author{Alessandro Toschi} 
\affiliation{Institute for Solid State Physics, TU Wien, 1040 Vienna, Austria}

\date{\today} 

\pacs{}

\begin{abstract}
We generalize the formalism of the dynamical vertex approximation (D$\Gamma$A)  -- a diagrammatic extension of the dynamical mean-field theory (DMFT)-- to treat magnetically ordered phases.
To this aim, we start by concisely illustrating the many-electron formalism for performing ladder resummations of Feynman diagrams in systems with broken SU(2)-symmetry, associated to ferromagnetic (FM) or antiferromagnetic (AF) order. 
We then analyze the algorithmic simplifications introduced by taking the {\sl local} approximation of the two-particle irreducible vertex functions in the Bethe-Salpeter equations, which defines the ladder implementation of D$\Gamma$A for magnetic systems.  The relation of this assumption with the DMFT limit of large coordination-number/ high-dimensions is explicitly discussed. 
As a last step, we derive the expression for the ladder D$\Gamma$A self-energy in the FM- and AF-ordered phases of the Hubbard model.  The physics emerging in the AF-ordered case is explicitly illustrated by means of approximated calculations  based on a {\sl static} mean-field input for the D$\Gamma$A equations.  The results obtained capture fundamental aspects of both metallic and insulating ground states of two-dimensional antiferromagnets,  providing a reliable compass for future, more extensive applications of our approach. Possible routes to further develop diagrammatic-based treatments of magnetic phases in correlated electron systems are briefly outlined in the conclusions.
\end{abstract}\maketitle

\section{Introduction}
The algorithmic treatment of correlations effects in many fermion systems still poses one of the hardest challenges to condensed matter theory.
This is especially true in the parameter regimes most interesting from the physical point of view, where unconventional magnetic or superconducting phenomena are often observed\cite{Blundell2001,Bednorz1986,Chen2019,Chubukov2015,Li2019,Wilson2019}: Intermediate-to-strong coupling, proximity to classical/quantum phase transitions\cite{Loehneysen2007,Brando2016} and to Mott-Hubbard insulating phases\cite{Imada1998,Hansmann2013,DeMedici2011,Isidori2019,Springer2019}, reduced dimensionality (i.e., confinement of electrons in layers or of ultra-cold atoms in optical lattices).
In such cases, conventional weak-coupling approaches, such as band-theory, DFT\cite{Hohenberg1964}, GW\cite{Hedin1965,Aryasetiawan1998},  and FLEX\cite{Bickers1989,Bickers2004} typically yield rather poor results, calling for a full quantum many-body approach to the problem of interest. 

Among the cutting-edge schemes capable to treat electronic correlations over different space- and time-scales, we recall the determinant and the diagrammatic\cite{Kozik2010} Monte Carlo, and the extensions\cite{Maier2005,Rohringer2018} of the dynamical mean-field theory (DMFT)\cite{Georges1996}. 
Within the latter class of approaches, the most recent ones are the diagrammatic extensions\cite{Rohringer2018} of DMFT, which have shown a rapid development over the last decade.
Aside from specific details, all the diagrammatic extensions of DMFT share the same philosophy and the same goal: a systematic inclusion of non-local correlations on top of the purely local ones captured non-perturbatively by the DMFT. 
They are built as a two-step procedure\cite{Rohringer2018}: (i) calculation of a two-particle purely local, but dynamical, vertex function\cite{Rohringer2012}  from the auxiliary Anderson impurity model (AIM) associated to the DMFT; (ii) usage of this vertex as the effective dynamical interaction of a new Feynman diagrammatic expansion around the DMFT solution. 

This way, all the non perturbative, but purely local, information computed in DMFT, including the description of Mott-Hubbard metal-insulator transitions, will be -per construction- included from the very beginning in the subsequent diagrammatic treatment. The latter, typically consisting in ladder\cite{Kusunose2006,Toschi2007,Hafermann2009} or parquet\cite{Valli2015,Li2016,Krien2020} resummations build upon the DMFT vertex, will introduce the missing information about spatial correlations. 

Due to their diagrammatic nature, these approaches do not face intrinsic cluster size restrictions. They are,  thus, particularly suited to describe systems in the proximity of (quantum) phase-transition and bosonic collective modes in the non-perturbative regime. 
Recent applications and results range from the treatment of criticality\cite{Rohringer2011,Antipov2014,Hirschmeier2015,DelRe2019} and quantum criticality\cite{Schaefer2017,Schaefer2019}, the description of the quasi long-range antiferromagnetic order in 2D\cite{Schaefer2015,vanLoon2015-2} to the interplay between AF-fluctuations and superconductivity\cite{Kitatani2019,Kitatani2020,Astretsov2020}, see also Ref.~\onlinecite{Rohringer2018}.

Almost all applications mentioned above were restricted to single-orbital models with repulsive interaction in their {\sl paramegnetic} phase. While first generalizations of these schemes have been presented to treat multi-orbital physics (such as ab-initio D$\Gamma$A\cite{Toschi2011,Galler2016} or Dual Fermion for graphene\cite{Hirschmeier2015}) and attractive interactions\cite{DelRe2019}, to the best of our knowledge, none of these approaches has been hitherto extended to cases with spontaneous symmetry breaking. In fact, focused studies\cite{Kunes2019, Kunes2020, Kunes2021} of collective excitations in the broken symmetry phases of strongly correlated systems have been very few even at the ``simpler"  DMFT level, including the recent, pioneering DMFT analyses of the excitonic ordered phases\cite{Kunes2019, Kunes2020}. 

In this paper, we start filling this gap by extending the formalism of one of these approaches, the dynamical vertex approximation (D$\Gamma$A)  in its widespread flavor (i.e., based on {\sl ladder} diagram resummations in the dominant channels) to treat some of the most important symmetry-broken phases: {\sl ferromagnetic} (FM) and {\sl antiferromagnetic}  (AF) orders. 

To this aim two main ingredients are necessary: (i) the identification of the dynamical vertex to be extracted from the DMFT calculations\cite{Georges1996,Sangiovanni2006,Taranto2012,Hausoel2017} for the FM/AF 
ordered phases; (ii) the explicit expressions of the ladder diagram equations for the broken SU(2)-symmetry cases considered, namely the expressions for the physical susceptibilities and the self-energy. 

To achieve the first goal, we will take the DMFT limit of large coordination number/high dimensions\cite{Metzner1989,Georges1996} for the irreducible vertex functions of the symmetric as well as of the SU(2)-symmetry broken phase(s). The second task requires, instead, a generalization of the ladder expressions hitherto adopted\cite{Toschi2007,Katanin2009,Rohringer2016,Rohringer2018,DelRe2019} for the symmetric case to the FM or the AF long-range order.
Although this strategy is conceptually straightforward, the increased number of degrees of freedom to be considered --due to the
 lack of SU(2)-symmetry-- reflects in a relatively involved structure of the terms to be considered, especially for the AF case.
In particular, since many related derivations in the literature are restricted to specific aspects of the problems considered, 
we will provide in the first part of our work a concise illustration of the Bethe-Salpeter equations (BSE)
in magnetically ordered phases, and then discuss how the corresponding expressions and the underlying symmetry relations get simplified by assuming locality of the 2PI vertex function $\Gamma$ - a common feature of random phase approximation (RPA), DMFT and ladder D$\Gamma$A for models with on-site electronic interaction.  


Eventually, after merging the ingredient (i) and (ii), we derive the corresponding ladder D$\Gamma$A expression for the momentum-dependent self-energy in the broken symmetry phase considered.

The behavior of the different collective modes, as well as of their effect on the electronic scattering, will then be analyzed in selected realizations of AF-ordered phases exploiting a (static) mean-field-like simplification of the input for the corresponding D$\Gamma$A expressions.  In this context, we will illustrate the different  mechanisms driving the spectral properties of insulating and metallic ground states of two-dimensional antiferromagnets in terms of the  interplay between the (RPA-like) Higgs and Goldostone modes and the particle-hole continuum of the fermionic excitations. 
While the obtained results (for which we refer the interested reader to the corresponding Sec.~V) are of specific relevance on their own, they also outline a robust framework for the interpretation of future D$\Gamma$A calculations, allowing  to draw some general conclusions about the physics potentially accessible by diagrammatic extensions of DMFT for magnetically ordered systems.  


Eventually, it should be emphasized that the derivations presented in our paper will be useful also beyond the specific framework of the D$\Gamma$A approach. In fact, a similar formalism is directly applicable to the analysis of the collective modes in the magnetic phases of DMFT.  Further, very similar ladder structures will be  encountered by generalizing  to the broken SU(2)-symmetry case other diagrammatic extensions of DMFT based on ladder approximations, such as Dual Fermion\cite{Rubtsov2008}, Dual Boson\cite{Rubtsov2012}, 1PI\cite{Rohringer2013}, TRILEX\cite{Ayral2015}, TRILEX${^2\,}$ \cite{Stepanov2019}, and FLEX+DMFT\cite{Kitatani2015}, and, to some extent, DMF$^2$RG\cite{Taranto2014,Wentzell2015}, i.e. the merger of functional renormalization group\cite{Metzner2012} and DMFT, as well as the recently introduced Single Boson Exchange (SBE)\cite{Krien2019SBE}. 

 The two main advances  obtained in this work, i.e., the formal derivation of the ladder D$\Gamma$A equations for magnetically ordered phases, as well as the insight gained on the physics of correlated antiferromagnets by hands of a simplified application of the approach are clearly reflected in the structure of our paper: The formal derivations are presented in the following three sections (II-IV), while the reader mainly interested to the physical discussion can be directly referred to Sec.~V.

Specifically, the paper is organized as follows: In Sec.\ref{sec:form} we introduce the general formalism necessary for our diagrammatic treatment of the FM and AF phases in correlated systems.  In Sec.~\ref{sec:inf_dim}, we discuss the locality properties of the irreducible vertex functions to be considered in DMFT and (ladder) D$\Gamma$A, and their relation with the high-dimensionality/connectivity limit.  In Sec.~\ref{sec:dga}, we derive explicitly the  BSE of DMFT for the broken SU(2)-symmetry phases, as well as the corresponding self-energy expression in ladder D$\Gamma$A. Finally in Sec.~\ref{sec:RPA-DGA}, by hand of an approximated calculation, we illustrate the physical content of our extended D$\Gamma$A expressions, namely the main physical mechanisms at work (Goldstone, Higgs and density modes) as well as their expected effect on the spectral properties in D$\Gamma$A. Conclusions and outlook are presented in Sec.~\ref{sec:concl}.

\section{Formalism}\label{sec:form}

\subsection{General definitions}

Let us consider a fermionic system with $N$ internal degrees of freedom, as for example the spin and/or the orbitals of electrons in solids or the hyperfine levels of neutral atoms trapped in optical lattices etc. and let us associate the spin-orbital index $\alpha =1,..,N$ to such degrees of freedom. Observable operators can be constructed using the SU($N$) representations plus the identity matrix that we indicate as $T^{(a)}_{\alpha\beta}$,  with $a =1, ..., N^2$. The operator related to the $a$-th representation can be expressed in the Heisenberg picture  as $\hat{O}^{(a)}(x)=T^{(a)}_{\alpha\beta}\cd_{\alpha}(x)\cc_{\beta}(x)$, where we adopted a four-vectorial notation with $x\equiv (\bR,\tau)$ and $\cc_\alpha(x)=e^{\hat{H}\tau}\cc_{\bR\alpha}e^{-\hat{H}\tau}$, and a summation over repeated indices is intended. 

The one-particle Green's function is  $G^{\alpha \beta}(x_1,x_2)\equiv -T_{\tau}\left<\cc_{\alpha}(x_1)\cd_\beta(x_2)\right>$, defining the quantum statistical average of $\hat{O}^{(a)}(x)$ as $\left<\hat{O}^{(a)}(x)\right>=T^{(a)}_{\alpha\beta}G^{\beta\alpha}(x,x^+)$.

Correlation functions, e.g.~in the particle-hole sector, can be formally derived introducing the bilinear action $S_{ext} = -\int\int dx dy\, h^{\alpha\beta}(x,y)\overline{c}_\alpha(x)c_{\beta}(y)$ as an external source added to the original action of the system (here the integral symbol is a short-hand notation standing for $\int dx \equiv \sum_\bR\int_0^\beta$, with $\beta = \frac{1}{k_b T}$ being the inverse of the temperature).
The generalized susceptibility in the particle-hole notation reads:

\begin{equation}\label{def:gen}
\left.\chi^{\alpha\beta}_{\gamma\delta}(x_1,x_2,x_3,x_4) \equiv \frac{\delta G^{\beta\alpha}(x_2,x_1)}{\delta h^{\gamma\delta}(x_3,x_4)}\right|_{h=0},
\end{equation}

Carrying out the functional derivative (eventually evaluated at zero external field) yields the expression for generalized susceptibility in terms of the two-particle and one-particle Green's functions, i.e. $\chi^{\alpha\beta}_{\gamma\delta}(x_1,x_2,x_3,x_4) = G^{\alpha\beta}_{\gamma\delta}(x_1,x_2,x_3,x_4)-G^{\beta\alpha}(x_2,x_1)G^{\delta\gamma}(x_4,x_3)$, where $G^{\alpha\beta}_{\gamma\delta}(x_1,x_2,x_3,x_4) \equiv T_{\tau}\left<\cd_{\alpha}(x_1)\cc_\beta(x_2)\cd_{\gamma}(x_3)\cc_\delta(x_4)\right>$.

We recall that it is sometimes useful to change the representation of the generalized susceptibilities. This can be done by expressing the external source as $S_{ext} = -\int\int dx dy \sum_{a}h^{(a)}(x,y)\,T^{(a)}_{\alpha\beta}\,\overline{c}_\alpha(x)c_{\beta}(y)$. 
 The corresponding expression for the generalized susceptibility reads: 
\begin{eqnarray}\label{def:phys}
\chi^{ab}(x_1,x_2,x_3,x_4) &\equiv& \frac{\delta \mathcal{G}^{(a)}(x_1,x_2)}{\delta h^{b}(x_3,x_4)} 
\end{eqnarray}
where $\mathcal{G}^{(a)}(x,y) \equiv T^{(a)}_{\alpha\beta}G^{\beta\alpha}(y,x)$. 
The susceptibilities in the two different representations are related to each other by the following relation:
\begin{equation}\label{def:rot}
 \chi^{ab}(x_1,x_2,x_3,x_4) = T^{(a)}_{\alpha\beta}\,\chi^{\alpha\beta}_{\gamma \delta}(x_1,x_2,x_3,x_4)\,T^{(b)}_{\gamma\delta}.
\end{equation}
As the generalized susceptibility as defined in Eq.(\ref{def:phys}) depends on the representation indices rather than  the spin-orbital ones, this basis is often more suitable for the physical interpretation,  because the operators  $\hat{O}^{(a)}$ are observables in quantum mechanics \footnote{From now on we will assume that $T^{(a)}$ is a set of hermitian and trace normalized generators of SU(N), i.e. Tr$\left[T^{(a)}\cdot T^{(b)}\right] = \delta_{ab}$.}. On the other hand, the definition in Eq.(\ref{def:gen}) remains useful for practical purposes, since most numerical calculations are performed using this ``computational'' basis. 
{
\par Working with the physical basis is particularly useful when the system possesses some symmetries. In fact, let us consider the case where the Hamiltonian commutes with an operator $\hat{O}$, i.e. $[\hat{H},\hat{O}] = 0$, where $\hat{O} = \sum_{\bR}\psi^\dag_\bR \, O_\bR \, \psi_\bR$.  In this case, by performing a a change of basis in the spinor representation defined by  $U_\bR = \exp\left[i\, O_\bR \,\lambda\right]$ for some real parameter $\lambda$, and recognizing that the hamiltonian does not change in the new representation, we obtain the following relation for the generalized     susceptibility:
\begin{eqnarray}\label{eq:relation}
\chi^{ab}(x_1,x_2,x_3,x_4) =  F^{a^{\,}a^\prime}_{\bR_1 \bR_2}\, \chi^{a^\prime b ^\prime} (x_1,x_2,x_3,x_4) F^{b^{\,}b^\prime}_{\bR_3\bR_4}, \nonumber \\
\end{eqnarray}
where  a summation over repeated indices is intended  and: 
\begin{eqnarray}\label{eq:coeff}
F^{ab}_{\bR\bR^\prime} = \mbox{Tr}\left[T^{(b)}\, U^\dag_{\bR}\, T^{( a )}\,  U^{\,}_{\bR^\prime} \right] .
\end{eqnarray}
Sometimes the hamiltonian possesses particle-hole symmetry, i.e. $\hat{H}$ is left unchanged after the following canonical transformation $\psi^{\,}_\bR \to U_\bR\psi^\dag_\bR$, with $U_\bR$ being a unitary matrix. In this case, the relation for the generalized susceptibility is more complicated and reads:
\begin{eqnarray}\label{eq:relation_ph}
\chi^{ab}(x_1,x_2,x_3,x_4) =  \widetilde{F}^{a^{\,}a^\prime}_{\bR_1 \bR_2}\, \chi^{a^\prime b ^\prime} (x_2,x_1,x_4,x_3) \widetilde{F}^{b^{\,}b^\prime}_{\bR_3\bR_4}, \nonumber \\
\end{eqnarray}
where:
\begin{eqnarray}\label{eq:coeff_ph}
\widetilde{F}^{ab}_{\bR\bR^\prime} = \mbox{Tr}\left[T^{(b)}\left( U^\dag_{\bR}\, T^{( a )}\,  U^{\,}_{\bR^\prime}\right)^T \right] .
\end{eqnarray}


}

In Appendix \ref{sec:app_a}, we present an explicit derivation of the coefficients defined in Eqs.(\ref{eq:coeff},\ref{eq:coeff_ph}).

\par 
In this work, we will explicitly consider the single band Hubbard model on a square ({\sl two}-dimensional) lattice:
\begin{equation}
H =  \sum_{ij,\sigma}t_{ij}\,\cd_{i\sigma}\cc_{j\sigma}+U\sum_{i\sigma}\hat{n}_{i\uparrow}\hat{n}_{i\downarrow},
\end{equation}
where $t_{ij}$ is the electronic hopping amplitude between the  sites $i$ and $j$ and $U$ is the local Coulomb repulsion.
For the explicit calculations shown in Secs.~IV and V, we will consider the unfrustrated case (i.e., only with a nearest-neighbour hopping $t$) as well as the inclusion of a next-to-nearest hopping term $t^\prime$.

When the system does not break {\sl translational} symmetry, as in the paramagnetic (PM) or ferromagnetic (FM) cases, a complete set of representations is provided by the SU(2) generators, i.e. $T^{(a)} = \sigma_a/\sqrt{2}$,  where $\sigma_0 = \mathbbm{1}_{2\times 2} $ is the identity, and  $\sigma_{a = \{1,2,3\}} = \{\sigma^{(x)},\sigma^{(y)},\sigma^{(z)}\}$ are the Pauli matrices.
Hence, one can identify $16$ possible components for the physical susceptibility $\chi^{ab}$. 
However,  the symmetry relations holding for the PM and FM lower down such a number to $2$ (that is more generically valid for a system with SU$(N\geq 2)$ symmetry ~\cite{Delre2018})  and $4+2 =6$  respectively.

In the antiferromagnetic  (AF) case, instead, the full translational symmetry of the original lattice is broken, and sub-lattice indices must be taken into account. As we will see in the next sections, the number of possible independent, non-vanishing components of $\chi$ increases correspondingly to $32+16 =48$.
However, in the DMFT limit of infinite dimensions, as well as in the ladder D$\Gamma$A, such a number is reduced down to $8+4 =12$.

\subsection{Bethe-Salpeter equations}\label{sec:BS_form}

In this section, we illustrate the general expressions of the Bethe-Salpeter equations (BSE) for the generalized susceptibilities of the FM- and AF-ordered systems and briefly discuss the associated symmetry properties.

By carrying out the functional derivative in Eq.(\ref{def:gen}), the following expression for the generalized susceptibility in the ``computational'' basis is obtained:
\begin{eqnarray}\label{BSE:gen:ind}
\chi_{1234} &=&  -G(2,3)G(4,1) + \nonumber \\ &&\int\prod_{i = 1}^4 di^\prime\, G (2,2^\prime)G(1^\prime,1)\Gamma_{1^\prime 2^\prime 3^\prime 4^\prime}\,\chi_{3^\prime 4^\prime 3\,4},
\end{eqnarray} 
where we adopt generalized indices $i = (x_i,\alpha_i)$, and the irreducible vertex  in the corresponding channel is $\Gamma_{1234} = \frac{\delta \Sigma(2,1)}{\delta G(3,4)}$.
By expanding all the space-time functions  $f$ (with translational invariance properties) in terms of their Fourier components, we get
\begin{eqnarray}
f(x_1,x_2,x_3,x_4) &=&  \\ &\frac{1}{\left(\beta V\right)^3}&\sum_{kk^\prime q} e^{i[kx_1 -(k+q)x_2 + (k^\prime + q)x_3 -k^\prime x_4]}f^{kk^\prime q}, \nonumber 
\end{eqnarray}
where $\beta = (k_B T)^{-1}$  and $V$ is the volume of the system.
The Fourier expansion of the Green's function reads: $G^{\alpha\beta}(x_1,x_2) = \frac{1}{V\beta}\sum_k e^{-ik(x_1-x_2)}G^{\alpha\beta}(k)$.
Hence, the Fourier transformed expression for the BSE of $\chi$ reads:
\begin{eqnarray}\label{eq:BSE:generic}
\widetilde{\chi}^{\alpha\,\beta}_{\gamma\,\delta}(q) &=& \widetilde{\chi}^{\hspace{0.25cm}\alpha\,\beta}_{0,\,\gamma\,\delta}(q) 
\nonumber \\
&-& \frac{1}{(\beta V)^2}\sum_{\alpha^\prime\beta^\prime\gamma^\prime\delta^\prime}\widetilde{\chi}_{0,\,\beta^\prime\alpha^\prime}^{\hspace{0.25cm}\alpha\,\beta}(q)\cdot\Gamma^{\alpha^\prime\beta^\prime}_{\gamma^\prime\delta^\prime}(q)\cdot\widetilde{\chi}^{\,\gamma^\prime\delta^\prime}_{\,\gamma \delta}(q),\nonumber \\
\end{eqnarray}
where the tilde symbol represents matrix in the space of the four-momenta $k$ and $k^\prime$, "$\cdot$" represents the matrix product, and we defined the disconnected susceptibility as: 
\begin{equation}
\widetilde{\chi}^{\hspace{0.25cm}\alpha\,\beta}_{0,\,\gamma\,\delta}(q) \equiv -(V\beta)\, \delta_{kk^\prime}\,G^{\beta\gamma}(k)G^{\delta \alpha}(k+q)\,.
\end{equation} 

The BSE in Eq.~(\ref{eq:BSE:generic}) can be formally rewritten\cite{Abrikosov1975,Bickers2004,Rohringer2012,Rohringer2018} in terms of the full vertex function $\mathcal{F}$ as: 
\begin{eqnarray}\label{def:full:vertex}
\chi_{1234} &=& \chi_{0,1234}- \nonumber \\ 
&& \int\prod_{i = 1}^4di^\prime G(1^\prime,1)G(2,2^\prime)\,\mathcal{F}_{1^\prime 2^\prime 2^\prime 4^\prime}\,G(3,3^\prime)G(4,4^\prime). \nonumber \\
\end{eqnarray}
Hence, substituting the definition in Eq.(\ref{def:full:vertex}) into Eq.(\ref{BSE:gen:ind}), the explicit expression of $\mathcal{F}$ in Fourier space reads:
\begin{eqnarray}
\widetilde{\mathcal{F}}^{\,\alpha\beta}_{\,\gamma\delta}(q) &=& \widetilde{\Gamma}^{\,\alpha\beta}_{\,\gamma\delta}(q) -
\nonumber \\
&\displaystyle \frac{1}{(V\beta)^2}&\sum_{\alpha^\prime\beta^\prime\gamma^\prime\delta^\prime}
\widetilde{\Gamma}^{\,\alpha\beta}_{\gamma^\prime\delta^\prime}(q)\cdot
\widetilde{\chi}^{\hspace{0.25cm}\delta^\prime\,\gamma^\prime}_{0,\,\beta^\prime\,\alpha\prime}\,(q)\cdot\widetilde{\mathcal{F}}^{\,\alpha^\prime\beta^\prime}_{\,\gamma\delta}(q).\nonumber \\
\end{eqnarray}
In order to lighten the notation, in the subsequent sections we shall drop the tilde symbol for indicating matrices in the four-momenta space, also omitting the explicit dependence on transferred four-momentum $q$:  $\widetilde{\mathcal{F}}^{\alpha\beta}_{\gamma\delta}(q) \equiv \mathcal{F}^{\,\alpha\beta}_{\,\gamma\delta}$ .

\subsubsection{Ferromagnetic order}

In a ferromagnet, the SU(2) symmetry is spontaneously broken into a reduced U(1) symmetry, where the angular momentum along the $z$-axis is conserved. Hence, in this case, we have $6$ independent components of generalized susceptibility that can be expressed either in the computational or physical basis defined in Eqs.(\ref{def:gen},\ref{def:rot}), respectively. 
The conservation of $\hat{S}^z$ implies that we can group the representations into {\sl spin longitudinal} (i.e., associated to the diagonal operators $\{\mathbbm{1},\sigma^{(z)}\}$) and {\sl spin transverse} (i.e., those corresponding to the remaining operators $\{\sigma^{(x)},\sigma^{(y)}\}$). Any component $\chi^{ab}$ mixing operators belonging to the two different sectors vanishes, e.g. $\chi^{zx}=\frac{1}{2}\sigma^{(z)}_{\alpha\beta}\chi^{\alpha\beta}_{\gamma\delta}\sigma^{(x)}_{\gamma\delta} = 0$, that can be seen by computing the coefficients defined in Eq.(\ref{eq:coeff}). Indeed, if we set $U_\bR = \sigma^z$,  $F^{z b}_{\bR \bR^\prime} = \frac{1}{2}\mbox{Tr}[\sigma_b \sigma^z\sigma^z\sigma^z] = \delta_{z b}$, while $F^{x b}_{\bR \bR^\prime} = \frac{1}{2}\mbox{Tr}[\sigma_b \sigma^z\sigma^x\sigma^z] =-\delta_{x b}$, that when substituted in Eq.(\ref{eq:relation}) yields the following relation $\chi^{zx} = -\chi^{zx}$. 
As another consequence of the U($1$) symmetry, not all the transverse spin sector components are independent from each other. In particular, we have that $\chi^{xx}=\chi^{yy}$ and $\chi^{xy}=-\chi^{yx}$. This can be seen by setting $U_\bR = \exp[-i \sigma^z \frac{\pi}{4}]$ and realizing that in this case we have $F^{xb} = -\delta_{yb}$ and $F^{yb} = \delta_{xb}$.
These symmetry properties, which are evident in the physical basis, reduce the number of possible spin indices combinations in the computational basis.
Hence, a more compact notation can be introduced for the computational basis: $\chi^{\alpha \alpha}_{\beta\beta} \equiv \chi_{\alpha\beta}$, $\chi^{\alpha\beta}_{\beta \alpha} \equiv \chi_{\overline{\alpha\beta}}$. 
We list the $6$ independent susceptibility components of the physical basis in Table \ref{tab:ferro} in terms of their explicit expressions in the computational representation.
\renewcommand{\arraystretch}{1.5}
\begin{table}[t!]
\centering
  \begin{tabular}{ | c | }
  \hline
	Longitudinal Channel \\
	\hline
	$\chi^{\rho\rho} = \frac{1}{2}\left(\chi_{\uparrow\uparrow} + \chi_{\uparrow\downarrow} + \chi_{\downarrow\uparrow} + \chi_{\downarrow\downarrow}\right)$ \\
	$\chi^{zz} = \frac{1}{2}\left(\chi_{\uparrow\uparrow}  -\chi_{\uparrow\downarrow} - \chi_{\downarrow\uparrow}  + \chi_{\downarrow\downarrow}\right)$ \\
	$\chi^{\rho z} = \frac{1}{2}\left(\chi_{\uparrow\uparrow} - \chi_{\uparrow\downarrow}  +\chi_{\downarrow\uparrow}  - \chi_{\downarrow\downarrow}\right)$ \\
	 	$\chi^{z\rho} = \frac{1}{2}\left(\chi_{\uparrow\uparrow} + \chi_{\uparrow\downarrow}  -\chi_{\downarrow\uparrow}  - \chi_{\downarrow\downarrow}\right)$\\ 
	 	\hline
	 	Transverse Channel\\
	 	\hline
	 	$\chi^{xx} = \frac{1}{2}\left(\chi_{\overline{\uparrow\downarrow}} + \chi_{\overline{\downarrow\uparrow}}\right)$\\
	 	$\chi^{xy}  =\frac{i}{2}\left(  \chi_{\overline{\uparrow\downarrow}} \,-\, \chi_{\overline{\downarrow\uparrow}}\right)$ \\
	 	\hline
  \end{tabular}
  \caption{Non-vanishing correlators of the FM phase  in the transverse/longitudinal channels. Their expression in the physical basis is explicitly expanded in the computational one.}
  \label{tab:ferro}
\end{table}
\renewcommand{\arraystretch}{1.}

As expected, the SU(2)/magnetic degeneracy between two-particle correlation functions (i.e., between the components of the physical susceptibilities) is broken in the ferromagnet, i.e.  $\chi^{zz}\not=\chi^{xx}$.
Further, one must also notice the emergence of {\sl mixed} correlators, namely $\chi^{\rho z}$, $\chi^{z\rho}$ and $\chi^{xy}$, that were identically zero in the SU(2)-symmetric (paramagnetic) case.
The first two terms describe a linear coupling of the density (magnetization along the $z$-axis) to an external field along the $z$-axis (chemical potential)\cite{kuboki2017}. Such a linear dependence obviously vanishes in the paramagnetic case, where, given the isotropy of the system, the density can be expanded only in even powers of the external field.  
It is worth to notice the mixed physical correlator $\chi^{z\rho}$ vanishes in the case of particle-hole symmetry. For a  ferromagnetic system that conserves $\hat{S}^z$ the ph-symmetry transformation is given by $\psi^{\,}_\bR \to U_\bR\psi^\dag_\bR$, with $U_\bR = e^{i\Pi\bR}\sigma^x$. Therefore, in this case we have $\widetilde{F}^{z b}_{\bR\bR^\prime} =e^{-i\Pi(\bR_1-\bR_2)} \frac{1}{2}\mbox{Tr}\left[\sigma_b(\sigma^x\sigma^z\sigma^x)^T\right]  =  -e^{i\Pi(\bR-\bR^\prime)}\delta_{zb} $ and analogously  $\widetilde{F}^{\rho b}_{\bR\bR^\prime} =e^{-i\Pi(\bR-\bR^\prime)} \delta_{\rho b} $. Substituing the coefficients in Eq.(\ref{eq:relation_ph}) we obtain the following relation $\chi^{z\rho}(x_1,x_2,x_3,x_4) = -e^{-i\Pi(x_1-x_2+x_3-x_4)} \chi^{z\rho}(x_2,x_1,x_4,x_3)$, that in four momentum space reads:
\begin{equation}\label{eq:relation_rho_z}
    \chi^{z\rho}(k,k^\prime,q) \stackrel{\text{ph}}{=} -\chi^{z\rho}(-\Pi-k-q,-\Pi-k^\prime-q,q).
\end{equation}
The last equation implies that the physical susceptibility $\chi^{z\rho}(q) = \frac{1}{(V\beta)^2}\sum_{kk^\prime} \chi^{z\rho}(k,k^\prime,q) = 0$ when the system is particle-hole symmetric. We notice, that this does not apply to the mixed correlators in the transverse channel because, even if $\sigma^y$ is odd under $U^{\,}_\bR$, it picks up an additional minus sign after the transposition, i.e. $(\sigma^x\sigma^y\sigma^x)^T = \sigma^y$. 
\par The appearance of a non vanishing mixed-correlator $\chi^{xy}$ in presence of a finite magnetization along $z$ is intrinsically rooted into the quantum nature of the spin operators. 
For instance, one can consider the physical susceptibility $\chi^{xy}(\bq , \tau) = 2\,T_\tau\left<\hat{S}^x_\bq(\tau)\hat{S}^y_{-\bq}(0)\right>$ for $\tau \to 0$,
 where $\hat{S}^{a}_\bR = \frac{1}{2}\sigma^{(a)}_{\alpha\beta}\cd_{\bR\alpha} c^{\ }_{\bR\beta}$, $\hat{S}_\bq^a = \frac{1}{\sqrt{V}}\sum_\bR e^{i\,\bq\cdot\bR}\,\hat{S}^{a}_\bR$ is the Fourier transform of the angular momentum component along the $a$-axis.
In a FM, this does no longer vanish and displays a discontinuity in its imaginary part at $\tau = 0$ \cite{KrienThesis}, because of the commutation relations between the angular momentum components, i.e.:
\begin{equation}\label{eq:disc}
\chi^{xy}(\mathbf{q},0^+)-\chi^{xy}(\mathbf{q},0^-) = 2\left<\left[\hat{S}_\bq^x,\hat{S}_{-\bq}^y\right]\right> =i  \, m^z,
\end{equation}
with $m^z = \frac{1}{V}\sum_\bR\left<2\,\hat{S}^z_\bR\right>$. 
Such a discontinuity at $\tau = 0$ is reflected into a power-law decay of the Fourier transform of $\chi^{xy}(\bq ,\tau)$. Its specific expression can be directly derived from the corresponding Lehemann representation:
\begin{eqnarray}\label{eq:xy_susc}
\chi^{xy}(\bq,\omega) =\frac{2}{Z} \sum_{mn}\frac{\omega\,b_{mn}\,\mbox{Im}\left[\bra{m}\hat{S}^x_\bq\ket{n}\bra{n}\hat{S}^y_{-\bq}\ket{m}\right]}{\omega^2 + (E_m-E_n)^2}, \nonumber \\
\end{eqnarray}
where  $b_{mn} = \left(e^{-\beta E_n}-e^{-\beta E_m}\right)$ (see Appendix \ref{app:corr:func} for a more generic discussion). 
From Eq.(\ref{eq:xy_susc}) one can identify\cite{KrienThesis} the asymptotic behavior of $\chi^{xy}$ at large frequencies, that is: 
\begin{equation}\label{eq:xy_asympt}
    \chi^{xy}(\bq,\omega \to \infty) = -\frac{ m^z }{\omega}.
\end{equation}
As we will show in Sec.\ref{sec:RPA-DGA}, this mixed correlator directly appears in the definition of the asymptotics of the electronic self-energy of the (anti)ferromagnetically ordered phase.
Consistently with its high-frequency asymptotics, the mixed susceptibility in Eq.(\ref{eq:xy_susc}) is an {\sl odd} function of $\omega$ \footnote{This condition and, more in general, Eq.~(\ref{eq:xy_susc}) holds, if the eigenstates of the hamiltonian are real in the real-space (lattice) basis and the system possesses inversion symmetry. However, also in the most generic case, where $\chi^{xy}(\bq,\omega)$ is not an odd function of $\omega$, Eq.~(\ref{eq:xy_asympt}) remains valid.}.


Finally, it is worth recalling that when the $SU(2)$-symmetry gets restored (e.g.,for $T > T_c$ or by driving the system through a QCP),  only two independent susceptibility components survive, namely the magnetic susceptibility $\chi_m = \chi^{zz} = \chi^{xx} = \chi^{yy}$ and the charge susceptibility $\chi_c = \chi^{\rho\rho}$, with all other mixed correlators vanishing. \\

We write the BSE for a FM in a compact way, exploiting block-wise spinorial matrices. 
In particular, it can be seen that the BSE for the ferromagnetic case are decoupled for the longitudinal and transverse channels defined in the previous section, and are given respectively by:
\begin{eqnarray}\label{BSE:ferro:para}
\mathcal{F}_{\parallel} &=& \Gamma_{\parallel} - \frac{1}{(V\beta)^2}\,\Gamma_\parallel\cdot \chi_{0\,\parallel}\cdot\mathcal{F}_{\parallel},\\
\label{BSE:ferro:perp}
\mathcal{F}_{\perp} &=& \Gamma_{\perp} - \frac{1}{(V\beta)^2}\,\Gamma_\perp\cdot \chi_{0\,\perp}\cdot\mathcal{F}_{\perp},
\end{eqnarray}  
where:
\begin{eqnarray}
\mathcal{A}_{\parallel} =
\left(\begin{array}{cc}
\mathcal{A}_{\uparrow\uparrow}&\mathcal{A}_{\uparrow\downarrow}\\
\mathcal{A}_{\downarrow\uparrow}&\mathcal{A}_{\downarrow\downarrow}
\end{array}\right)&,&
\chi_{\parallel} = 
\left(\begin{array}{cc}
\chi_{0\uparrow\uparrow}&0\\
0&\chi_{0\downarrow\downarrow}
\end{array}\right),
\end{eqnarray}\\
\begin{eqnarray}
\mathcal{A}_{\perp} =
\left(\begin{array}{cc}
\mathcal{A}_{\overline{\uparrow\downarrow}}&0\\
0&\mathcal{A}_{\overline{\downarrow\uparrow}}
\end{array}\right)&,&
\chi_{0\perp} = 
\left(\begin{array}{cc}
\chi_{0\uparrow\downarrow}&0\\
0&\chi_{0\downarrow\uparrow}
\end{array}\right),
\end{eqnarray}
where $\mathcal{A}$ represents both vertices $\Gamma$ and $\mathcal{F}$, $\chi_{0\,\sigma\sigma^\prime} \equiv -(V\beta)\,\delta_{kk^\prime}\,G_{\sigma}(k)G_{\sigma^\prime}(k + q)$.

Evidently, the computational basis offers a convenient representation for the transverse channel (which describes the Goldstone modes), because the corresponding BSE does not mix with other channels. 
Instead, in the longitudinal channel, to which the (gapped) Higgs mode belongs, it is not possible anymore to decouple the charge from the spin degrees of freedom using spin diagonalization. This is due to the emergence of the mixed correlators shown in Table \ref{tab:ferro} that introduces an interaction between the charge and the spin sectors. 
 
\renewcommand{\arraystretch}{1.5}
\begin{table}[htbp!]
    \centering
    
    \begin{tabular}{|p{1.25cm}|c c c c c c c c|}
    \hline
    
        & $\chi^{AA}_{\uparrow\uparrow}$
         & $\chi^{AB}_{\uparrow\uparrow}$
         & $\chi^{AA}_{\uparrow\downarrow}$
         & $\chi^{AB}_{\uparrow\downarrow}$
         & $\chi^{AB}_{\downarrow\downarrow}$
         & $\chi^{AA}_{\downarrow\downarrow}$
         & $\chi^{AB}_{\downarrow\uparrow}$
         & $\chi^{AA}_{\downarrow\uparrow}$
         \\[1.5ex] 
   \hline
          \centering $\chi^{\rho\rho}$
         &+&+&+&+&+&+&+&+\\[1.5ex]  
        \centering $\chi^{\bar{z}\bar{z}}$
         &$+$&$-$&$-$&$+$&$-$&$+$&$+$&$-$\\[1.5ex] 
         \centering $\chi^{\rho\bar{z}}$
         &$+$&$-$&$-$&$+$&$+$&$-$&$-$&$+$\\[1.5ex] 
        \centering $\chi^{\bar{z}\rho}$
        &$+$&$+$&$+$&$+$&$-$&$-$&$-$&$-$\\[1.5ex] 
    \hhline{|=|========|}
    
      \centering $\chi^{\bar{\rho}\bar{\rho}}$
         &$+$&$-$&$+$&$-$&$-$&$+$&$-$&$+$\\[1.5ex]  
        \centering $\chi^{zz}$
         &$+$&$+$&$-$&$-$&$+$&$+$&$-$&$-$\\[1.5ex] 
         \centering $\chi^{\bar{\rho}z}$
         &$+$&$+$&$-$&$-$&$-$&$-$&$+$&$+$\\[1.5ex] 
        \centering $\chi^{z\bar{\rho}}$
        &$+$&$-$&$+$&$-$&$+$&$-$&$+$&$-$\\[1.5ex] 
    \hline
    \end{tabular}
    \caption{Non-vanishing correlators of the AF phase in the limit of infinite dimensions in the longitudinal even/odd channels. Their expression in the physical basis is explicitly expanded in terms the computational one.  To be consistent with the definitions given in the main text,   all  symbols $\pm$ must be multiplied times a factor $1/2$.}
    \label{tab:longitudinal_AF}
    \vspace{0.5cm}
    \centering
    
    \begin{tabular}{|p{1.25cm}|cccc|}
    \hline
    &$\chi^{AA}_{\overline{\uparrow\downarrow}}$  
    &$\chi^{AB}_{\overline{\uparrow\downarrow}}$ 
    &$\chi^{AA}_{\overline{\downarrow\uparrow}}$
    &$\chi^{AB}_{\overline{\downarrow\uparrow}}$ \\
    \hline
    \centering$\chi^{xx}$& + &+ &+ & + \\
    \centering$\chi^{x\bar{y}}$& $i$ &$-i$ &$-i$ & $i$ \\
    \hhline{|=|====|}
    \centering$\chi^{\bar{x}\bar{x}}$& + &$-$ &+ & $-$ \\
    \centering$\chi^{\bar{x}y}$& $i$ &$i$ &$-i$ & $-i$ \\
    \hline
    \end{tabular}
    \caption{Non-vanishing correlators of the AF phase in the limit of infinite dimensions in the transverse even/odd channel. Their expression in the physical basis is explicitly expanded in the computational one. To be consistent with the definitions given in the main text,  all  symbols $\pm$ and $\pm i$ must be multiplied times a factor $1/2$.}
    \label{tab:transverse_AF}
    
\end{table}
\renewcommand{\arraystretch}{1.}

\subsubsection{Antiferromagnetic order}

In the case of antiferromagnetism, the system does not posses anymore the full translational invariance of the original lattice model. When the original lattice ${\cal L}$ has a  bipartite structure, we have that ${\cal L}= {\cal L}_A \cup {\cal L}_B$, with ${\cal L}_A\cap {\cal L}_B = 0$, and the Hamiltonian is invariant under discrete translations belonging to the sub-lattice ${\cal L}_A$, which contains the origin. A prototypical case, relevant for our analysis, is a tight-binding model in a hypercubic lattice in presence of a staggered magnetic field, that reads:
\begin{eqnarray}
H_0 &=&-t\sum_{<ij>,\sigma} \cd_{iA\sigma}c^{\ }_{jB\sigma}+ \mbox{H.c.}  \nonumber \\
&+&h_S\sum_{a}(-1)^{a}\sum_{i\sigma}{\sigma} \cd_{ia\sigma}c^{\ }_{ia\sigma}
\nonumber \\
&=&\sum_{\bk\in \text{MBZ}}\sum_{\sigma}\sum_{ab}\cd_{\bk a\sigma}\mathcal{H}^\sigma_{ab}(\bk) c^{\ }_{\bk b \sigma },
\end{eqnarray}
where MBZ is the Brillouin zone of the sublattice $A$, whose measure is half of the original lattice Brillouin zone, and $\mathcal{H}^\sigma(\bk) = 
\left(\begin{array}{cc}
\sigma h_S&\epsilon(\bk)\\
\epsilon(\bk)    & -\sigma h_S  
\end{array}\right)$, with $\epsilon(\bk)$  being the  lattice dispersion relations. The staggered magnetization of the system is defined as $m_S = \frac{1}{V}\sum_{\bk\sigma}\frac{\sigma}{2} \left<n_{\bk A\sigma} - n_{\bk B\sigma}\right> $, and the energy gap is given by $2\Delta$, where $\Delta = \frac{U}{2}m_S$.
The loss of the full translational invariance is taken into account by the sub-lattice indices $\{a\} = \{A,B\}$, which can be regarded -to some extent- as orbital indices. 
Hence, one finds more independent representations for the AF than in FM case. Formally, the 4 spin-orbital internal degrees of freedom $\alpha = (a,\sigma)$ correspond to 16 operator representations: These are defined by the outer product of $\frac{1}{\sqrt{2}}\{\mathbbm{1}_{2\times 2},\sigma^{(x)},\sigma^{(y)},\sigma^{(z)}\}\otimes\frac{1}{\sqrt{2}} \{\mathbbm{1}_{2\times 2},\tau^{(x)},\tau^{(y)},\tau^{(z)}\} $, where $\sigma^{(i=x,y,z)}$ and $\tau^{(i = x,y,z)}$ are Pauli matrices acting respectively on the spin and orbital Bloch spheres.   
Exploiting the conservation of the total spin operator along the z-axis, whose representation is given by $\sigma^{(z)}\otimes \mathbbm{1}$, we can group the representations into two different channels that are the spin longitudinal channel, defined by $\{\mathbbm{1}_{2\times 2},\sigma^{(z)}\}\otimes \{\mathbbm{1}_{2\times 2},\tau^{(x)},\tau^{(y)},\tau^{(z)}\} $  and the spin transverse  one given by  $\{\sigma^{(x)},\sigma^{(y)}\}\otimes \{\mathbbm{1}_{2\times 2},\tau^{(x)},\tau^{(y)},\tau^{(z)}\} $  as in the ferromagnetic case. 
Furthermore, given that  the following operator $\sigma^{(x)}\otimes\tau^{(x)}$ is conserved, the number of independent correlators are 32 and 16  in the longitudinal and transverse channels respectively. 

We will see in the next section, that in the DMFT limit of infinite dimensions it is possible to neglect 
the correlation function arising from the off-diagonal representations acting on the orbital Bloch sphere $\tau^{(x)}$ and $\tau^{(y)}$. This amounts in a considerable reduction of the non-trivial correlators, that drop to $8$ and $4$  in the longitudinal and transverse spin sectors respectively.
In such case, the representations we need to retain are: $T^\rho \equiv \frac{1}{2}\,\mathbbm{1}\otimes \mathbbm{1}$,   $T^{\overline{\rho}} \equiv \frac{1}{2}\,\mathbbm{1}\otimes \tau^{(z)}$,  $T^z \equiv \frac{1}{2}\,\sigma^{(z)}\otimes \mathbbm{1}$, $T^{\overline{z}} \equiv \frac{1}{2}\,\sigma^{(z)}\otimes \tau^{(z)}$ in the longitudinal spin channel and $T^x \equiv \frac{1}{2}\,\sigma^{(x)}\otimes \mathbbm{1}$,   $T^{\overline{x}} \equiv \frac{1}{2}\,\sigma^{(x)}\otimes \tau^{(z)}$,  $T^y \equiv \frac{1}{2}\,\sigma^{(y)}\otimes \mathbbm{1}$, $T^{\overline{y}} \equiv \frac{1}{2}\,\sigma^{(y)}\otimes \tau^{(z)}$ in the transverse spin channel. We can further classify these representations according to their properties under the unitary transformation $\mathcal{U} \equiv \sigma^{(x)}\otimes \tau^{\,(x)}$, whose associated operator commutes with the Hamiltonian. In particular, the representations listed here are or even or odd\footnote{A representation $T$ is even/odd under $\mathcal{U}$ if $\mathcal{U}\, T\, \mathcal{U} = \pm T$.}  matrices under $\mathcal{U}$.   According to this definition, we have that $T^\rho$, $T^{\overline{z}}, T^x, T^{\overline{y}}$ are even representations under $\mathcal{U}$, while  $T^{\overline\rho}$, $T^{z}, T^{\overline{x}}, T^{y}$ are odd under $\mathcal{U}$. As long as $\mathcal{U}$ is related to a symmetry of the system, i.e.~its associated operator commutes with the Hamiltonian, all the correlators involving an even operator and an odd operator vanish \footnote{Let us note that, if an external homogeneous magnetic field along the z-axis, represented by $T^z$, was added in the Hamiltonian, ${\cal U}$ would not represent a symmetry of the system anymore and even- and odd-sectors would interact between each other.}. 
For example, let us consider the transformation under such a transformation of $\chi^{\bar{z}\bar{\rho}}$, that measures the interaction between the staggered field and the charge density wave fluctuations.  Using the definition in Eq.(\ref{eq:coeff})  we have that $F^{\bar{z}b} = \delta_{\bar{z}b}$, while $F^{\bar{\rho}b} = -\delta_{\bar{\rho}b}$, that after we substitute the coefficents in Eq.(\ref{eq:relation}) leads to the identity $\chi^{\bar{z}\bar{\rho}} = -\chi^{\bar{z}\bar{\rho}}$.  
Therefore, to identify the symmetry properties of a representation in the AF case, one needs  to specify if it belongs to the longitudinal or transverse spin sector and if it is even or odd under $\mathcal{U}$.
The list of the independent susceptibility components in the DMFT limit of infinite dimensions is given in Tables (\ref{tab:longitudinal_AF},\ref{tab:transverse_AF}) .
\par When the system has particle-hole symmetry, i.e. $\hat{H}$ is left unchanged under the transformation $\psi^{\,}\to \sigma^{(x)}\otimes\tau^{(z)}\psi^\dag$, we obtain a very similar relation for the mixed susceptibilities in the longitudinal channel as the one in Eq.(\ref{eq:relation_rho_z}) for the ferromagnetic case, e.g.
\begin{eqnarray}\label{eq:rho_zeta_ph}
\chi^{\bar{z}\rho}(k,k^\prime,q)\stackrel{\text{ph}}{=}-\chi^{\bar{z}\rho}(-k-q,-k^\prime-q,q)\,.
\end{eqnarray}
This implies that the physical mixed susceptibilities, that are obtained after averaging the generalized ones over the fermionic indices, vanish exactly in this case.
We notice that this does not apply to the transverse sector. In fact, even if $T^{\bar{y}}$ is odd under the ph-transformation, it picks up an additional minus sign after transposing it, because it is an antisymmetric representation.

It is worth briefly mentioning how the paramagnetic solution is recovered, when $T> T_N$ and both spin and translational symmetries are fully restored. Here, one must pay additional attention w.r.t. the FM case:  Expressing the resulting susceptibilities in terms of the components listed in Table \ref{tab:longitudinal_AF} yields an apparent doubling of the surviving correlators in the $SU(2)$-symmetric case: $\chi^{\rho\rho}$,   $\chi^{\bar{\rho}\bar{\rho}}$, $\chi^{zz}$,$\chi^{\bar{z}\bar{z}}$. 
This simply reflects the mismatch between the RBZ adopted in the AF phase and the conventional BZ exploited to express the susceptibilities in the conventional charge/magnetic sectors.
Specifically, the magnetic/charge susceptibility of the paramagnetic phase as function of the crystal momentum ${\bQ}$ of the full BZ are recovered as:
\begin{eqnarray}
\chi_{m}(\bQ) &=& \left\{
\begin{array}{cc}
\chi^{zz}(\bQ),& \mbox{ if } \bQ \in RBZ \\ \\
\chi^{\bar{z}\bar{z}}(\bQ-\boldsymbol{\Pi}),& \mbox{ otherwise } 
\end{array}\right. , \nonumber \\   \\
\chi_{c}(\bQ) &=& \left\{
\begin{array}{cc}
\chi^{\rho\rho}(\bQ),& \mbox{ if } \bQ \in RBZ \\ \\
\chi^{\bar{\rho}\bar{\rho}}(\bQ-\boldsymbol{\Pi}),& \mbox{ otherwise } 
\end{array}\right. .\nonumber
\end{eqnarray} 
\\
\par

Turning to the explicit expression of the BSE, we note that, as in the FM case, the spin component along the $z$-axis remains conserved. Therefore, we can use the same block-wise spinorial representation introduced in the previous subsection. In addition, we have to consider the sub-lattice indices of the AF. The BSE of the longitudinal and transverse spin sectors read:
\begin{widetext}
\begin{eqnarray}\label{BSE:AF:para}
\mathcal{F}_{cd}^{ab}|_{\sigma\sigma^\prime}(kk^\prime q)&=& {\Gamma}_{cd}^{ab}|_{\sigma\sigma^\prime}(kk^\prime q)
-\frac{1}{(V\beta)^2}\sum_{k_1,k_2}\sum_{\sigma_1\sigma_2}\sum_{a^\prime b^\prime c^\prime d^\prime}  {\Gamma}_{c^\prime d^\prime}^{a\,\,b}|_{\sigma\sigma_1}(kk_1\,q) {\chi}^{\ d^\prime\,c^\prime}_{0\,b^\prime\,a\prime}|_{\sigma_1\sigma_2}(k_1k_2\,q) \mathcal{F}_{c\,\,d}^{a^\prime b^\prime}|_{\sigma_2\sigma^\prime}(k_2k^\prime q), \\
\label{BSE:AF:perp}
\mathcal{F}_{cd}^{ab}|_{\overline{\sigma\bar{\sigma}}}(kk^\prime q)&=& {\Gamma}_{cd}^{ab}|_{\overline{\sigma\bar{\sigma}}}(kk^\prime q) 
-\frac{1}{(V\beta)^2}\sum_{k_1k_2}\sum_{a^\prime b^\prime c^\prime d^\prime}  {\Gamma}_{c^\prime d^\prime}^{a\,\,b}|_{\overline{\sigma\bar{\sigma}}}(kk_1\,q) {\chi}^{\hspace{0.3cm}d^\prime\,c^\prime}_{0,\,b^\prime\,a\prime}|_{\overline{\sigma\bar{\sigma}}}(k_1k_2\,q) \mathcal{F}_{c\,\,d}^{a^\prime b^\prime}|_{\overline{\sigma\bar{\sigma}}}(k_2 k^\prime q),
\end{eqnarray}
\end{widetext}
where Latin letters refer to sub-lattice indices, the summations over momenta are restricted to the MBZ, and 
\begin{eqnarray}
\chi_{0, cd}^{\ \  ab}|_{\sigma\sigma^\prime}(kk^\prime q) &\equiv& -(\beta V)\delta_{kk^\prime}\delta_{\sigma\sigma^\prime}\,G_\sigma^{bc}(k)\,G_{\sigma}^{da}(k+q),\nonumber \\
\chi_{0, cd}^{\ \  ab}|_{\overline{\sigma\bar{\sigma}}}(kk^\prime q)& \equiv &-(\beta V)\delta_{kk^\prime}\,G_\sigma^{bc}(k)\,G_{\bar{\sigma}}^{da}(k+q)\,.
\end{eqnarray}
Eventually, we note that Eqs.(\ref{BSE:AF:para},\ref{BSE:AF:perp}) correspond to two independent systems of coupled linear equations with $16\times 2 \times N^2$  and $16\times  N^2$ unknowns respectively, where $N$ is the size of the Matsubara frequency box.

 \section{Vertex functions in $d = \infty$}\label{sec:inf_dim}
 
 The core idea of the D$\Gamma$A\cite{Toschi2007,Rohringer2018} is to generalize the DMFT approximation at the two-particle level:
 D$\Gamma$A, hence, lifts the DMFT assumption of pure locality of all 1PI irreducible (skeleton) diagrams, allowing for a non-local self-energy, but it
  keeps the very same locality conditions of DMFT for the two-particle vertex functions. 
 
 As stated in the literature\cite{Georges1996},  by taking the $d \rightarrow \infty$  (DMFT) limit\cite{Metzner1989}  all the fully 2PI vertex diagrams ($\Lambda$) become completely {\sl local}.   
 The vertex irreducible in a specific channel $r$ ($\Gamma_r$) displays, instead, a residual momentum dependence, but  {\sl only} for special {\bf k}-points [such as $(0,0,0,0, \ldots)$,  $(\pi,\pi,\pi,\pi, \ldots), $ etc.] whose relative measure in the Brillouin zone scales to zero for  $d \rightarrow \infty$. 
 As a result of this peculiar momentum-dependence, in DMFT {\sl only} the {\sl purely local} part of $\Gamma$ contributes to the physical susceptibilities/collective modes, once these are computed by performing the internal momentum integrals of the corresponding BSE. 
 
In this section, we will demonstrate how the DMFT locality of the vertex functions $\Lambda$ and $\Gamma$ is generalized to FM- and AF-ordered phases, taking explicitly into account, for the latter, the corresponding doubling of the unit cell.
The very same locality assumption for the two vertex classes ($\Lambda$ and $\Gamma$) will be made, then, for the D$\Gamma$A of the magnetic phases in its full parquet or in its ladder version, respectively. 

We exploit this occasion also to illustrate the $d \rightarrow \infty$ scaling of two-particle vertex diagrams. In particular, we will show how the vertex $\Lambda$ truly collapses to a purely local quantity for $d \rightarrow \infty$ and discuss the specific residual momentum dependence of the remnant classes of diagrams. Such a detailed diagrammatic discussion, to best of our knowledge, is not explicitly addressed in the literature\cite{Georges1996,Zlatic1990}. We recall, nonetheless, the complementary derivations based on the 
local Baym-Kadanoff functional of Ref.~[\onlinecite{Janis1999}] and of Ref.~[\onlinecite{Hettler2000}].
 
 \subsection{Fully irreducible vertex}\label{sec:fully_dmft}
 
We start by considering the $d \rightarrow \infty$ limit of the most fundamental diagrammatic building-block on the two-particle level: the fully 2PI vertex $\Lambda$, defined by the subset of all vertex diagrams which cannot be split by cutting $2$ fermionic lines\cite{Bickers2004,Rohringer2012}.  We briefly recall that to study such limit one needs  to:  (i) properly rescale the hopping $t \rightarrow   \frac{t}{\sqrt{d}}$; (ii) consider the dimensional contributions of all summations on the lattice site-indexes required by the Fourier transform and/or internal index contractions (of order $d$, if a summation is performed on the nearest-neighboring sites); (iii) compare the contribution of purely local diagrams (of $\mathcal{O}(1)$ in $d \rightarrow \infty$) w.r.t. their first non-local corrections.

The  high-dimensional scaling of $\Lambda$ is controlled by the (compact) topology of its diagrams: One easily observes that each of the (four) incoming/outgoing external lines of the fully 2PI diagrams are connected to {\sl three} internal lines (otherwise the diagram would be two-particle reducible). Hence, any insertion of a neighboring site (e.g., $j \neq i$) in a purely local vertex diagrams for $\Lambda$  will scale -at least- as  $\frac{1}{\sqrt{d}}$ w.r.t. its purely local counterpart: Due to the diagram topology, the leading order corrections for $d \rightarrow \infty$  will originate from a {\sl triple} nearest-neighboring propagation (of ${\mathcal O}(\frac{1}{d^{3/2}}$)) and {\sl one} single sum of the next-neighboring sites (${\mathcal O}(d)$).

\begin{figure}
\includegraphics[width=0.4\textwidth]{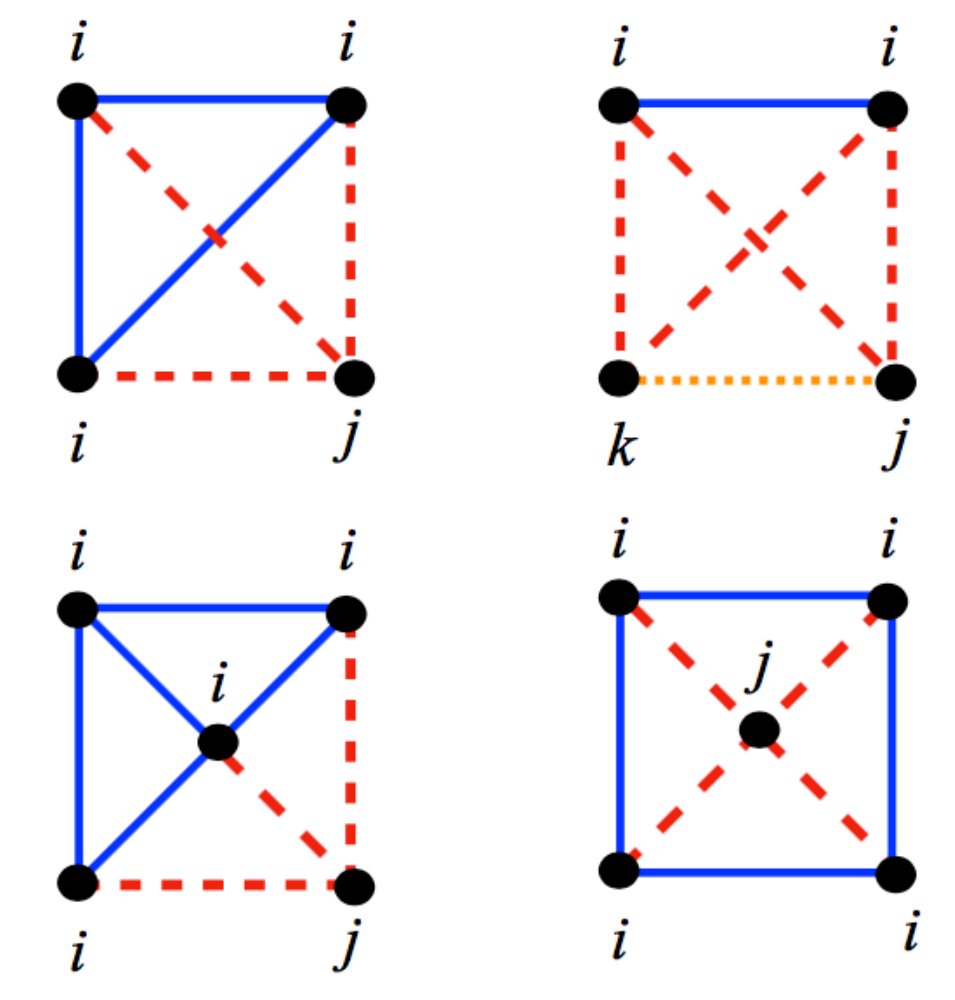}
\caption{Dimensional scaling properties of representative diagrams for the 2PI vertex $\Lambda$ (see text). The font of the different lines indicates the scaling of the corresponding propagation: {\sl blue solid} lines mark a local propagation [${\mathcal O}(1)$], {\sl red dashed} a propagation between nearest neighboring sites [${\mathcal O}(\frac{1}{d^{1/2}})$], and {\sl orange dotted} a propagation between (at least) next-to-nearest neighboring sites [at most of ${\mathcal O}(\frac{1}{d})$].}
\label{fig:Lambda}
\end{figure}

We illustrate the above-mentioned scaling properties by hands of a couple of  representative diagrams for $\Lambda$, taken from the lowest orders of its perturbative expansion: the envelope diagram (${\mathcal O}(U^4)$) and the envelope diagram with a ``seal" (${\mathcal O}(U^5)$), depicted in Fig.~\ref{fig:Lambda}.

In the first case (first diagram on the left), we immediately recognize how the leading-order correction ${\mathcal O}(\frac{1}{\sqrt{d}})$ arises from the insertion of a single nearest-neighboring site ($j \neq i$). We note here that {\sl no} other corrections to the corresponding local diagram can scale more slowly for $d \rightarrow \infty$:  (i) neither those arising by the Fourier summation over sites at larger distances (e.g., considering a next-to-nearest neighboring site for $j$  would  "cost" an additional $\frac{1}{\sqrt{d}}$ scaling factor overall), (ii) nor those including more neighboring sites ($k \neq j$) in the diagram. As for the latter case (shown in  Fig.~\ref{fig:Lambda} right), one exploits the property that if $j$ and $k$ are both nearest-neighbor of $i$, they cannot be also nearest-neighbors to each other, resulting in a doubled Manhattan distance and, thus, in a faster  $d \rightarrow \infty$-scaling of the propagation $j \leftrightarrow k$.

As for the second example (bottom panels of Fig.~\ref{fig:Lambda}), it is clear that if a neighboring site is inserted in one of the external vertices of the diagram, the very same considerations as above apply.
Instead, if a nearest neighboring vertex $j$ is inserted in the center of the diagram, the corresponding contribution displays an even faster scaling for $d \rightarrow \infty$, because one has {\sl four} non-local propagations  ${\mathcal O}(\frac{1}{d^2})$ and only {\sl one} internal summation (${\mathcal O}(d)$).
The same scaling consideration evidently applies to all higher-order contributions to $\Lambda$, where neighboring sites are added in the internal part of the diagrams. 

Hence, {\sl all} non-local corrections to any purely local 2PI vertex diagram becomes negligible in  $d \rightarrow \infty$: The fully 2PI vertex $\Lambda$ is thus {\sl purely  local}
in DMFT, in perfect analogy with the 1PI self-energy. As this result originates from the basic ``topological" properties of the 2PI diagram only, it is immediately applicable also to the SU(2)-symmetry-broken cases (FM or AF in bipartite lattices) relevant for our work. 
 We note, in passing, that the topologically ``compact" structure of the 2PI diagrams also ensures that  $\Lambda_{\uparrow \downarrow}$ decays to the bare interaction $U$ at high-frequencies in the {\sl whole} $(\nu,\nu^\prime,\omega)$-space\cite{Rohringer2012,Wentzell2020}.

 \subsection{The full scattering amplitude}
 
 The analysis of the high-dimensional scaling of the full scattering amplitude ${\mathcal F}$ requires a more detailed inspection. 
In fact, ${\mathcal F}$,  which is defined by the sum of all {\sl connected}  diagrams,  contains also all possible two-particle {\sl reducible} contributions in the different $ph$, $\overline{ph}$ and $pp$ channels. 
The ``topology"  of these terms is opposite to that of the 2PI vertex described above, as all elements of the subclasses of  reducible diagrams ($\Phi_{ph}$, $\Phi_{\overline{ph}}$ and $\Phi_{pp}$) entail diagrammatic structures, where pairs of external incoming/outgoing lines are coupled to a {\sl pair} of internal lines. For this reason, the leading order of non-local reducible diagrams with $i \neq j$ does {\sl not} decay faster than its local counterpart,  surviving in $d \rightarrow \infty$ limit. A clear example is provided by the second-order polarization diagrams, whose leading non-local contribution (e.g., with its second site $j$ taken as a nearest neighbor of the first one ($i$)),  is of the {\sl same} order as its purely local counterpart, not vanishing in $d \rightarrow \infty$.

These polarization diagrams are responsible for the residual momentum-dependence of the full scattering amplitude ($F$) in DMFT. In fact, all  momentum-dependent contributions of the two-particle vertices in DMFT originate from the internal bubbles present as building blocks of reducible diagrams, i.e.~the frequency/momentum convolution of two Green's functions (e.g.~${\mathcal B}(q) \! \propto \! \sum_{k_1} G(k_1) G(k_1+q)$) . Therefore, differently from above discussion on
$\Lambda$, specific information on the properties of the lattice, for which the limit of $d \rightarrow \infty$ is taken, is necessary to derive an explicit expression of $F$.

 \subsection{Hypercubic lattice - PM  case} 
 
 For an hypercubic lattice, in Ref.~\onlinecite{Georges1996}, it has been shown that $\Phi_{ph}$ only deviates from its local counterpart, if  the transfer momentum {\bf q} equals one of the special vectors mentioned before. Formally one has:  $\Phi_{\rm ph}^{\nu\nu^\prime\omega}({\bf q}, {\bf k}, {\bf k}^\prime) \rightarrow  \Phi_{\rm ph}^{\nu\nu^\prime\omega}(X({\bf q}))$ where, following  the notation of Ref.~\onlinecite{Georges1996}, we introduce the variable $X({\bf q}) = \frac{1}{d} \sum_{\alpha=1}^d \cos (q_\alpha)$ assuming non zero values only for special values of the momentum, such as  $X=1$ for {\bf q}= $(0,0,0,0, \ldots)$,  $X=-1$ for  $(\pi,\pi,\pi,\pi, \ldots)$, etc. Hence, for a generic values of the momentum, $X \equiv 0$, and $\Phi_{\rm ph}$ reduces to its purely local part, computable directly from the auxiliary AIM of DMFT [$\Phi_{\rm ph}(\nu, \nu^\prime \omega; X=0) = \Phi_{\rm ph}^{AIM}(\nu, \nu^\prime \omega)$]. 
  
 By generalizing their argument to other channels (in fact: to their  corresponding ``bubble terms")  and exploiting the results of Sec. IIIA, one can write the explicit parquet decomposition of the full scattering amplitude in DMFT as follows:
 \begin{eqnarray}
 {\mathcal F}^{\rm DMFT}(k,k^\prime,q)  =  \Lambda^{\rm AIM}(\nu,\nu^\prime, \omega) \! + \!  \Phi_{\rm ph}(\nu, \nu^\prime, \omega; X({\bf q})) \, + & & \nonumber \\  + \Phi_{\rm {\overline{ph}}}(\nu,\nu^\prime,\omega; X({\bf k}^\prime \! - \! {\bf k})) + \Phi_{\rm pp}(\nu, \nu^\prime, \omega; X({\bf k} \!+\! {\bf k^\prime} \! + \! {\bf q})), & &  \nonumber\\ 
 & &
 \label{eqn::fdmft}
 \end{eqnarray}
where the fully local 2PI vertex $\Lambda$ can be extracted directly from the inverse parquet equation of the auxiliary AIM of DMFT, and {\sl all} reducible components $\Phi$ depends only on the transfer momentum in the corresponding channel\cite{Georges1996,Janis1999}, through the function $X$. 
Due to Eq.~\ref{eqn::fdmft}, whose validity is discussed in the appendix by hands of representative diagrams, the calculations of the BSE in DMFT gets considerably simplified.
In particular, any irreducible vertex in a specific channel only depends on the transfer momenta of the {\sl other} channels. As an example, for the (longitudinal) ph channel, one has:
\begin{widetext}
\begin{eqnarray}
\Gamma^{\rm DMFT}_{\rm ph}(k,k^\prime,q) & = & \Lambda^{\rm AIM}(\nu,\nu^\prime,\omega) + \Phi_{\rm \overline{ph}}(\nu, \nu^\prime, \omega ;  X({\bf k}^\prime -{\bf k})) + \Phi_{\rm pp}(\nu,\nu^\prime,  \omega; X({\bf k}+ {\bf k}^\prime  +{\bf q})) \nonumber\\
&  =  & \Gamma_{\rm ph}(\nu, \nu^\prime,\omega; X({\bf k}^\prime -{\bf k}), X({\bf k} + {\bf k^\prime} +{\bf q})) 
\label{eqn::gammadmft}
 \end{eqnarray}
 \end{widetext}
Analogous expressions hold for $\Gamma_{\rm \overline{ph}}, \Gamma_{\rm pp}$.  

When inserting any of these DMFT irreducible vertex functions $\Gamma$ in a lattice BSE, all terms  corresponding to special momentum realizations will {\sl not} contribute to the internal momentum summations (e.g., for the ph channel: over {\bf k} and {\bf k'}), because they are defined over a zero-measure subset of the $d=\infty$ Brillouin zone. Hence, when computing any physical susceptibility/collective modes in $d \rightarrow \infty$, the assumption of a {\sl full locality} of $\Gamma$ yields the exact result.
Hence, in ladder D$\Gamma$A, where one keeps the same locality assumption of the DMFT at the level of the BSE, one can compute the irreducible vertices in the channel of interest directly inverting the BSE of the auxiliary AIM: $\Gamma_{\rm ph}^{\rm D\Gamma A} \equiv \Gamma_{ \rm ph}^{\rm AIM}(\nu, \nu^\prime, \omega) = \left[\chi_{\rm ph} \right]^{-1}_{\nu \nu^\prime}(\omega) -    [\chi_{\rm ph}^0 ]^{-1}_{\nu \nu^\prime}(\omega)$, where $[\chi_{\rm ph}^0]_{\nu \nu^\prime}(\omega) = -\beta G_{\rm AIM}(\nu) G_{\rm AIM}(\nu+ \omega) \delta_{\nu \nu^\prime}$.
 The same arguments also apply to the irreducible vertex functions and the BSE of the FM case, since the FM-ordering is not associated to any change of the PM-BZ.

 \subsection{Hypercubic lattice - AF case}\label{sec:hyper_dmft}

 In the AF case, where the unit cell doubles, every Green's and vertex function acquire an explicit dependence on the two inequivalent sub-lattice $A$, $B$. This modification does not affect, in any respect, the general arguments given for the fully 2PI vertex $\Lambda$ in Sec. \ref{sec:fully_dmft}, as those do not rely on specific details of their underlying lattices. Hence, also in the AF, the 2PI vertex of DMFT remains fully local in spatial coordinates, which means:
 \begin{equation}
 \Lambda^{ab}_{cd}(k,k',q) = \Lambda_{a}(\nu,\nu^\prime,\omega) \; \delta_{ab} \, \delta_{bc} \, \delta_{cd} \,,
 \label{eqn::lambdaAF}
\end{equation}
where we have used the compact notation $\Lambda_a \equiv \Lambda^{aa}_{aa}$.

The generalization to the AF case is less obvious for the corresponding reducible contribution ($\Phi$'s).
In fact, the structure of the two particle diagrams is more complex than in the PM case, since a further dependence on sub-lattice indices arises. 
However, the inspection of the relevant diagrams presented in App.~C allows to generalize the Eqs.~(\ref{eqn::fdmft})-(\ref{eqn::gammadmft}) to the AF case:
\begin{widetext}
 \begin{eqnarray}
 {\mathcal F}^{ab}_{cd}(k,k^\prime,q) &  =  & \Lambda_{a}(\nu,\nu^\prime, \omega) \,  \delta_{ab} \delta_{bc}\delta_{cd} + \Phi_{{\rm ph} \, ac}^{\phantom{aa}}(\nu, \nu^\prime ,\omega; X({\bf q})) \, \delta_{ab} \delta_{cd} \nonumber  \\
  & + &  \Phi_{{\rm \overline{ph}} \, ab}^{\phantom{ab}\, }(\nu,\nu^\prime, \omega; X({\bf k^\prime} -{\bf k})) \, \delta_{ad} \delta_{bc} + \Phi_{{\rm pp} \,  ab}^{\phantom{ab} }(\nu, \nu^\prime, \omega; X({\bf k} + {\bf k^\prime}  + {\bf q}))\, \delta_{ac} \delta_{bd} ,
 \label{eqn::fdmftAF}
 \end{eqnarray} 
\end{widetext}
where we used the following compact notation for the sub-lattice indices: $\Phi_{ph\, ab} \equiv \Phi^{\phantom{{aa}}\,aa}_{ph \,bb} $, $\Phi_{\overline{ph}\, ab} \equiv \Phi^{\phantom{{ab}}\,ab}_{\overline{ph} \,ba} $, $\Phi_{pp\, ab} \equiv \Phi^{\phantom{{aa}}\,ab}_{pp \,ab} $.
From Eq.~(\ref{eqn::fdmftAF}), which is the generalization of the  parquet equation to the AF case, the corresponding irreducible vertex in the different channels is derived:
\begin{widetext}
 \begin{equation}
 \Gamma_{{\rm ph} \, ab}^{\phantom{ph}\, {cd}}(k,k^\prime,q)   = \Lambda_{a}(\nu,\nu^\prime, \omega) \,  \delta_{ab} \delta_{bc}\delta_{cd} + 
     \Phi_{{\rm \overline{ph}} \, ab}^{\phantom{ab}\, }(\nu,\nu^\prime, \omega; X({\bf k^\prime} -{\bf k})) \, \delta_{ad} \delta_{bc} + \Phi_{{\rm pp} \,  ab}^{\phantom{ab} }(\nu, \nu^\prime, \omega; X({\bf k} + {\bf k^\prime}  +  {\bf q}))\, \delta_{ac} \delta_{bd} 
 \label{eqn::gdmftAF}
 \end{equation} 
\end{widetext}
Similarly as in the PM case, the residual non-local structure of the $\Phi$'s survives only along the special lines. For instance,  one finds  $\Phi_{ph\,ab}(\nu,\nu^\prime,\omega,X) =  \phi^{(1)}_{a}(\nu,\nu^\prime,\omega,X) \,\delta_{ab} + \phi^{(2)}(\nu,\nu^\prime,\omega,X)\,\delta_{a\bar{a}}$, and when $\bq$ is a generic point, $\phi_a^{(1)}(\nu,\nu^\prime,\omega,0) = \Phi_{ph\,a}^{AIM}(\nu,\nu^\prime,\omega)$ while $\phi^{(2)}(\nu,\nu^\prime,\omega,0) = 0$.
Eventually, by inserting these expressions in the BSE for the physical response functions/collective modes, one gets that, also in the presence of an AF magnetic order, the only not-vanishing contribution of $  \Gamma^{ab}_{cd}(k,k^\prime,q)$ for $d \rightarrow \infty$ is originated by its purely local part:

 \begin{equation}
\Gamma_{{\rm ph} \, ab}^{\phantom{ph} \, {cd}}(k,k^\prime,q) = \Gamma_{{\rm ph}\, a}^{{\rm AIM} \, } (\nu,\nu^\prime, \omega) \,  \delta_{ab} \delta_{bc}\delta_{cd} 
\label{eqn::gammaAFloc}
 \end{equation}

 Such a locality condition will be, eventually, exploited for defining the ladder D$\Gamma$A equations for the AF ordered phase.  As we will see in the following, this allows for considerable simplifications in treating the ladder equations in the broken-symmetry cases.

 \section{The D$\Gamma$A expressions}\label{sec:dga}
 
We will now explicitly derive the equations for the ladder D$\Gamma$A in the broken symmetry (FM and AF) phases, exploiting the properties of the DMFT vertex functions illustrated in the previous section.
We stress that all analytical expressions reported below are valid for any ladder approximation based on the {\sl locality} of the 2PI vertex $\Gamma$ in a given channel including, among others, also the basic case of the RPA.

Specifically, we will discuss: (i) how to extract the input of the ladder D$\Gamma$A equations from a DMFT solution in a broken symmetry (FM or AF) phase; (ii) which expressions must be used to compute the corresponding momentum dependent-response functions, through the lattice BSE of DMFT\cite{Georges1996,KrienThesis,stepanov2018}, and, eventually; (iii) how the D$\Gamma$A, momentum dependent self-energy is obtained through the corresponding Schwinger-Dyson equation.
 
 \subsection{Irreducible vertex functions}
 
In ladder D$\Gamma$A, all 2PI vertex functions entering in the BSE coincide to those of DMFT and, hence, according to the results of Sec.\ref{sec:inf_dim}
, can be approximated to their purely local counterpart

This allow for several algorithmic simplifications, the first of which is, evidently, the possibility to extract the input\footnote{It has been found\cite{Schaefer2013,Janis2014,Schaefer2016c,Ribic2016,Thunstroem2018,Chalupa2018,Springer2020} that the 2PI vertex functions of an AIM and/or a DMFT solution display multiple divergences in intermediate-to-strong coupling regime, explainable as non-perturbative footprints\cite{Chalupa2021} of the formation and the screening of local moments onto the other scattering sectors\cite{Gunnarsson2017,Springer2020,Reitner2020}. However, while their occurrence might have a relevant impact\cite{Kozik2015,Stan2015,Gunnarsson2016,Gunnarsson2017,Tarantino2018,Vucicevic2018} on several advanced many-electron schemes, it does not directly affect the ladder D$\Gamma$A approach, since all ladder-diagram resummations based on local 2PI vertices can be equivalently rewritten\cite{Rohringer2013,Rohringer2013a,Rohringer2018,vanLoon2020} in terms of the full local vertex $F$ (which diverges only in correspondence of a phase-transition\cite{Schaefer2016c}) and of the purely non-local Green's function $G({\bf k},\nu) - G_{AIM}(\nu)$. This procedure is applicable\cite{KrienThesis} in the 
broken symmetry phases, too.}
 for computing the D$\Gamma$A ladder directly from the auxiliary AIM associated to the DMFT solution in the broken-symmetry phase.  
In practice, {\sl both} for the FM and the AF case, the 2PI vertex in the $ph$ sector can be extracted by inverting the BSE for the impurity site (``$A$") of the auxiliary AIM in the longitudinal channel: 
\begin{equation}
\left(
\begin{array}{cc}
\Gamma^{A}_{\uparrow\uparrow}&\Gamma^{A}_{\uparrow\downarrow}\\ \\
\Gamma^{A}_{\downarrow\uparrow}&\Gamma^{A}_{\downarrow\downarrow}\end{array} 
\right) =  \left(
\begin{array}{cc}
\chi^{A}_{\uparrow\uparrow}&\chi^{A}_{\uparrow\downarrow}\\ \\
\chi^{A}_{\downarrow\uparrow}&\chi^{A}_{\downarrow\downarrow}\end{array} 
\right)_{\rm AIM}^{-1} \! - \left(
\begin{array}{cc}
\chi^{A,0}_{\uparrow\uparrow}&0\\ \\
0&\chi^{A,0}_{\downarrow\downarrow}\end{array} 
\right)_{\rm AIM}^{-1}
\label{eqn::gamma-aim}
\end{equation}
and in the transverse channel:
\begin{equation}
\left(
\begin{array}{cc}
\Gamma^{A}_{\overline{\uparrow\downarrow}}&0\\ \\
0&\Gamma^{A}_{\overline{\downarrow\uparrow}}\end{array} 
\right) =  \left(
\begin{array}{cc}
\chi^{A}_{\overline{\uparrow\downarrow}}&0\\ \\
0&\chi^{A}_{\overline{\downarrow\uparrow}}\end{array} 
\right)_{\rm AIM}^{-1} \! - \left(
\begin{array}{cc}
\chi^{A,0}_{\uparrow\downarrow}&0\\ \\
0&\chi^{A,0}_{\downarrow\uparrow}\end{array} 
\right)_{\rm AIM}^{-1}
\label{eqn::gamma-aim_tr}
\end{equation}
where $\chi^{A,0}_{\sigma\sigma^\prime} = -\beta G^{AA, AIM}_{\sigma} G^{AA, AIM}_{\sigma^\prime}$ defines the corresponding bubble term.
 The analogous expression for the $B$ site does not require further calculations, as it can be directly obtained by flipping the corresponding spin directions $\Gamma^B_{\sigma, \sigma^\prime} = \Gamma^{A}_{\bar{\sigma}, \bar{\sigma}^\prime} $, $\Gamma^B_{\overline{\sigma, \bar{\sigma}}} = \Gamma^{A}_{\overline{\bar{\sigma}, {\sigma}}} $. 
 \par The same applies to the FM case, where one has just to drop the extra sub-lattice indices. 
 
 While hitherto almost all vertex calculations within DMFT have been performed for SU(2)-symmetric situations, extending them to magnetically ordered phases, in order to extract the vertex input for the D$\Gamma$A will be certainly possible:
 First Extended-DMFT\cite{PhysRevB.52.10295,Si1996} calculations of the full local vertex $F$ in the AF-ordered phase of the extended Hubbard model have been presented in Ref.~\onlinecite{stepanov2018}.
 \subsection{BSE equations}
 As already disclosed in Eq.~(\ref{eqn::gamma-aim}), the BSE within the ladder D$\Gamma$A get simplified in several ways. 
 \\
 \noindent
 {\sl Ferromagnet -} 
The D$\Gamma$A BSE of a FM look formally similar to the exact ones except for the prescription of considering Eq.~(\ref{BSE:ferro:para},\ref{BSE:ferro:perp}) as a spinorial block-wise equation in the space of the Matsubara frequencies rather than in the four-momentum space, to substitute the factor $1/(V\beta)\to1/\beta$ and $\chi_{0\,\sigma\sigma^\prime} \to -\delta_{\nu\nu^\prime}\frac{\beta}{V}\sum_{\bk}G_\sigma(k)G_{\sigma^\prime}(k + q)$.

Differently from the PM case, it is no longer possible to perform the spin diagonalization\cite{Bickers2004} of the BSE, due to the intrinsic interdependence of spin and the charge degrees of freedom in the magnetic phase (cf. Table \ref{tab:ferro}).
\par
{\sl Antiferromagnet -} 
The local assumption made at the level of 2PI vertex ($\Gamma^{ab}_{cd} = \delta_{ab}\delta_{cd}\delta_{ac}\Gamma_a$) implies for a ladder approximation that ${\mathcal F}^{ab}_{cd} = \delta_{ab}\delta_{cd} F^{ab}$, as it can be seen from Eq.~(\ref{eqn::fdmftAF}).
This approximation, translated in the physical basis, corresponds to neglect a specific class of correlators, namely those involving $\tau^x$ and $\tau^y$ acting in the sub-lattice space. 
 The list of the non-trivial correlators surviving in such a ladder approximation is given in Tables \ref{tab:longitudinal_AF},\ref{tab:transverse_AF}.

Within this approximation the BSEs take the following form:
\begin{widetext}
\begin{eqnarray}\label{BSE:AF:ladder:para}
{F}^{ab}_{\sigma\sigma^\prime}(kk^\prime q)&=& {\Gamma}^{a}_{\sigma\sigma^\prime}(kk^\prime q)\delta_{ab}
-\frac{1}{(V\beta)^2}\sum_{k_1,k_2}\sum_{\sigma_1\sigma_2}\sum_{cd} {\Gamma}^{a}_{\sigma\sigma_1}(kk_1\,q)\delta_{ac}\, {\chi}^{\ \ c\, d}_{0,\sigma_1\sigma_2}(k_1k_2\,q) {F}^{d\,b}_{\sigma_2\sigma^\prime}(k_2k^\prime q), \\
\label{BSE:AF:ladder:perp}
{F}^{ab}_{\overline{\sigma\bar{\sigma}}}(kk^\prime q)&=& {\Gamma}^{a}_{\overline{\sigma\bar{\sigma}}}(kk^\prime q)\delta_{ab}
-\frac{1}{(V\beta)^2}\sum_{k_1,k_2}\sum_{cd} {\Gamma}^{a}_{\overline{\sigma\bar{\sigma}}}(kk_1\,q)\delta_{ac}\, {\chi}^{\ \ c\, d}_{0,\overline{\sigma\bar{\sigma}}}(k_1k_2\,q) {F}^{d\,b}_{\overline{\sigma\bar{\sigma}}}(k_2k^\prime q), 
\end{eqnarray}  
\end{widetext}
where ${\chi}^{\ \ a\, b}_{0,\sigma\sigma^\prime}(kk^\prime\,q) \equiv  \chi_{0, \, bb}^{\ \  aa}|_{\sigma\sigma^\prime}(kk^\prime q) $ and ${\chi}^{\ \ a\, b}_{0,\overline{\sigma\bar{\sigma}}}(kk^\prime\,q) \equiv  \chi_{0, \, bb}^{\ \  aa}|_{\overline{\sigma\bar{\sigma}}}(kk^\prime q) $.
We note that the BSE in Eq.(\ref{BSE:AF:ladder:para}) can be now expressed using $4\!\times\!4$ matrices in the spin-orbital indices $\alpha = (a,\sigma)$, and Eq.(\ref{BSE:AF:ladder:perp}) using $2\!\times\!2$ matrices in the sub-lattice index:
\begin{eqnarray}\label{eq:BSE_long_1}
\bar{\bar{F}}_{\parallel}^{\,kk^\prime q} &=& \bar{\bar{\Gamma}}_{\parallel}^{\,kk^\prime q} -\frac{1}{(V\beta)^2}\sum_{k_1k_2}\bar{\bar{\Gamma}}_\parallel^{\,k k_1 q}\cdot \bar{\bar{\chi}}_{0,\parallel}^{\,k_1 k_2 \,q}\cdot  \bar{\bar{F}}_{\parallel}^{\,k_2 k^\prime q}, \nonumber \\ \\
\label{eq:BSE_transv}
\bar{\bar{F}}_{{\overline{\sigma\bar{\sigma}}}}^{\,kk^\prime q} &=& \bar{\bar{\Gamma}}_{{\overline{\sigma\bar{\sigma}}}}^{\,kk^\prime q} -\frac{1}{(V\beta)^2}\sum_{k_1k_2}\bar{\bar{\Gamma}}_{\overline{\sigma\bar{\sigma}}}^{\,k k_1 q}\cdot \bar{\bar{\chi}}_{0,{\overline{\sigma\bar{\sigma}}}}^{\,k_1 k_2 \,q}\cdot  \bar{\bar{F}}_{{\overline{\sigma\bar{\sigma}}}}^{\,k_2 k^\prime q}, \nonumber \\
\end{eqnarray}
where $\bar{\bar{F}}_{\parallel},\,\bar{\bar{\Gamma}}_{\parallel}$ and $\bar{\bar{\chi}}_{0,\parallel}$ are $4\!\times\!4$ matrices whose explicit expression is given in Appendix \ref{app:matrices}, 
and:
\begin{eqnarray}
\bar{\bar{F}}_{{\overline{\sigma\bar{\sigma}}}}^{kk^\prime q} = 
\left(
\begin{array}{cc}
F^{AA}_{\overline{\sigma\bar{\sigma}}}&F^{AB}_{\overline{\sigma\bar{\sigma}}}\\ \\
F^{BA}_{\overline{\sigma\bar{\sigma}}}&F^{BB}_{\overline{\sigma\bar{\sigma}}}\end{array}
\right),\\
\bar{\bar{\Gamma}}_{{\overline{\sigma\bar{\sigma}}}}^{kk^\prime q} = 
\left(
\begin{array}{cc}
\Gamma^{AA}_{\overline{\sigma\bar{\sigma}}}&0\\ \\
0&\Gamma^{AA}_{\overline{\bar{\sigma}{\sigma}}}\end{array}
\right),\\
\bar{\bar{\chi}}_{0,{\overline{\sigma\bar{\sigma}}}}^{kk^\prime q} = 
\left(
\begin{array}{cc}
\chi^{AA}_{0,\overline{\sigma\bar{\sigma}}}&\chi^{AB}_{0,\overline{\sigma\bar{\sigma}}}\\ \\
\chi^{BA}_{0,\overline{\sigma\bar{\sigma}}}&\chi^{BB}_{0,\overline{\sigma\bar{\sigma}}}\end{array}
\right).  
\end{eqnarray}
While Eq.~(\ref{eq:BSE_transv}) is already expressed in a convenient basis, we can simplify the BSE in the longitudinal channel in Eq.~(\ref{eq:BSE_long_1}) by exploiting the symmetry of the problem.
In particular, we observe that all matrices in Eqs.~(\ref{matrices:AF:para1},\ref{matrices:AF:para2},\ref{matrices:AF:para3}) commute with the matrix $\sigma^x\otimes\tau^x$ and that once we rotate them using the unitary transformation constructed with the eigenvectors of $\sigma^x\otimes\tau^x$, we obtain a block-diagonal representation of all the three matrices that are split into six $2\!\times\!2$ matrices. Hence, this rotation represents a suitable basis to reduce the complexity of the BSEs in the spin-longitudinal channel. They get split into two independent channels that we call {\sl longitudinal-even} and {\sl longitudinal-odd}: 
\begin{eqnarray}\label{eq:BSE_long_odd_even}
\bar{\bar{F}}_{\parallel,\pm}^{\,kk^\prime q} &=& \bar{\bar{\Gamma}}_{\parallel,\pm}^{\,kk^\prime q} -\frac{1}{(V\beta)^2}\sum_{k_1k_2}\bar{\bar{\Gamma}}_{\parallel \pm}^{\,k k_1 q}\cdot \bar{\bar{\chi}}_{0,\parallel,\pm}^{\,k_1 k_2 \,q}\cdot  \bar{\bar{F}}_{\parallel,\pm}^{\,k_2 k^\prime q}, \nonumber \\
\end{eqnarray}
where we indicate with "+" the even sector, while with "-" the odd one and:
\begin{eqnarray}\label{eq:fpm}
\bar{\bar{F}}_{\parallel,\pm}^{kk^\prime q} &=& 
\left(
\begin{array}{cc}
F^{AA}_{\uparrow\uparrow}\pm F^{AB}_{\uparrow\downarrow}&F^{AA}_{\uparrow\downarrow}\pm F^{AB}_{\uparrow\uparrow}\\ \\
F^{AA}_{\downarrow\uparrow}\pm F^{AB}_{\downarrow\downarrow}&F^{AA}_{\downarrow\downarrow}\pm F^{AB}_{\downarrow\uparrow}\end{array}
\right),\\ 
\bar{\bar{\chi}}^{kk^\prime q}_{0,\parallel,\pm}  &=& \left(
\begin{array}{cc}
\chi^{AA}_{0,\uparrow\uparrow}&\pm \chi^{AB}_{0,\uparrow\uparrow}\\ \\
\pm \chi^{AB}_{0,\downarrow\downarrow}&\chi^{AA}_{0,\downarrow\downarrow}\end{array}
\right), \\
\bar{\bar{\Gamma}}^{kk^\prime q}_{\parallel,\pm}  &=&\left(
\begin{array}{cc}
\Gamma^{A}_{\uparrow\uparrow}&\Gamma^{A}_{\uparrow\downarrow}\\ \\
\Gamma^{A}_{\downarrow\uparrow}&\Gamma^{A}_{\downarrow\downarrow}\end{array}
\right).\hspace{3.5cm}\end{eqnarray}
By inspection of Table \ref{tab:longitudinal_AF}, we can see that the elements of the matrix in Eq.(\ref{eq:fpm}) are actually given by linear combinations of terms respectively in the even and odd longitudinal sectors. 
\par
We refer to this procedure to simplify the BSE of the AF-phase into a block-diagonal representation as ``spin-orbital diagonalization'', in analogy with the spin diagonalization of the paramagnetic case\cite{Bickers2004}, where the two independent channels are the charge and the spin one.

\subsection{Equation of motion}

 The final step for completing the ladder D$\Gamma$A formalism for the magnetically ordered phases is to derive the expression of the corresponding self-energy. This is done in D$\Gamma$A exploiting the Schwinger-Dyson (SD) equation, which relates the self-energy with the full scattering amplitude.

We report here the SD equation for a generic model with on-site density-density interactions $\hat{H}_{int} = \frac{1}{2}\sum_i U_{\alpha\beta}\,\hat{n}_{i\alpha}\hat{n}_{i\beta}$, where Greek letters represent generic spin-orbital indices:
\begin{eqnarray}
\Sigma^{\alpha\beta}_k  & \equiv  &  -\rho_{\alpha\beta}U_{\alpha\beta} + \delta_{\alpha\beta}\sum_\gamma\, U_{\alpha\gamma }\,\rho_{\gamma\gamma} +      \nonumber \\ 
 & - &  \frac{1}{(V\beta)^2}\sum_{k^\prime\,q}\sum_{\gamma\beta^\prime \gamma^\prime\delta^\prime}U_{\alpha\gamma}\,G^{\alpha\beta^\prime}_{k+q}\,\mathcal{F}^{\beta\,\beta^\prime}_{\gamma^\prime\delta^\prime}(kk^\prime q)\,G^{\gamma^\prime\gamma}_{k^\prime + q}G^{\gamma\delta^\prime}_{k^\prime}, \nonumber \\
\label{eqn:SDgeneric}
\end{eqnarray}
with $\rho_{\alpha\beta} \equiv  \frac{1}{V}\sum_i\left<\cd_{i\beta}c^{\ }_{i\alpha}\right>$ in the first (Hartree) term on r.h.s. of the expression.
For the specific single orbital Hubbard cases, which we are going to consider explicitly, the expression gets further simplified, since $U_{\alpha\beta} =U \delta_{\alpha\bar{\beta}}$. \\

\noindent
{\sl Ferromagnet} - In the case of a FM order  $G^{\alpha\beta}_k = \delta_{\alpha\beta}G^{\alpha}_k$, therefore the Hartree terms reads  $U\rho_{\bar{\alpha}\bar{\alpha}}\delta_{\alpha\beta} \equiv U   n_{\bar{\alpha}}\delta_{\alpha\beta}$ and the higher-order term: $\delta_{\alpha\beta}\frac{U}{(\beta V)^2}\sum_{k^\prime q}G_{k+q}^{\alpha} \, \mathcal{F}^{\alpha\alpha}_{\bar{\alpha}\bar{\alpha}}(kk^\prime q) G^{\bar{\alpha}}_{k^\prime + q} G^{\bar{\alpha}}_{k^\prime}$. 
The corresponding equation of motion for the self energy reads:
\begin{equation}\label{eq:mot:ferro}
\Sigma_\sigma(k) = Un_{\bar{\sigma}}-\frac{U}{(\beta V)^2}\sum_{k^\prime q}G_{k+q}^{\sigma}\,\mathcal{F}_{\sigma\bar{\sigma}}(kk^\prime q) G^{\bar{\sigma}}_{k^\prime + q} G^{\bar{\sigma}}_{k^\prime}.
\end{equation}
In order to obtain  a transparent expression for the scattering amplitude ${\cal F}$ for the ladder D$\Gamma$A, we  start\cite{Rohringer2016,DelRe2019} from its parquet decomposition\cite{Bickers2004,Gunnarsson2016}, i.e.:
\begin{equation}\label{parquet:exact}
\mathcal{F} = \Lambda + \Phi_{ph} + \Phi_{\overline{ph}} + \Phi_{pp},
\end{equation}
where $\Lambda$ is the 2PI vertex function and $\Phi_{ch}$, with $ch = \{ph,\overline{ph},pp\}$, represent scattering processes that are two particle reducible in the $ph,\overline{ph}, pp$ channels respectively, defined through the relation
\begin{equation}
  \mathcal{F}= \Phi_{r} + \Gamma_{r},  
\end{equation}
where $\Gamma_{r}$ is the 2PI vertex function in the given channel $r$.
Within D$\Gamma$A all 2PI-vertices are fully local, therefore $\Lambda \sim \Lambda^{loc}(\nu,\nu^\prime,\omega)$ is a function of the Matsubara frequencies only\cite{Toschi2007,Valli2015,Rohringer2018,Kauch2020}.
Within ladder-D$\Gamma$A\cite{Toschi2007,Katanin2009,Rohringer2016,Rohringer2018}, the same assumption applies {\sl also} to the 2PI vertices in all channels $\Gamma_{r}\sim\Gamma^{loc}_{r}(\nu\nu^\prime\omega)$.
It is useful, thus, to
introduce the auxiliary quantities ${F}_{r}= \Phi_{r} + \Gamma^{loc}_{r}$ which we will refer to, generically, as {\sl ``ladders"}, in the corresponding channel. 
Further,  for models with an on-site repulsion, we can also decide to neglect\cite{Toschi2007,Katanin2009,Rohringer2016,Rohringer2018} -as a further simplification- the non-local contributions in the particle-particle sector $\Phi_{pp}\sim\Phi^{loc}_{pp}(\nu\nu^\prime\omega)$.

Finally, exploiting the following crossing relation: $$\Phi_{\overline{ph},\,\sigma\bar{\sigma}}(kk^\prime q) = -\Phi_{ph,\,\overline{\sigma\bar{\sigma}}}(k,k+ q,k^\prime -k), $$ we can explicitly write the ladder D$\Gamma$A formula for the 1PI vertex function as defined in Eq.~(\ref{parquet:exact}),
\begin{widetext}
\begin{eqnarray}\label{1PI:ferro}
\mathcal{F}_{\sigma\bar{\sigma}}(kk^\prime q)&\sim &-\mathcal{F}^{loc}_{\sigma\bar{\sigma}}(\nu\nu^\prime \omega) + F_{\sigma\bar{\sigma}}(kk^\prime q) - F_{\overline{\sigma\bar{\sigma}}}(k,k+q,k^\prime - k) \nonumber \\
&=& -\mathcal{F}^{loc}_{\sigma\bar{\sigma}}(\nu\nu^\prime \omega)+ \frac{1}{2}\left[F^{\rho\rho} - F^{zz}+\sigma\left(F^{z\rho}-F^{\rho z}\right) \right](kk^\prime q)+
\frac{1}{2}\left[-2F^{xx}+ 2i\sigma F^{xy} \right](k,k+q,k^\prime -k),
\end{eqnarray}
\end{widetext}
where we have expressed the ladders in the physical basis, by inverting the relations in Table \ref{tab:ferro}, i.e. $F_{\sigma\bar{\sigma}} = \frac{1}{2}(F^{\rho\rho} -F^{zz} + \sigma F^{z\rho} -\sigma F^{\rho z})$ and $F_{\overline{\sigma\bar{\sigma}}} = F^{xx} -i\sigma F^{xy}$.
We can now substitute the approximated 1PI vertex function in Eq.~(\ref{1PI:ferro})  into Eq.~(\ref{eq:mot:ferro}) and we obtain the following SD explicit  expression for the self-energy: 
\begin{widetext}
\begin{eqnarray}
\Sigma_\sigma(k) -Un_{\bar{\sigma}}&\sim& \frac{U}{2V\beta^2}\sum_{q\nu^\prime} G^{\sigma}_{k+q}\,\chi^0_{\bar{\sigma}\bar{\sigma}}(q\nu^\prime)\left[F^{\rho\rho} - F^{zz}+\sigma(F^{z\rho}-F^{\rho z}  )\right](\nu\nu^\prime q) +\nonumber \\ 
&+& \frac{U}{2V\beta^2}\sum_{q\nu^\prime}G^{\bar{\sigma}}_{k+q}\,\chi^0_{\sigma\bar{\sigma}}(q\nu^\prime)\left[-2F^{xx}+ 2i\sigma F^{xy} \right](\nu\nu^\prime q) +
\nonumber \\
&-&\frac{U}{V\beta^2}\sum_{q\nu^\prime}G^{\sigma}_{k+q}\,\chi^0_{\bar{\sigma}\bar{\sigma}}\,\mathcal{F}^{loc}_{\sigma\bar{\sigma}}(\nu\nu^\prime q),
\end{eqnarray}
\end{widetext}
where we defined $\chi^0_{\sigma\sigma^\prime}(q\nu^\prime)\equiv -\frac{1}{V}\sum_{\bk^\prime} G^\sigma_{k^\prime}G^{\sigma^\prime}_{k^\prime + q}$. \\

{\sl Antiferromagnet -} In the case of an antiferromagnet (AF) in a bipartite lattice, the composite spin-orbital index is given by $\alpha =(a,\sigma) $, where $a$ and $\sigma$ represent the sub-lattice and spin indices respectively: 
 
\begin{table}
\centering
\begin{tabular}{|c|c|}
\hline
$\alpha$ & $(a,\sigma)$ \\ 
\hline
$\beta$ & $(b,\sigma^\prime)$ \\ 
\hline
$\gamma$ & $(c,\sigma_1)$ \\ 
\hline
$\beta^\prime$ & $(b^\prime,\sigma_2)$ \\ 
\hline
$\gamma^\prime$ & $(c^\prime,\sigma_3)$ \\ 
\hline
$\delta^\prime$ & $(d^\prime,\sigma_4)$ \\ 
\hline
\end{tabular}
\caption{Relation between indices expressed in the compact and extended notations.}
\label{tab:dictionary}
\end{table}

 Following the ``dictionary" between the compact spin-orbital  and the expanded notations we reported in Table~\ref{tab:dictionary},  we can write the self-energy in the AF case as:
\begin{widetext}
\begin{eqnarray}\label{self:AF}
\Sigma^{ab}_\sigma(k) - U\,\delta_{ab}\, n_{a,\bar{\sigma}}=- \frac{U}{(V\beta)^2}\sum_{k^\prime q}\sum_{b^\prime c^\prime d^\prime} G_\sigma^{ab^\prime}(k+q)\,\mathcal{F}^{bb^\prime}_{c^\prime d^\prime}|_{\sigma\bar{\sigma}}(kk^\prime q)\,G^{c^\prime a}_{\bar{\sigma}}(k^\prime + q) G_{\bar{\sigma}}^{ad^\prime}(k^\prime). \nonumber \\
\end{eqnarray}
\end{widetext}
Similarly as in the FM case, we can express $\mathcal{F}$ using its corresponding parquet decomposition, i.e. $\mathcal{F} = \Lambda + \Phi_{ph} + \Phi_{\overline{ph}} + \Phi_{pp}$. We note that for a generic set of generalized space-time/spin-orbital indices, the crossing relation $\Phi_{\overline{ph}}(1234) = -\Phi_{ph}(1432) $ holds, where $i = (x_i, \alpha_i)$, that in the case of the AF in Fourier space, reads:
\begin{equation}\label{eq:crossing} 
\Phi_{\overline{ph},{cd}}^{\hspace{0.4cm}ab}|_{\sigma\sigma^\prime}(kk^\prime q) = - \Phi_{{ph},{cb}}^{\hspace{0.4cm}ad}|_{\overline{\sigma\sigma^\prime}}(k,k+q, k^\prime - k).
\end{equation}

We can now perform a ladder approximation as done for the FM case. We recall, that in the AF case the sub-lattice index dependence of ladders is simplified with respect to the exact solution. In particular, the locality of $\Gamma$ in DMFT and ladder D$\Gamma$A implies that $F^{ab}_{cd}|_{\sigma\sigma^\prime}=\delta_{ab}\delta_{cd}F^{aa}_{cc}|_{\sigma\sigma^\prime}\equiv F^{ac}_{\sigma\sigma^\prime}\delta_{ab}\delta_{cd}$.
Under these assumptions, the 1PI vertex function $\mathcal{F}$ assumes the following form:
\begin{eqnarray}\label{eq:1PI_AF}
\mathcal{F}^{ab}_{cd}|_{\sigma\sigma^\prime}(kk^\prime q) &\sim& -\delta_{ab}\delta_{bc}\delta_{cd}\,\mathcal{F}^{loc}_{a,\sigma\sigma^\prime}(\nu\nu^\prime \omega) + \delta_{ab}\delta_{cd}\,F^{ab}_{\sigma\sigma^\prime}(kk^\prime q) +
 \nonumber \\  
&-&\delta_{ad}\delta_{cb}F^{ac}_{\overline{\sigma\sigma^\prime}}(k,k+q,k^\prime - k),
\end{eqnarray}
where we used the following exact local relation $\Lambda^a_{\sigma\sigma^\prime} + \Phi^a_{pp,\sigma\sigma^\prime} - \Gamma^a_{ph,\sigma\sigma^\prime}  -\Gamma^a_{\overline{ph},\sigma\sigma^\prime} = \mathcal{F}^{loc}_{a,\sigma\sigma^\prime}$

Using the expression of the 1PI vertex function evaluated in the ladder approximation in Eq.~(\ref{eq:1PI_AF}) and substituting it into Eq.~(\ref{self:AF}), we obtain the following expression for the self-energy: 
\begin{widetext}
\begin{eqnarray}\label{ladder:sigma:AF}
\Sigma^{ab}_{\sigma}(k)-\delta_{ab}Un_{a,\bar{\sigma}} &\sim&-\frac{U}{(V\beta)^2}\sum_{k^\prime q} \sum_c G^{ab}_\sigma(k+q)F^{bc}_{\sigma\bar{\sigma}}(kk^\prime q)G^{ac}_{\bar{\sigma}}(k^\prime)G_{\bar{\sigma}}^{ca}(k^\prime+q) \nonumber \\
&+&\frac{U}{(V\beta)^2}\sum_{k^\prime q} \sum_c G^{ab}_{\bar{\sigma}}(k+q)F^{bc}_{\overline{\sigma\bar{\sigma}}}(kk^\prime q)G^{ac}_{{\sigma}}(k^\prime)G_{\bar{\sigma}}^{ca}(k^\prime+q)  \nonumber \\
&+&\frac{U}{(V\beta)^2}\sum_{k^\prime q} G^{ab}_\sigma(k+q)\mathcal{F}^{loc}_{b,\sigma\bar{\sigma}}(\nu\nu^\prime\omega)G^{ab}_{\bar{\sigma}}(k^\prime)G_{\bar{\sigma}}^{ba}(k^\prime+q)\,. \nonumber \\
\end{eqnarray}
\end{widetext}
We note that in the expression of the self-energy calculated within the ladder approximation in Eq.~(\ref{ladder:sigma:AF}), three main contributions arise. In the first two, non-local terms belonging to the spin longitudinal channel ($F^{bc}_{\sigma\bar{\sigma}}$) and to the spin transverse channel  ($F^{bc}_{\overline{\sigma\bar{\sigma}}}$) appear and there is a in internal summation over the sub-lattice index. 

\begin{figure}
\begin{center}
\includegraphics[width=\columnwidth]{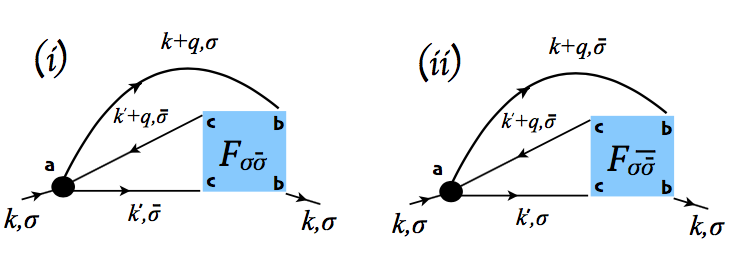} 
\caption{Diagrammatic representation of the non-local corrections to $\Sigma$ stemming from the fluctuations in longitudinal (i) and in the transverse (ii) channel within the ladder D$\Gamma$A for the AF phase.}
\label{fig::SD_AF}
\end{center}
\end{figure}

Conversely, in the third one, only local terms ($\mathcal{F}^{loc}_{b,\sigma\bar{\sigma}}$) are present and there is no internal summation over the sub-lattice index.  We show the diagrams corresponding to the longitudinal and transverse contributions in Fig.~\ref{fig::SD_AF}. 
Finally, it is worth noting, that it is possible to express the self-energy using the physical basis representation  for ladders by simply inverting the relations in Tables~ \ref{tab:longitudinal_AF},\ref{tab:transverse_AF}.

\section{AF-D$\Gamma$A results with a mean-field input}\label{sec:RPA-DGA}

To illustrate how the ladder-D$\Gamma$A equations derived in the previous sections work in practice, we present below a simplified, albeit fundamental application of our scheme  to AF-ordered phase of the $2D$-Hubbard model at $T=0$. 

Specifically, the approximated calculations of collective modes and the spectral properties presented in this section have been performed by evaluating all the D$\Gamma$A expressions for the AF phase (AF-D$\Gamma$A) starting from a {\sl static} mean-field input (instead of the DMFT one). 
Diagrammatically, this corresponds to retain the {\sl lowest} order contributions in $U$ for both the 1PI local self-energy and the 2PI local vertex appearing in the BSE and Schwinger-Dyson equations of the D$\Gamma$A for the AF-ordered system.

\begin{figure}[t!]
\includegraphics[width = 0.9\columnwidth]{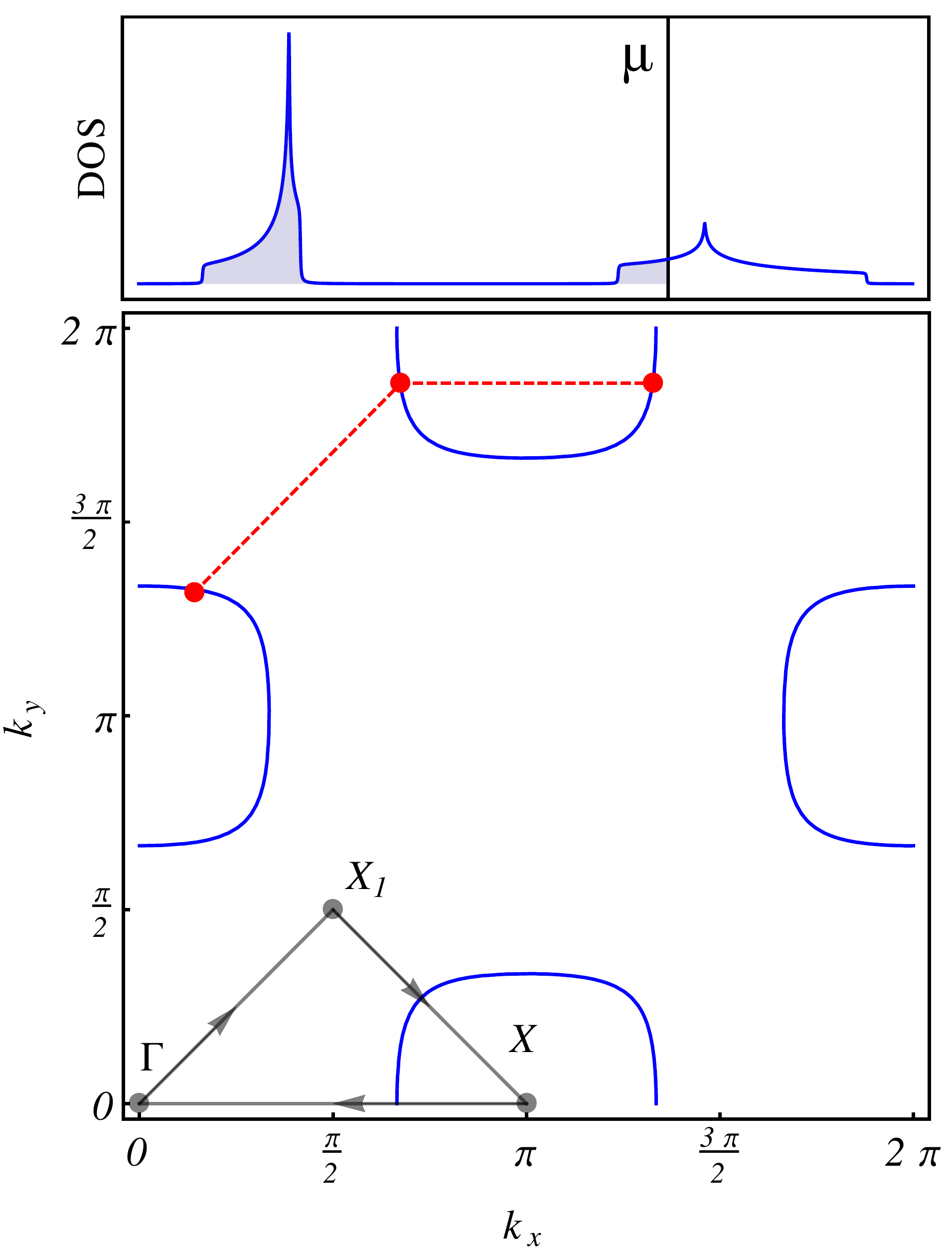}
\caption{Lower panel: Fermi surface of the mean-field solution of the AF-ordered
phase for the electron doped case (ii), with $n = 1.2$ and $t^\prime/t = 0.45$;  the path indicated by the gray arrows is the specific one exploited in the following Figures; the red dashed lines mark particle-hole excitations connecting the same/different pockets on the FS (s.~text). Upper panel: Corresponding density of states (DOS).   }
\label{fig:FS}
\end{figure}

\begin{figure*}
\centering
\includegraphics[width=0.3\textwidth]{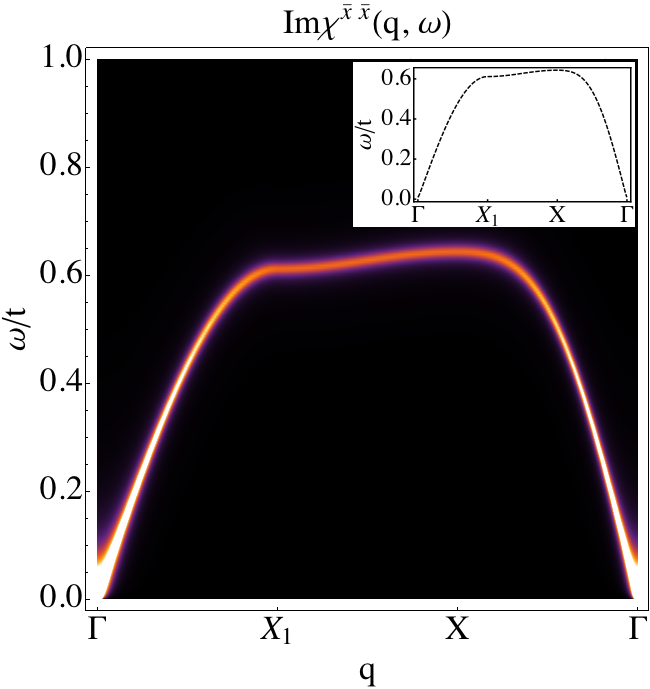} \hspace{4mm}
\includegraphics[width=0.3\textwidth]{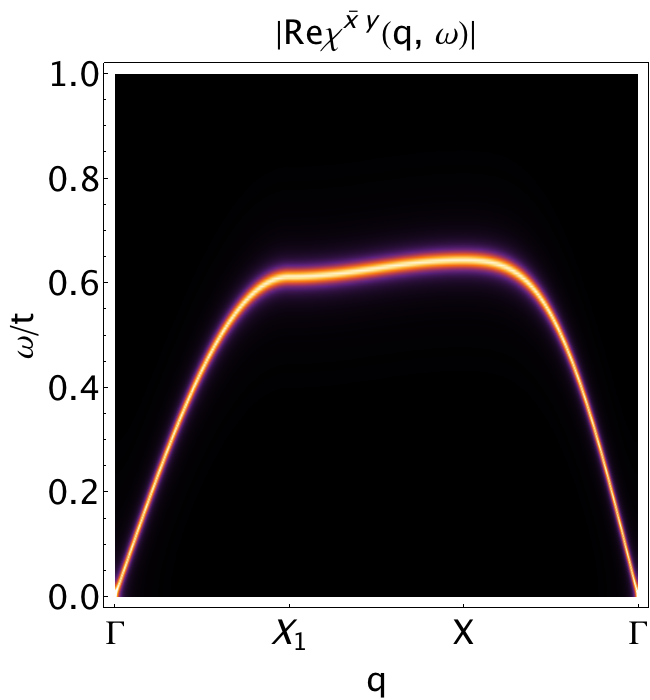} \hspace{4mm}
\includegraphics[width=0.3\textwidth]{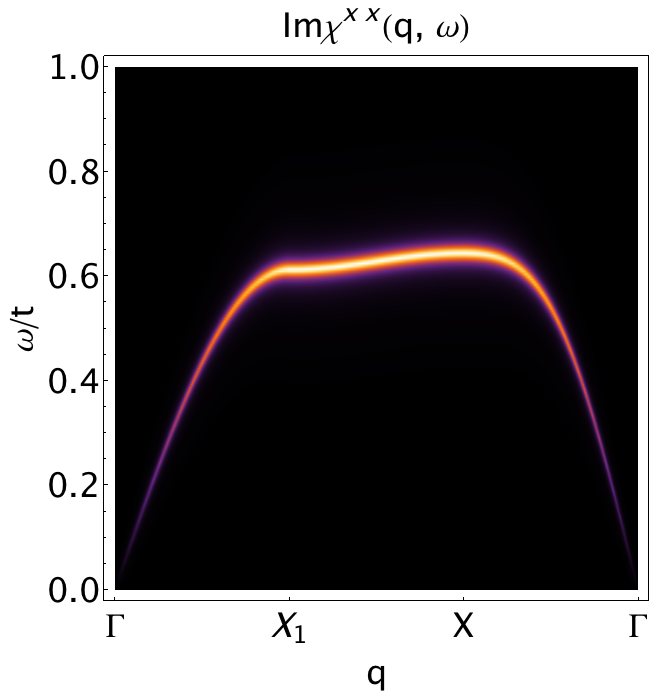}
\\
\centering
\includegraphics[width=0.3\textwidth]{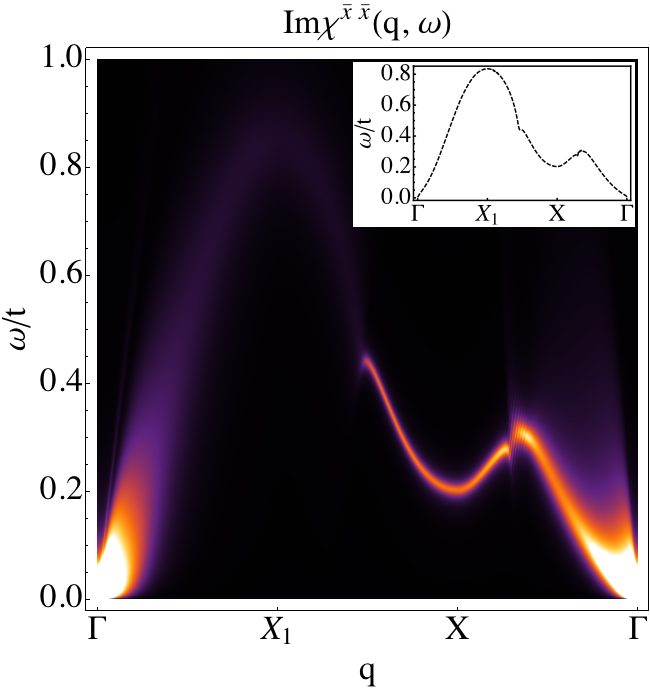}
\includegraphics[width=0.3\textwidth]{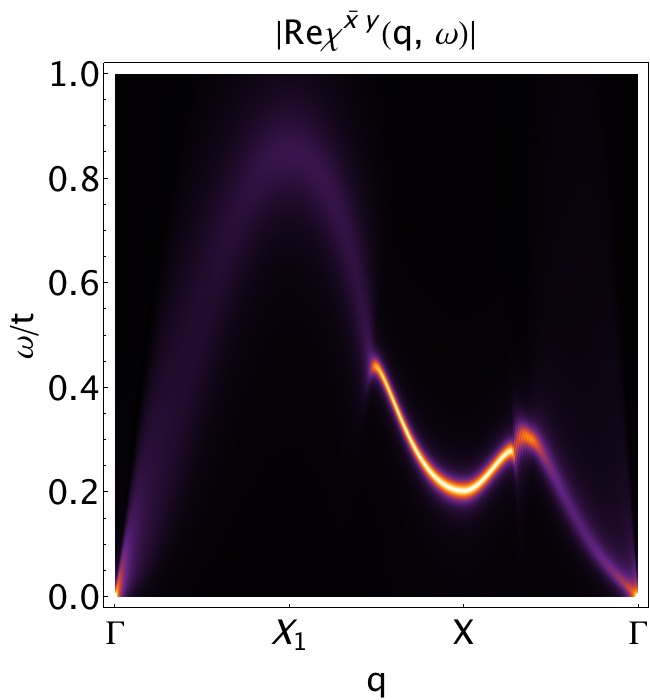} \hspace{4mm}
\includegraphics[width=0.3\textwidth]{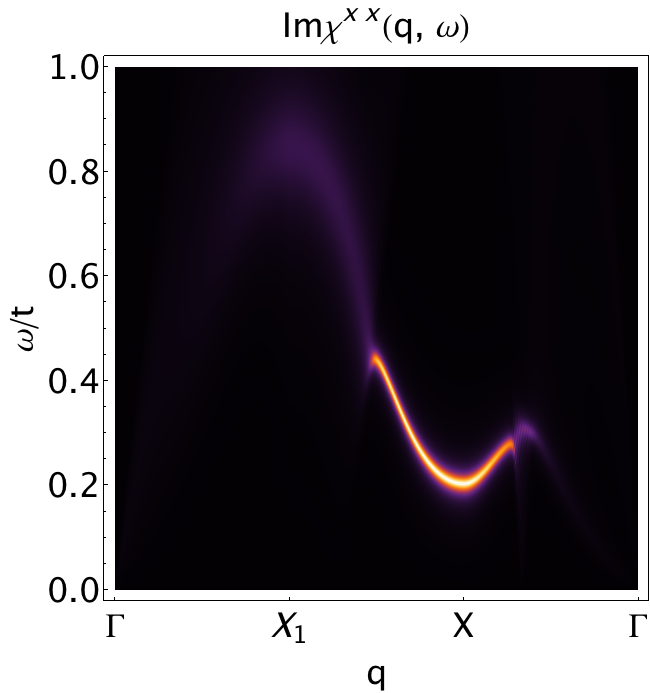}
\caption{Intensity plot of the imaginary/real (absorption) part of the transverse susceptibilities $\chi_{\bar{x}\bar{x}}$, $\chi_{\bar{x},y}$,$\chi_{xx}$ (in arbitrary units) of the AF phase as a function of momentum (along the path shown in Fig.~\ref{fig:FS}) and frequency computed at $T=0$ through  the RPA-expression Eq.~(\ref{eq:susc_transv_RPA}) for the particle-hole symmetric case (upper panels) and for the electron-doped case (lower panels). All expressions have been evaluated for a (retarded) frequency $z= \omega +i\delta$, with $\delta = 0.01$. The brightest colors refers to the highest intensity values, while the black colors indicates a vanishing intensity.  In the insets we show, as a guide to the eye, the dispersion relations of the corresponding magnons.}
\label{fig:tramod1}
\end{figure*}

\begin{figure}
\includegraphics[width=0.22\textwidth]{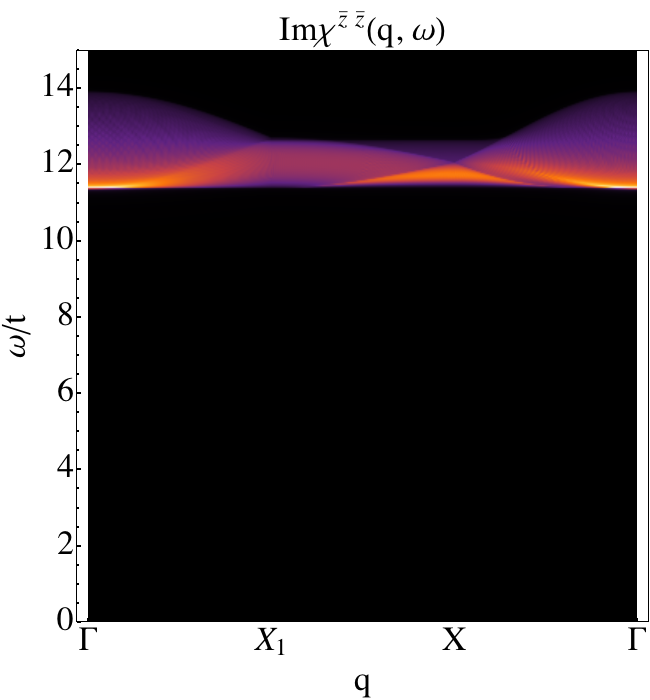}
\includegraphics[width=0.22\textwidth]{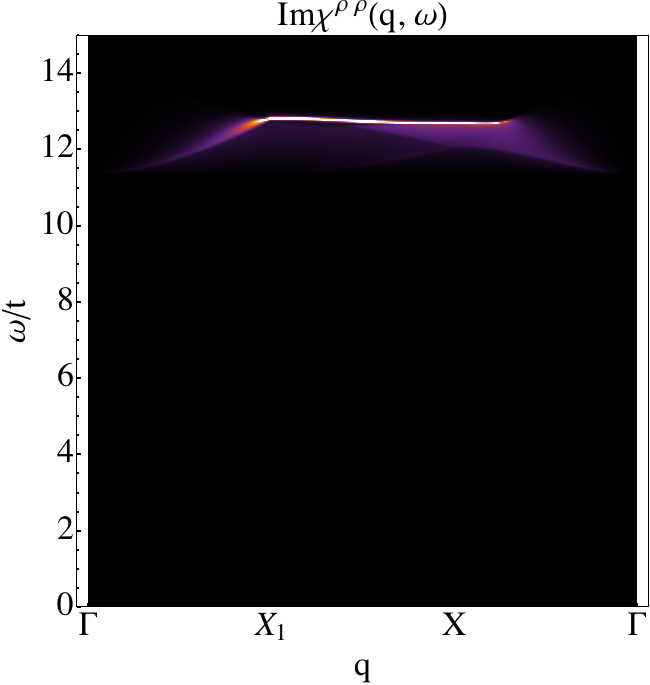} 
\includegraphics[width=0.45\columnwidth]{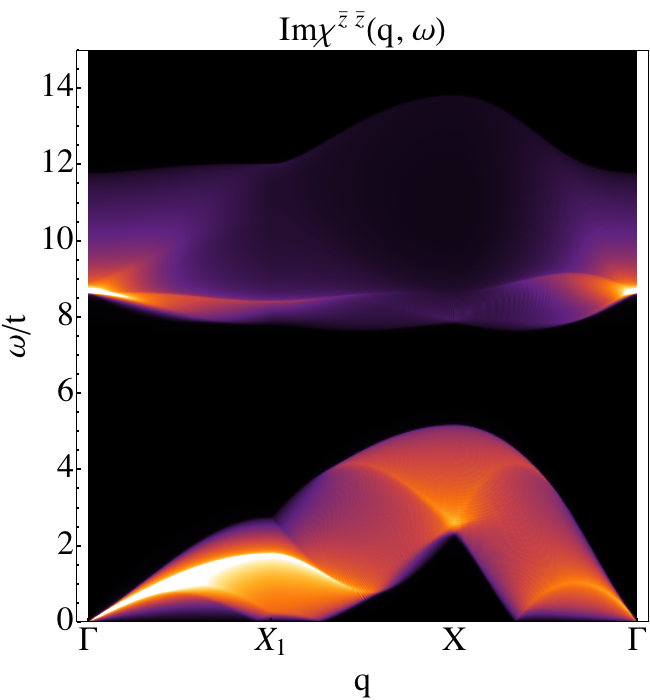}
\includegraphics[width=0.45\columnwidth]{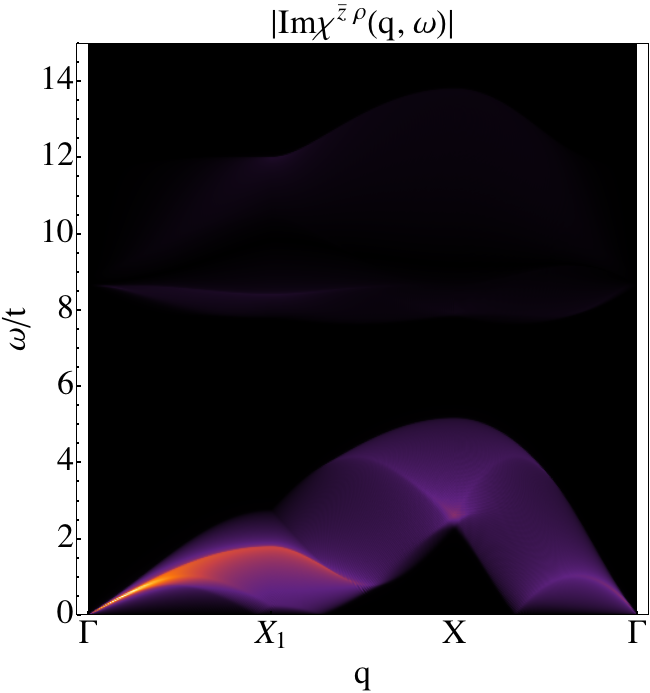}
\caption{Same as in Fig.~\ref{fig:tramod1} but for the spin longitudinal channel. (Upper panels)  $\chi_{\bar{z}\bar{z}}$, $\chi_{\rho \rho}$  are shown for the particle-hole symmetric case for the particle hole case.  (Lower panels) The susceptibilities $\chi_{\bar{z}\bar{z}}$, $\chi_{\bar{z} \rho}$ are displayed for the electron-doped case. 
The corresponding expressions have been evaluated for a (retarded) frequency $z= \omega +i\delta$, with $\delta = 0.01$.  }
\label{fig:lonmod1}
\end{figure}

Within this framework,
the irreducible vertex function reduces to the bare interaction, as in RPA. Hence, under this assumption and using Eqs.~(\ref{def:full:vertex},\ref{eq:BSE_transv},\ref{eq:BSE_long_odd_even}) the physical susceptibilities read: 
\begin{eqnarray}\label{eq:susc_transv_RPA}
\bar{\bar{\chi}}_{{\overline{\sigma\bar{\sigma}}}}(q) &=& \bar{\bar{\chi}}_{{0\overline{\sigma\bar{\sigma}}}}(q) -\bar{\bar{\chi}}_{0\overline{\sigma\bar{\sigma}}}(q)\cdot \bar{\bar{\Gamma}}_{{\overline{\sigma\bar{\sigma}}}}\cdot  \bar{\bar{\chi}}_{{\overline{\sigma\bar{\sigma}}}}(q),
\\ \nonumber \\
\label{RPA:parallel}
\bar{\bar{\chi}}_{\parallel,\,\pm}(q) &=& \bar{\bar{\chi}}_{0\parallel,\,\pm}(q) -\bar{\bar{\chi}}_{0\parallel,\,\pm}(q)\cdot \bar{\bar{\Gamma}}_{\parallel}\cdot  \bar{\bar{\chi}}_{\parallel,\,\pm}(q),
\end{eqnarray} 
with  $\bar{\bar{\chi}}_0(q) \equiv \frac{1}{(V\beta)^2}\sum_{k_1 k_2}\bar{\bar{\chi}}^{k_1 k_2 q}_0$, $\bar{\bar{\Gamma}}_{\overline{\sigma\bar{\sigma}}} = -U\mathbbm{1} $ and $\bar{\bar{\Gamma}}_{\parallel}  =U \tau^{(x)}$.
Within this scheme, we now proceed to explicitly calculate the D$\Gamma$A self-energy of the broken-symmetry phase. 
Because of the chosen mean-field input for the irreducible vertex of the BSEs, the full vertex of the SDE will depend on the exchanged four momentum only, i.e. $F(k,k^\prime,q) \sim F(q)$.
Moreover, if we are away from (quantum) critical points, one could argue that the most important contributions to the D$\Gamma$A self-energy originates from the {\sl transverse}  spin sector, in which the gapless Goldstone modes arise. Therefore, Eq.~(\ref{ladder:sigma:AF}) can be simplified into:

\begin{eqnarray}\label{eq:mot:RPA}
    \Sigma^{ab}_{\sigma}(k)-\delta_{ab}Un_{a,\bar{\sigma}} &\sim&\frac{U^2}{V\beta}\sum_{ q}  G^{ab}_{\bar{\sigma}}(k+q)\chi^{ab}_{\overline{\sigma\bar{\sigma}}}( q), \nonumber \\
    \end{eqnarray} 
where $\chi^{ab}_{\overline{\sigma \overline{\sigma}}}(q)$ is defined in Eq.~(\ref{eq:susc_transv_RPA}). 

The aim of the calculations presented below is -anyway- more ambitious than presenting a  mere {\sl proof-of-principle} of our scheme. 
On the contrary, our results, obtained in a precisely controlled framework, will provide  a reliable ``compass" for future computational benchmarks and, above all, for the physical interpretation of more complex developments and applications.

In the following, we examine two specific realizations of the AF order in a two-dimensional Hubbard model at $T=0$, corresponding to rather distinct physical situations: (i) a half-filling (particle-hole) symmetric case (with $U/t=12$,  $\mu  = \frac{U}{2}$, $\Delta/t  = 5.69 $) and (ii) a electron-doped case (with $U/t = 12 $, $\alpha = t^\prime/t = 0.45$, $n = 1.2$, that corresponds to  $\mu/t = 3.55$ and , $\Delta/t = 4.32$).
Consistent with the results by Igoshev {\sl et al.}\cite{Igoshev2010}, static mean-field (Hartree-Fock) calculations yield stable AF-order ground states for both parameter sets. The mean-field solutions of the two cases  differ qualitatively: The former is insulating, while the latter is metallic, with the Fermi surface shown in Fig.~\ref{fig:FS}.

Consistent with the derivations of Sec.~\ref{sec:dga}, we will first analyze the numerical results for the main physical ingredient of the ladder D$\Gamma$A, namely the collective modes in the magnetic sector, and, thereafter, we will discuss the corresponding effects on the electronic self-energy.   

\subsection{Collective modes}\label{sec:coll}




Within our simplified D$\Gamma$A framework, the expression of the collective modes (and of the associated BSEs) coincide to the RPA ones in the AF long-range ordered phase\cite{chubukov1992,rowe2012}.

We start by considering the half-filling  case (i). 
We show in Fig.~\ref{fig:tramod1} (upper panels for (i)) the results for absorption intensity of three independent transverse (Goldstone) modes: the odd $\chi_{\bar{x}\bar{x}}$, $\chi_{\bar{x}y}$ and the even $\chi_{xx}$, defined in Tab.~\ref{tab:transverse_AF} (see the Appendix for more details).
The results  can be easily interpreted by recalling that,  while all Goldstone modes share the same denominator (and, hence, the same dispersion), they differ in the numerators.
In particular, by performing a hydrodynamic expansion of the latter, one gets numerators which scale in frequency with different behaviors ($\omega^0$, $\omega$ and $\omega^2$ for $\bar{x}\bar{x}$, $\bar{x}y$ and $xx$, respectively). This makes, as one expects, the staggered $\bar{x}\bar{x}$  (non-staggered $xx$) Goldstone mode the most (least) dominant one at low-energies, with the $\bar{x}y$ displaying an intermediate behavior, as it can be readily seen in the intensity plots of Fig.~\ref{fig:tramod1}.

Not surprisingly for the particle-hole symmetric case under consideration, the intensity plots of the corresponding 
longitudinal modes ($\chi_{\bar{z}\bar{z}}$ and $\chi_{\rho \rho}$, in the upper panels of Fig.~\ref{fig:lonmod1}) are rather featureless, due to a significant
energy gap of $2\Delta$ controlled by the large value of the order parameter $m_S$. 


We turn now to analyze the results obtained for the electron-doped case (ii).

By comparing the intensity plots of transverse (lower panels in  Fig.~\ref{fig:tramod1}) and longitudinal modes (lower panels in  Fig.~\ref{fig:lonmod1}) to the corresponding half-filling results (upper panels), it is easy to visualize how the collective modes are affected by the low-energy fermionic quasi-particle excitations emerging in the doped case.

 In particular, we note the appearance of a significant absorption in the low-energy regime for all longitudinal susceptibilities, readily interpreted in terms of the continuum of particle-hole excitations  (lower panels in Fig.~\ref{fig:lonmod1}). With respect to this low-energy feature, the previously dominating  high-energy branches appear now significantly damped.
 The interplay with these particle-hole excitations is also responsible for a visible smearing out of all Goldstone modes over broad regions of the BZ (s.\ lower panels of  Fig.~\ref{fig:tramod1}). 
 
 In the lower panels of Fig.~\ref{fig:lonmod1}, we observe that the absorption intensity increases in the vicinity of the $\Gamma$ point: This is due to particle-hole excitations that connect two points of the same Fermi pocket (e.g., horizontal red dashed line in Fig.~\ref{fig:FS}). 
 The intensity also increases close to $X_1$: This is due, instead, to particle-hole excitations connecting two points lying on different Fermi pockets (e.g., oblique red dashed line in Fig.~\ref{fig:FS}).
 
 A noticeable exception is represented by the large-momenta interval around $X$: Given the geometry of the underlying FS, for these  values of ${\bf q}$ and $\omega$, it is {\sl not} possible to generate particle-hole excitations. 
 Furthermore, sizable change of slope of the Goldstone mode along the path $X  \rightarrow \Gamma$ can be observed by comparing the results of the insulating (upper panels in Fig.~\ref{fig:tramod1}) and the metallic AF (lower panels in Fig.~\ref{fig:tramod1}) . 
 

 The numerical results shown in the lower panels of  Fig.~\ref{fig:tramod1} can be rationalized by performing a hydrodynamic expansion of the susceptibility expressions. Precisely, we analyze their bubble-terms contributions which,  within this simplified D$\Gamma$A context, completely control the momentum/frequency dependence of the corresponding susceptibilities. 
 While referring to  the Appendix for details, we briefly discuss  here the main outcome. The bubble-terms in the AF phase consist of two contributions: 
 $$\chi_0(\bq,\omega) = \chi_{0,\mbox{\footnotesize inter}}(\bq,\omega)+ \chi_{0,\mbox{\footnotesize intra}}(\bq,\omega) $$
i.e., the interband  and the intraband terms. The former is always present for an AF-ordered system, while the latter becomes relevant for metallic solutions, e.g., in case (ii).
 In fact, while $\chi_{0,\mbox{\footnotesize inter}}(\bq,\omega)$ is responsible for the main structures of the Goldstone and the Higgs modes, shown in Figs.~\ref{fig:tramod1} and \ref{fig:lonmod1},  $\chi_{0,\mbox{\footnotesize intra}}(\bq,\omega)$ acquires a singular part in the presence of
a FS. Specifically,  this happens  when the hydrodynamic expansion of the quasi-particle dispersion along the Goldstone mode ($\omega = c |{\bf q}| $) intersects the FS. Quantitatively,  this corresponds to the condition   $c < 4\, \alpha t\, q_F$, where $q_F = \sqrt{(\mu + 4\alpha t-\Delta)/{2\alpha t}}$ is the absolute value of the Fermi momentum. The leading order contributions at low-energy is given by:
 \begin{eqnarray}
\chi_{0,\mbox{\footnotesize intra}}^{xx} &=& A\,q_F^4\, I^{xx}\left(\frac{\omega}{4\alpha  t\,q_F\,|\bq|},\cos 4\theta\right)
\end{eqnarray}
where $A=\frac{1}{[4\pi\,(\Delta/t)]^2\,\alpha t}$  , $\Delta = m_S U/2$, $m_S$ being the staggered magnetization.  $I^{\alpha,\beta}(x_1, x_2)$ (see Appendix \ref{app:intra}) are complex functions of their arguments (whereas $\theta$ is the angle defining a direction in the BZ). 

The {\sl non-vanishing} imaginary part of these functions is responsible for the broadening of the Goldstone modes discussed above, while their explicit dependence on $\theta$ reflects in a corresponding angular modulation of the modes. In the AF metallic case, thus, the angular modulation appears {\sl already} at the leading order in the hydrodynamic expansion, consistent with the numerical results shown in  the lower panels of  Fig.~\ref{fig:tramod1}.

Finally,  the continuum of particle-hole excitation is also responsible for a {\sl sizable} coupling between the modes in the longitudinal section, as evidenced by the intensity-plot in the second of the lower panels  of Fig.~\ref{fig:lonmod1}, referring to $\chi_{\bar{z}\rho}$. The intensity of such coupling, 
vanishing exactly for the particle-hole symmetric case [see Eq.(\ref{eq:rho_zeta_ph})], tends to increase by increasing interaction. Hence, 
 if not properly included in RPA calculations, it might yield significant corrections in the intermediate-to-strong coupling regime.

\subsection{The self-energy in the AF-phase}
\label{sec:self:RPA-DGA}

The expression of the self-energy in Eq.~(\ref{eq:mot:RPA}) has been exploited for the two selected cases considered above. In the particle-hole/half-filled situation, the mean-field solution is fully gapped and this considerably quenches the D$\Gamma$A self-energy, because no fermionic/quasiparticle singularity is coupled to the massless Goldstone modes. Hence, in this case, the D$\Gamma$A corrections to the mean-field expressions reduce essentially to a {\sl quantitative} renormalization of the staggered magnetization, as discussed in Refs.~[\onlinecite{sigh1990,chubukov1992}]. 
In particular, we found that, in the limit of a strong Coulomb interaction $U\gg t$, the quasi-particle  residue is given by $Z_{qp} = \frac{1}{2}\left(1+\frac{1}{\sqrt{1 + 2\kappa}}\right)$, with $\kappa = \frac{1}{V}\sum_\bq\frac{1-\sqrt{1-\gamma_\bq^2}}{\sqrt{1-\gamma_\bq^2}}\sim 0.39$, where $\gamma_\bq = \frac{1}{2}[\cos q_x + \cos q_y]$ and the staggered magnetisation is renormalized by a factor $2\,Z_{qp}-1$.  If we expand $Z_{qp}$ linearly in $\kappa$, the renormalization factor becomes $ 1-\kappa \sim 0.61$, consistent with Ref.~[\onlinecite{chubukov1992}] as well as the result obtained in spin-wave theory for the Heisenberg model. 

\begin{figure*}[htbp!]
\begin{center}
    
    \includegraphics[trim=35 0 50 0,clip,width=0.95\columnwidth]{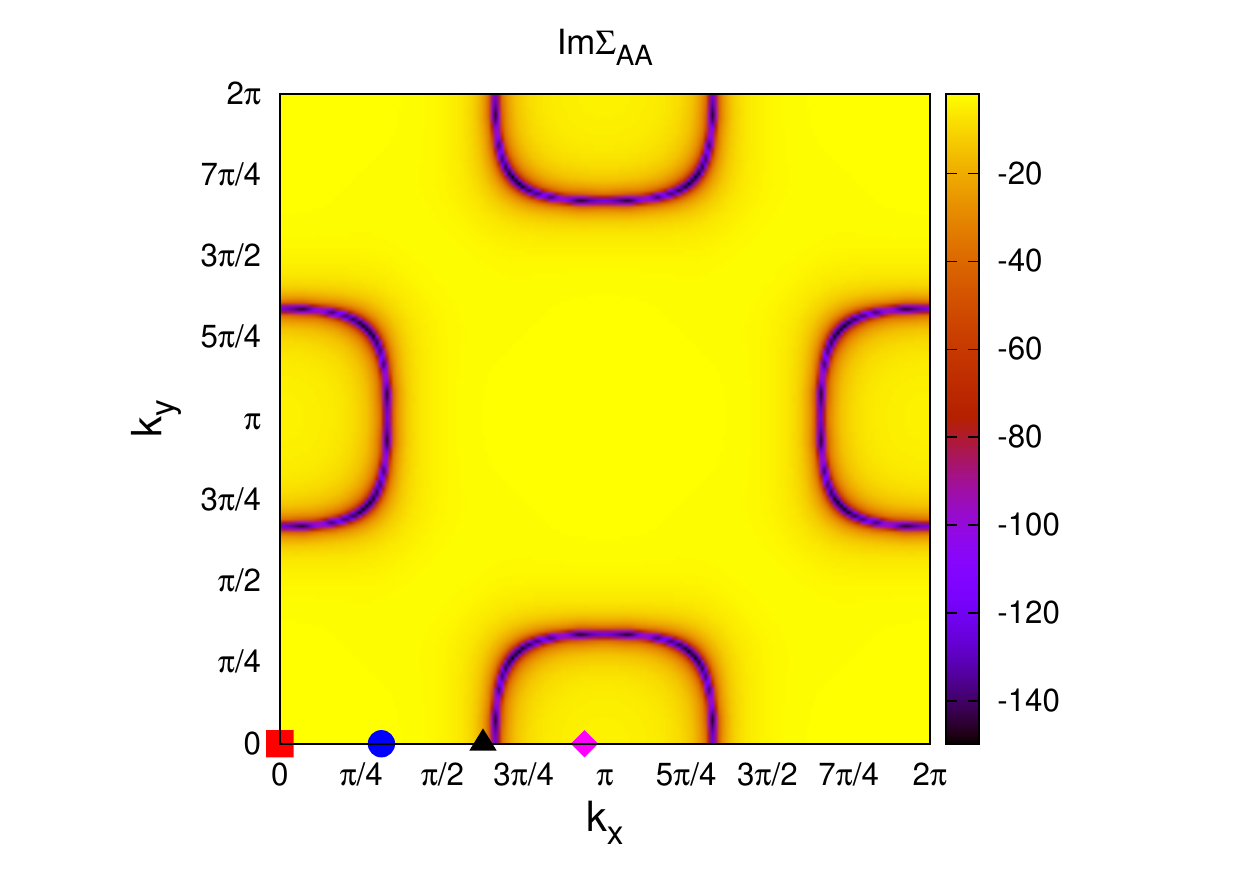}
     \includegraphics[trim=35 0 50 0,clip,width=0.95\columnwidth]{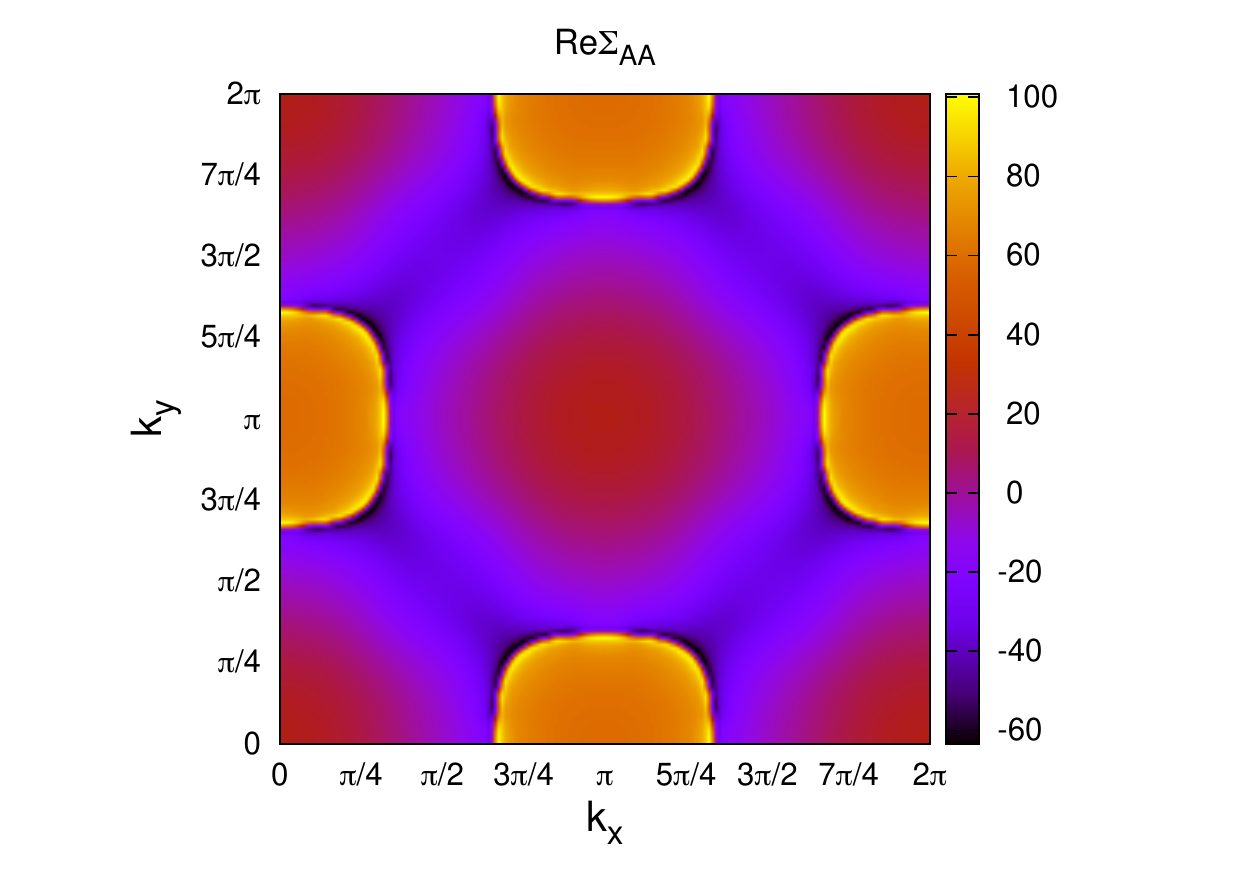}
    \end{center}

    \vspace{-10mm}
     \begin{center}

    \hspace{-7mm}
    \includegraphics[trim=25 3 20 15,clip,width=0.6\columnwidth]{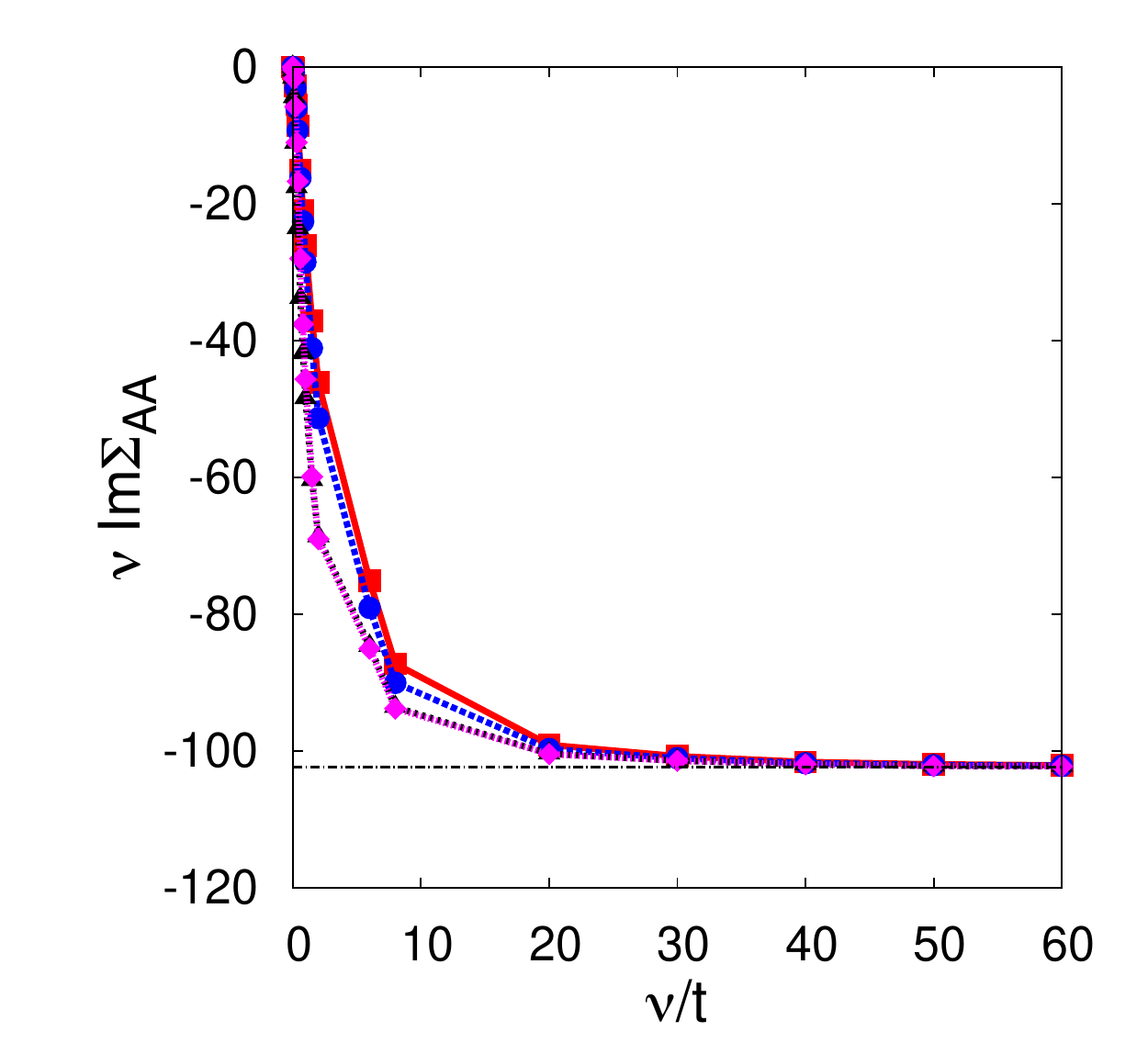}
    \includegraphics[trim=25 3 20 15,clip,width=0.6\columnwidth]{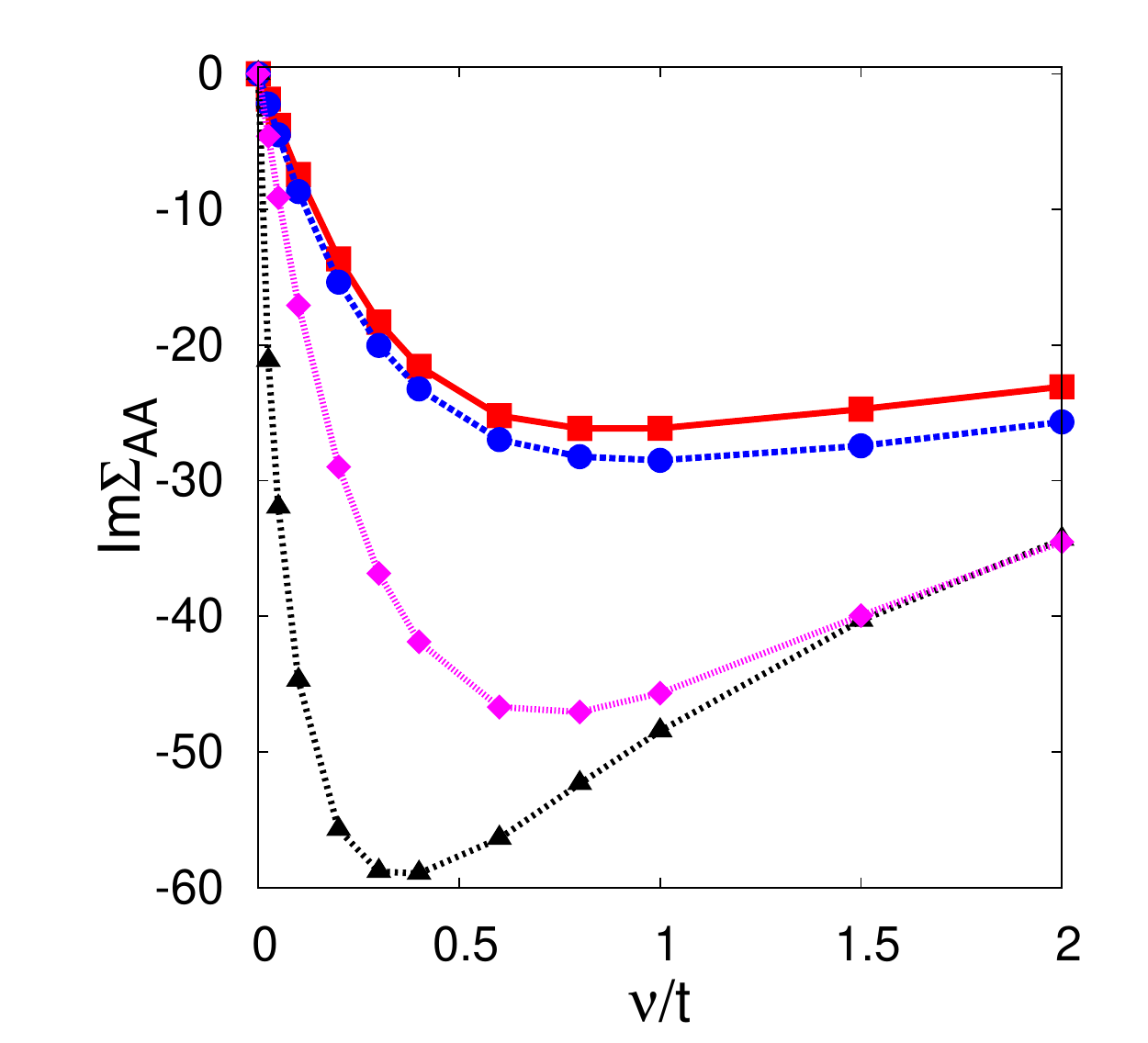}
    \includegraphics[trim=25 3 10 15,clip,width=0.6\columnwidth]{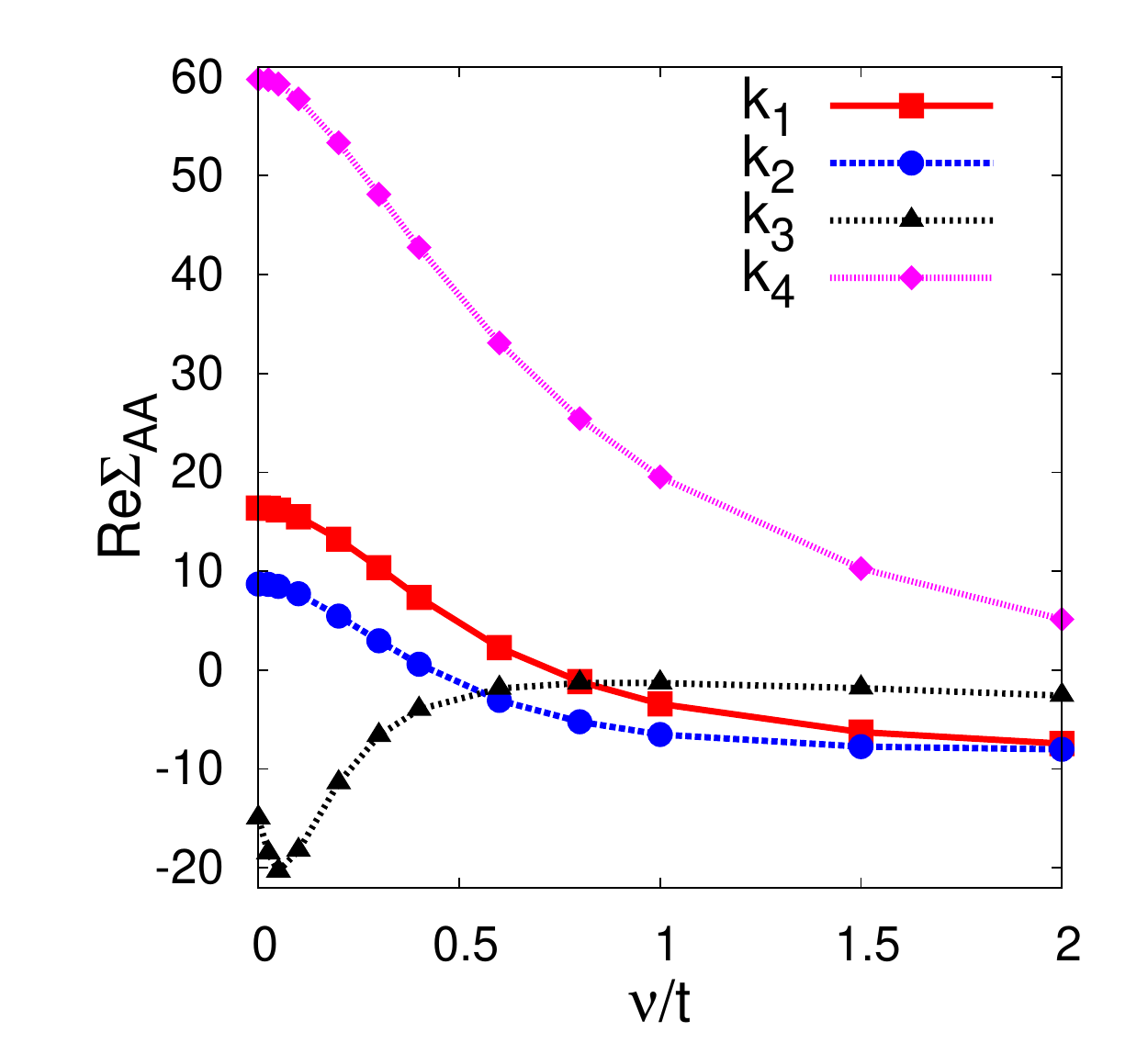}
     
    \end{center}

     \caption{(Upper Panels) Color intensity plots of the imaginary part (on the left) and real part (on the right) of the majority spin self-energy of the sublattice $A$  in the AF-phase as a function of the momenta calculated at the imaginary frequency $\nu/t = 0.025$, for $U/t = 12$, $n = 1.2$ and $\alpha = 0.45$. (Lower Panels) The
      the high frequency behavior of the self energy is reported in the leftmost panel, the thin line marking the asymptotic value defined in Eq.~(\ref{eq:asymAF}).
     The imaginary part (center) and real part (right) of the self-energy are shown as a function of the imaginary frequency $\nu$  for four  ${\bk}$-points lying on the $\Gamma$-X direction in the BZ and marked by different colors/symbols in the upper panels.  }
    \label{fig:self_k}
\end{figure*}

Much more interesting are the D$\Gamma$A results out-of-half filling. Here, the presence of an underlying Fermi-surface in the mean-field solution allows for important D$\Gamma$A self-energy corrections, originated by the {\sl combined} effect of bosonic (Goldstone) and fermionic (quasiparticles) excitations. 
Our numerical results are shown in Fig.~\ref{fig:self_k}, where we report explicitly the momentum dependence of the real and the imaginary part of the self-energy for the sub-lattice $A$ in the whole BZ (upper panels), and at the same time, its frequency dependence at four selected {\bf k}-points (lower panels). 

In order to clarify the overall behavior of the self-energy in the AF phase, it is convenient to consider separately the high-frequency and the low-frequency regime.

As we discuss below, the former is controlled by precise 
analytical relations, which extends the well known ones for the SU(2)-symmetric case.
By inverting the relations in  Table  \ref{tab:transverse_AF}, we can express the equation of motion in terms of the  susceptibilities in the physical basis as following:
\begin{eqnarray}\label{chi:phys:RPA}  \nonumber
 \chi^{ab}_{\overline{\sigma \overline{\sigma}}}(q) &  = &\frac{1}{2}\left[\chi^{xx}(q)+(-1)^{a + b}\chi^{\bar{x}\bar{x}}(q) \right]\nonumber \\
 & -& \delta_{ab}(-1)^a \,i\sigma \,\chi^{x\bar{y}}(q). 
 \end{eqnarray}  
Then, from Eqs.(\ref{eq:mot:RPA},\ref{chi:phys:RPA}), we can extrapolate the asymptotic behavior of $\Sigma$ in the limit of large frequencies. 
Specifically, for $\nu\to \infty$, we have:
\begin{widetext}
\begin{equation}
\nu \,\mbox{Im}\Sigma_\sigma^{ab}(k) =  \mbox{const.} = -U^2\delta_{ab}\frac{1}{V\beta}\sum_{\omega}\sum_{\mathbf{q}\in BZ}\left[\chi^{xx}( q)+ (-1)^a\,\sigma\,m_S
\,\mbox{Re}G^{aa}_{\bar{\sigma}}(q+k_0)\right],
\label{eq:asymAF}
\end{equation}
\end{widetext}
where we defined $k_0 = (\boldsymbol{0},\nu_0)$ for compactness of notation. 
We observe that, since the integrand function in Eq.(\ref{eq:mot:RPA}) is even under a shift  of $\Pi = (\pi,\pi)$ of the exchanged momentum, the summation in Eq.(\ref{eq:asymAF}) can be extended to the entire BZ. 
It is interesting to notice that, in the broken-symmetry phase, the high-frequency asymptotic behavior of $\Sigma$ depends both on the electronic density {\sl and} on the order parameter. This marks a qualitative difference from the normal phase, where the high-frequency asymptotics of $\Sigma$ is controlled by electronic density only. Specifically, the constant prefactor $m_S$ originates from the {\sl mixed} correlator $\chi^{x\bar{y}}$, which, as pointed out in  Sec.\ref{sec:BS_form},  is related to the order parameter through the commutation relation between the spin operators.
 The non-trivial match  between the analytic/expected expressions for the high-frequency behavior of the self-energy and our numerical calculations is explicitly shown for the case of the four selected {\bf k}-point in the leftmost bottom panel of Fig.~\ref{fig:self_k}. Such numerical agreement has been verified for all momenta, providing an useful  double-check for the algorithmic implementation of the AF-D$\Gamma$A expression.


Let us now focus on the low-energy properties of $\Sigma$.
 In the presence of an underlying Fermi surface, one expects that the most important information will be encoded in the corresponding Fermi energy and momenta.
The data shown in Fig.~\ref{fig:self_k} appear consistent with such expectation: One can readily identify the region of the Brillouin-zone, where the D$\Gamma$A corrections induce the strongest momentum dependence in the low-frequency self-energy. 
In particular, the data reported in the upper panels of Fig.~\ref{fig:self_k} clearly show how the {\sl largest} variation of both real and imaginary parts of $\Sigma_{AA}$ over the whole Brillouin zone occurs in the proximity of the underlying Fermi-surface of the mean-field solution (cf.~Fig.~\ref{fig:FS}). Specifically, by crossing the FS, both Im$\Sigma_{AA}$ (left upper panel of Fig.~\ref{fig:self_k}) and Re$\Sigma_{AA}$ (right upper panel of Fig.~\ref{fig:self_k}) get strongly enhanced in absolute value, whereas Re$\Sigma_{AA}$ also displays an evident change-of-sign. More quantitatively, we observe over the whole FS  a simultaneous divergence of both imaginary and real part of $\Sigma_{AA}$ in the zero-frequency limit, though with a different degree of severity. It should be also noted that the additional, and rather weak, sign-structures (oblique blue linear-shaped regions in the intensity plot of Re$\Sigma_{AA}$) essentially reflect the halving of the BZ due to the AF-order.

 As we detail in the following, the self-energy behavior shown in Fig.~\ref{fig:self_k} is the direct consequence of the {\sl combined action} of the massless Goldstone modes and quasiparticle excitations at the Fermi level, which we mentioned before. 
The physical mechanism is indeed similar to the one triggering the enhanced scattering rate observed in several D$\Gamma$A studies of the SU(2)-symmetric phases in the proximity of phase-transitions (in $d=3$)\cite{Rohringer2011,Rohringer2016,Rohringer2018,DelRe2019} and/or for very large value of the magnetic correlation length (in $d=2$) \cite{Schaefer2015,Schaefer2015-3}.
In the broken-symmetry phase, however, the analogy is {\sl not} complete. In fact, due to the finite value of the order parameter and of the corresponding doubling of the unit cell, the large self-energy corrections, evidenced by the color changes in Fig.~\ref{fig:self_k} do not correspond to a significant loss of coherence in the low-energy fermionic excitations.

In order to elucidate the physics encoded in the D$\Gamma$A self-energy of the AF-phase, we explicitly analyze the zero-energy poles of the corresponding Green's function as well as the the associated quasi-particle renormalization.  For capturing the low-frequency behavior of the self-energy we can keep just  $\chi^{\bar{x}\bar{x}}$ in Eq.~(\ref{chi:phys:RPA}), that accounts for the leading divergent orders when $\omega \sim 0$. Furthermore, we can approximate $G_{\bar\sigma}^{ab}(k+q)\sim G_{\bar\sigma}^{ab}(\bk,\omega + \nu)$, as we want to keep only the  $\bq\sim 0$ contributions in the integral in Eq.(\ref{eq:mot:RPA}).  Under these assumptions the equation of motion for $\bar{\bar{\Sigma}}_{\sigma}(k)-\delta_{ab}Un_{a,\bar{\sigma}}$ reads:
\begin{eqnarray}
 &\sim& \frac{U^2}{(2\pi)^{d+1}}\int_{-\infty}^{+\infty}d\omega \int_{BZ} d\bq \,\chi^{\bar{x}\bar{x}}(q) \,\tau^{(y)} \cdot \bar{\bar{G}}_{\sigma}(\bk,\omega + \nu)\cdot \tau^{(y)}. \nonumber
\end{eqnarray}
We can now express the self-energy in the basis of the quasi-particles, i.e.  
\begin{widetext}
\begin{equation}\label{self:bog}
\bar{\bar{\Sigma}}_{\sigma}(k)-\delta_{ab}Un_{a,\bar{\sigma}} = 
e^{-i\tau^{(y)}\theta_{\bk \sigma}} \tau^{(y)} \left(
\begin{array}{cc}
\sigma^+_\bk(\nu) & 0 \\
0 & \sigma^-_\bk(\nu)
\end{array}
\right)\tau^{(y)}\,
  e^{i\tau^{(y)}\theta_{\bk \sigma}} 
 = 
  e^{-i\tau^{(y)}\theta_{\bk \sigma}} 
 \left(
\begin{array}{cc}
\sigma^-_\bk(\nu) & 0 \\
0 & \sigma^+_\bk(\nu)
\end{array}
\right)
  e^{i\tau^{(y)}\theta_{\bk \sigma}},
\end{equation}  
\end{widetext}
where: 
\begin{eqnarray}\label{eq:poles_1}
\sigma_\bk^{\pm}(\nu) =\frac{U^2}{(2\pi)^{d+1}} \int_{-\infty}^{+\infty}d\omega\int_{BZ} d\bq\,\,
\frac{\chi^{\bar{x}\bar{x}} (q)}{i\nu +i\omega -\omega^\pm_\bk}, \nonumber \\
\end{eqnarray}
and after carrying out the integrals we have:
\begin{equation}
    \sigma^\pm_\bk(\nu\to 0) \propto -\mbox{sign}\left(\omega^\pm_\bk\right) \mbox{ln}\left(\frac{1}{\left|\omega_\bk^\pm\right|}\right) - i\, \frac{\nu}{\left|\omega^\pm_\bk\right|}.
\end{equation}


Hence, for $d = 2$, logarithmic and power-low divergences appear in the real and imaginary part and the self-energy, respectively, when $\bk \in$ FS.
 It is important to note, however, that the doubling of the unit cell associated to the AF phase {\sl does prevent} the quasi-particle excitations to be washed out by such divergences.

To explain this, let us first notice that 
the self-energy in Eq.~(\ref{self:bog}) is diagonal in the HF quasi-particles basis. Therefore, in this reference frame, the Dyson equation reads:
\begin{equation}
    \bar{\bar{G}}^{-1}_\bk(\nu) \sim \left(\begin{array}{cc}
        i\nu-\omega^+_\bk-\sigma^-_\bk(\nu) &0  \\
        0 & i\nu-\omega^-_\bk -\sigma^+_\bk(\nu) 
    \end{array}
    \right).
\end{equation}
The conduction band is dressed with the self-energy $\sigma^-_\bk(\nu)$ that depends on the valence electrons energy $\omega^-_\bk$. The presence of a gap prevents $\sigma_\bk^-(\nu)$ to diverge and the FL excitations are stable. 

Conversely, the valence electrons are dressed with the self-energy $\sigma_\bk^+(\nu)$, which diverges when $\nu \to 0$ at the FS. 

\begin{figure}[hbtp!]
    \includegraphics[width=\columnwidth]{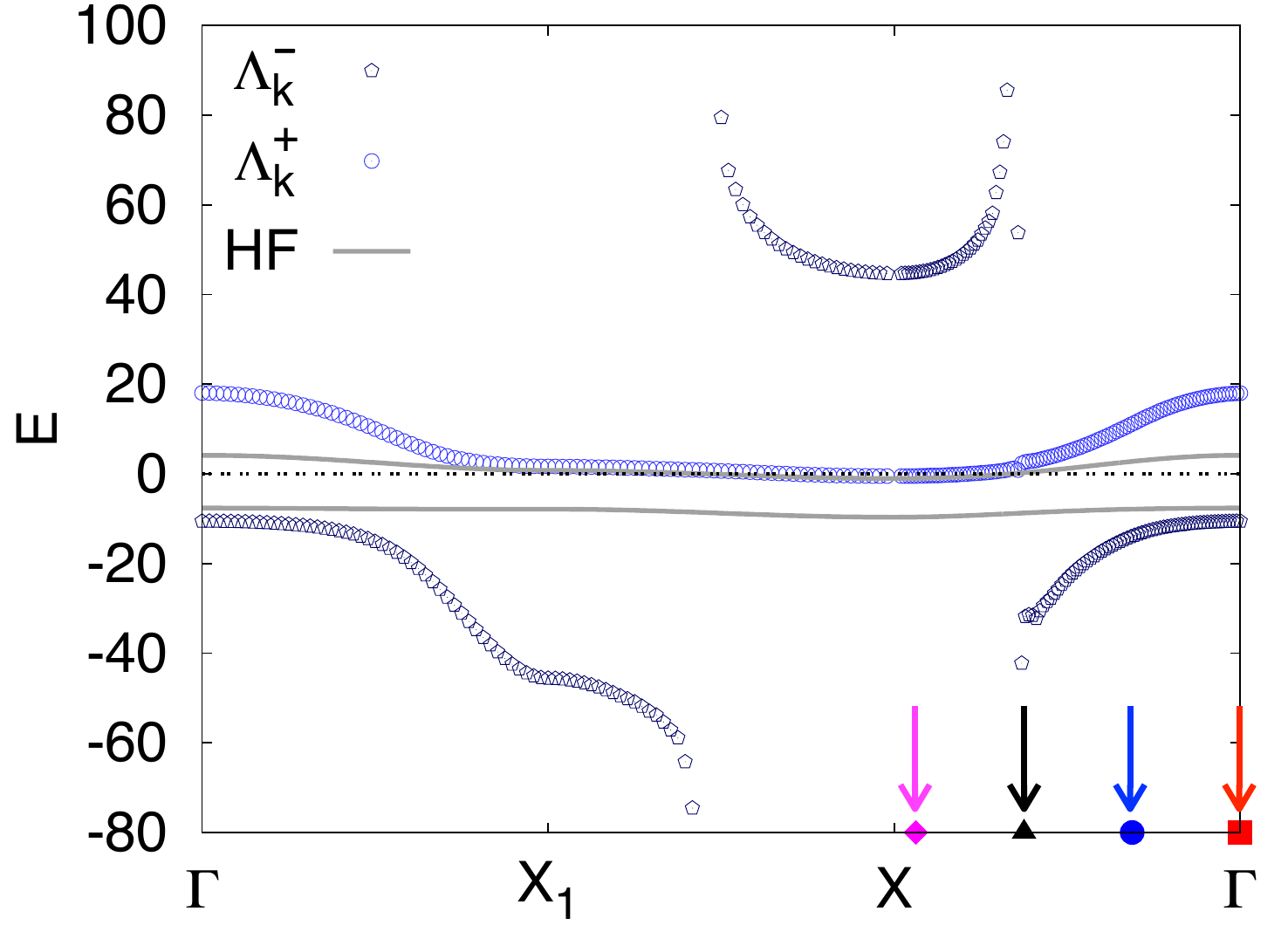}
    \caption{Plot of the lower and upper eigenvalue $\Lambda^\pm_\bk$ represented respectively by  pentagons and circles as a function of the momentum. We plotted the bands obtained in Hartree-Fock (gray solid lines) as a reference. The four arrows pointing onto the $x$-axis mark the corresponding momenta selected in Fig.~\ref{fig:self_k}.}
    \label{fig:evals}
\end{figure}

We support our analytical findings by showing in Fig.~\ref{fig:evals} the numerical values of the Green's function zero-energy poles  $\Lambda^\pm_\bk$, which defines two bands. The first one, $\Lambda^+_\bk$  is a smooth function of the crystal momentum and represents a {\sl conduction} band, which emerges from a sizable reshaping of the corresponding one in the HF solution.
In fact, it is only the second (valence) band ($\Lambda^-_\bk$) to be affected by  logarithmic singularities, which precisely appear when HF-conduction band crosses the Fermi level, as it is clearly shown in Fig.~\ref{fig:evals}.

On the basis of these considerations and of our numerical results, we can conclude that the conduction band is {\sl stable} under non-local quantum fluctuations: In spite of the large (or even diverging) values of Im $\Sigma$, the metallic coherence of the corresponding low-energy quasi-particle excitations is preserved.

At the same time, this analysis does not provide specific information about what happens at higher energy. In fact, to understand properly how the valence band is dressed by the conduction electrons we should numerically evaluate the Green's function on the real frequency axis.
While this is beyond the aim of the present work, we expect -in general- that such  corrections will affect the way how the high-energy spin excitations, such as the sharp spin-polarons\cite{Strack1992,Sangiovanni2006} clearly visible at DMFT level\cite{Strack1992,Sangiovanni2006,Taranto2012}, reshape the spectral functions and  the charge/optical response of the system. 

 In this perspective, future applications of our ladder D$\Gamma$A approach exploiting a full DMFT input (possibly directly computed on the real frequency axis\cite{Kugler2021,Seung2021}) could provide a very powerful set-up to investigate the spin-polaron physics in  realistic three- or two-dimensional cases.
In particular, one would aim at estimating the broadening of the  spin-polaron peaks induced by non-local correlations beyond DMFT and at identifying fingerprints of these excitations in the physics of bulk and layered antiferromagnets.

On a broader perspective, the results of this section shed light on the general physical behavior to be expected in metallic systems in the presence of magnetic order and/or strong magnetic fluctuations.
 The onset of a long-range AF order shifts the major effects of the magnetic correlations on the electronic spectra from low- to higher frequencies. Even when such effects are -per se- strong, as it happens in correlated metals due to the cooperative action of Goldstone modes and quasi-particles, the coherent nature of the underlying Fermi-liquid excitations remains preserved, if the order parameter is large enough. Hence, the largest quasi-particle scattering rates is expected to occur in critical or quantum critical regimes of (here: magnetic) phase-transitions. 
The possible occurrence of a {\sl minimal} metallic coherence at the phase-transition is compatible with the results of previous numerical studies\cite{Rohringer2016} performed in the proximity of a AF-transition of the Hubbard model in $d\! =\!3$. It is also consistent with several spectroscopic/transport observations made in the (almost bidimensional) cuprates when cooling the compounds below their superconducting transition temperature in the underdoped/optimally doped regime.

\section{Conclusions and Outlook}\label{sec:concl}

In this work, we have illustrated how to extend the ladder D$\Gamma$A approach, hitherto restricted to the SU(2) symmetric case, to the treatment of electronic correlations in magnetic systems, with ferromagnetic or antiferromagnetic long-range order. 

In particular, starting by general considerations on the two-particle vertex functions in the limit of infinite dimensions/coordination of the lattice, we have first generalized the condition of pure locality for the irreducible vertices of DMFT to solutions with magnetic order. Secondly, we have exploited the corresponding Bethe-Salpeter/ladder equations, which describe the longitudinal/transverse collective modes at the level of DMFT, to derive, through the Schwinger-Dyson relations, the ladder D$\Gamma$A expression for the electronic self-energy of the magnetically ordered phases.

To demonstrate the applicability of our extended D$\Gamma$A approach, we have exploited it to study a couple of simplified, but representative model cases with AF-order, where we approximated the DMFT input of the D$\Gamma$A equations to its static mean-field counterpart. 
In this framework, the collective modes inducing the non-local D$\Gamma$A correlation reduces  to the corresponding RPA ones. The reported results represents a solid benchmark for future, more demanding calculations. At the same time, the thorough analysis of the self-energy results in the non-particle hole symmetric-case allows us to outline important physical effects to be expected in the correlated magnetic systems, in particular concerning the coherence of the underlying electronic excitations.

Analogously as in the first ladder D$\Gamma$A derivation for the SU(2)-symmetric case\cite{Toschi2007}, we have considered here the most fundamental implementation of the approach. This consists in a single-shot correction of the DMFT results originated by the scattering with the corresponding magnetic modes. 
The question arises whether it is possible and/or convenient to implement a self-consistent version of the ladder D$\Gamma$A equations in the broken SU(2)-symmetric case.
While this issue certainly calls for a dedicated study, analogous to Refs.~[\onlinecite{Katanin2009,Rohringer2016,Rohringer2018,Kaufmann2020}], we observe here that additional constraints must be taken into account in the magnetically ordered phase. 
For example, one could try extend the so-called $\lambda$-correction, introduced within a' la Moriya schemes\cite{Katanin2009,Rohringer2016}, to the FM/AF case. The underlying idea, which can be regarded\cite{DelRe2019} -to some extent- as a dynamical version of the Two Particle Self-Consistent (TPSC) approach\cite{Vilk1997}, consists in constraining some important parameter of the theory (e.g., the mass of the spin propagator) to fulfill physically relevant relations, such, e.g., the asymptotic high-frequency behavior of the D$\Gamma$A self-energy.
Evidently, this task gets significantly harder in the  broken SU(2)-symmetry phase, because of the increased number of independent degrees of freedom (see Tabs.~\ref{tab:ferro}- \ref{tab:longitudinal_AF}-\ref{tab:transverse_AF}), and the precise interrelations between them which need to be preserved\footnote{For instance, correcting the mass of the spin-propagator would be no longer sufficient in a magnetically-ordered phase. A mass-correction might work for the Higgs/transverse mode, while the Goldstone mode(s) needs to be preserved in the entire ordered phase, allowing only for a correction of its(their) velocity. In this respect, one should also guarantee that, in spite of the different nature of the mass/velocity corrections, all the magnetic modes merge together when the order parameter vanishes. Similar challenges are faced in the context of fRG algorithms\cite{Vilardi2020}.}.
At the same time, the identification of the physical relations to be enforced 
is less obvious than in the symmetric case, as  well exemplified by the expression for high-frequency asymptotics of the self-energy in the AF-phase given in Eq.~(\ref{eq:asymAF}).

Another possibility would be to implement a true self-consistent loop, by inserting the momentum-dependent expression of the D$\Gamma$A self-energy back in the Bethe-Salpeter equations, while keeping fixed\footnote{Of course, a self-consistent adjustment of the local/impurity input\cite{Ribic2018,Ayral2016} could be also considered, but it would imply an even bigger increase of the numerical effort.}
the (local) irreducible vertices, in a similar spirit of the so-called ``internal'' self-consistency of the DF approach\cite{Rubtsov2008,Rohringer2018} and of the most recent algorithmic development\cite{Kaufmann2020} of  the ladder D$\Gamma$A in the SU(2)-symmetric case.
This route would avoid all the preliminary, physically-motivated {\sl ad hoc} implementations of {\sl a la Moriya} correction schemes, requiring, however, a higher numerical effort.
In any case, it needs to be verified, whether such self-consistent 
ladder implementation can guarantee a coherent description of the 
instability driven by an external parameter ($T$, $n$, $h$, ...) from both sides (ordered and disordered) of the magnetic transition. From the technical point of view, one should also ensure that the massless nature of the Goldstone modes remains preserved at every iteration.

While these considerations might serve as  guidance for future methodological advancements in the description of correlated magnetic systems, our preliminary study suggests that interesting results can be obtained by means of the one-shot ladder D$\Gamma$A scheme presented here, especially for investigating the non trivial behavior of the spectral properties of correlated magnets in the proximity of their classical or quantum phase-transitions. 

Finally, the explicit analytical expressions for the collective modes  given in Tabs.~\ref{tab:ferro}-\ref{tab:longitudinal_AF}-\ref{tab:transverse_AF} and the self-energy of the magnetically ordered phases could be quite inspiring for future extensions of the recently introduced fluctuation diagnostics post-processing methods\cite{Gunnarsson2015,Rohringer2020,Schaefer2021} and fluctuation analysis of the two-particle irreducible vertex function\cite{delre2021} to the broken SU(2)-symmetry phases. \\

\acknowledgements 
We thank  G.~Rohringer, M.~Capone, G.~Sangiovanni, V.~Zlatic, K. Held, A. N.  Rubtsov, F. Krien and E. A. Stepanov for insightful discussions. LDR and AT also thank the Simons Foundation for the great hospitality at the CCQ of the Flatiron Institute. 
The present work was supported by the Austrian Science Fund (FWF) through the project I 2794-N35 (AT) and by the U.S. Department of Energy, Office of Science, Basic Energy Sciences, Division of Materials Sciences and Engineering under Grant No. DE-SC0019469 (LDR).

\appendix
\section{Change of coordinates for four-point correlation functions}\label{sec:app_a}
In this section we formally derive Eqs.(\ref{eq:relation},\ref{eq:coeff},\ref{eq:relation_ph},\ref{eq:coeff_ph}).  Let us first consider the case where $\hat{H}$ is left invariant under the unitary transformation $\psi_\bR \to U^{\,}_\bR\,\psi_\bR$.
Therefore, by expressing the Fermi fields in the new basis we can rewrite Eq.(\ref{def:gen}) as following:
\begin{eqnarray}\label{eq:gen:transf}
 [U^\dag_{\bR_1}]_{\alpha^\prime \alpha}[U^{\, }_{\bR_2}]_{\beta\beta^\prime}[U^\dag_{\bR_3}]_{\gamma^\prime \gamma}[U^{\, }_{\bR_4}]_{\delta\delta^\prime}   \chi^{\alpha^\prime\beta^\prime}_{\gamma^\prime\delta^\prime}(x_1,x_2,x_3,x_4),\nonumber \\
\end{eqnarray}
where a summation over repeated indices is intended.
Substituing Eq.(\ref{eq:gen:transf}) into Eq.(\ref{def:phys}) we obtain:
\begin{eqnarray}\label{eq:phys:transf}
V^{a}_{\alpha\beta}(\bR_1,\bR_2)V^{b}_{\gamma\delta}(\bR_3,\bR_4)\chi^{\alpha\beta}_{\gamma\delta}(x_1,x_2,x_3,x_4),
\end{eqnarray}
where $V^a(\bR,\bR^\prime)  =  U^\dag_{\bR}\,T^{(a)}\,U^{\,}_{\bR^\prime} $. 
Choosing $T^{(a)}$ 
as a complete  set of hermitian generators satisfying the orthonormality condition 
Tr$\left[T^{(a)}T^{(b)}\right] = \delta_{ab}$,
we can express the $V^{a}$ matrix as a linear combination of the generators in the following way
\begin{eqnarray}\label{eq:expansion}
V^{a}(\bR,\bR^\prime) = \sum_{b}
\overbrace{\mbox{Tr}\left[T^{(b)}V^{a}(\bR,\bR^\prime)\right]}^{F_{\bR\bR^\prime}^{ab}}
T^{(b)},
\end{eqnarray}
where the coefficents of the expansions are the same as those defined in Eq.(\ref{eq:coeff}).
Substituting the last equation into Eq.\ref{eq:phys:transf} we finally obtain Eq.(\ref{eq:relation}).
\par Let us now consider a particle-hole transformation, i.e. $\psi^{\,}_\bR\to U_\bR^{\,}\psi^{\dag}_\bR$. Expressing the fermi operators in the new basis, we can rewrite Eq.(\ref{def:gen}) in the following way:
\begin{eqnarray}\label{eq:transf_ph:gen}
&&[U^\dag_{\bR_1}]_{\alpha^\prime \alpha}[U^{\, }_{\bR_2}]_{\beta\beta^\prime}[U^\dag_{\bR_3}]_{\gamma^\prime \gamma}[U^{\, }_{\bR_4}]_{\delta\delta^\prime}   \chi^{\beta^\prime\alpha^\prime}_{\delta^\prime\gamma^\prime}(x_2,x_1,x_4,x_3) = 
\nonumber \\
 &&[U^\dag_{\bR_1}]_{\beta^\prime \alpha}[U^{\, }_{\bR_2}]_{\beta\alpha^\prime}[U^\dag_{\bR_3}]_{\delta^\prime \gamma}[U^{\, }_{\bR_4}]_{\delta\gamma^\prime}   \chi^{\alpha^\prime\beta^\prime}_{\gamma^\prime\delta^\prime}(x_2,x_1,x_4,x_3).
\nonumber \\
\end{eqnarray}
Substituting Eq.(\ref{eq:transf_ph:gen}) in Eq.(\ref{def:phys}) yields:
\begin{eqnarray}
\widetilde{V}^a_{\alpha\beta}(\bR_1,\bR_2) \widetilde{V}^b_{\alpha\beta}(\bR_3,\bR_4)
\chi^{\alpha\beta}_{\gamma\delta}(x_2,x_1,x_4,x_3),
\end{eqnarray}
where $$\widetilde{V}^a_{\alpha\beta}=\sum_{\alpha^\prime\beta^\prime}[U^\dag_{\bR}]_{\beta \alpha^\prime}T^{(a)}_{\alpha^\prime\beta^\prime}[U^{\, }_{\bR^\prime}]_{\beta^\prime\alpha} = \left[U^\dag_{\bR} \,T^{(a)}\,U_{\bR^\prime}^{\,}\right]^T_{\alpha\beta}.$$
We can express this matrix using the same expansion as in Eq.(\ref{eq:expansion}), i.e.
\begin{eqnarray}
\widetilde{{V}}^{a}(\bR,\bR^\prime) = \sum_b\mbox{Tr}\overbrace{\left[T^{(b)}\widetilde{V}^a(\bR,\bR^\prime)\right]}^{\widetilde{F}^{ab}_{\bR\bR^\prime}}T^{(b)},
\end{eqnarray}
whose coefficents are the same as those appearing in Eq.(\ref{eq:coeff_ph}). After we substitute the last equation in Eq.(\ref{eq:transf_ph:gen}), we obtain Eq.(\ref{eq:relation_ph}).
\section{Generic properties of mixed and non-mixed correlators}\label{app:corr:func}
For completeness, here we show some generic properties of correlation functions evaluated along the imaginary axis. Let $\hat{A}$ and $\hat{B}$ be generic two-particle operators,  whose time-evolution in the Heisenberg representation is driven by a time-independent Hamiltonian, and let us define the correlation function between them:
\begin{equation}\label{eq:def:corr}
C_{AB}(\tau) = T_\tau\left<\hat{A}(\tau)\hat{B}(0)\right>,
\end{equation}
and its Fourier transform as: $C_{AB}(\omega) = \int_0^\beta d\tau\, C_{AB}(\tau)\,e^{i\omega \tau}$,  with $\omega = 2n\,\pi\,T$ being a bosonic Matsubara frequency. 
\renewcommand{\arraystretch}{1.5}

\begin{table}[htbp!]
    \centering
    \begin{tabular}{c|c|c|l}
    & $\hat{A}$ & $\hat{B}$& \\ \hline 
 (i)  & $\hat{A}^\dag$ & $\hat{B}^\dag$&  \ \ 
    $\begin{array}{ccc}
    C_{AB}^*(\tau) &=& C_{AB}(-\tau) \\
    C_{AB}^*(\omega) &=& C_{AB}(\omega)
    \end{array}$
     \\
  \hline
  (ii) &  $\hat{B}^\dag$& $\hat{A}^\dag$&
    \ \ $\begin{array}{ccc}
    C^*_{AB}(\tau) &=& C_{AB}(\tau) \\
    C^*_{AB}(\omega)&=& C_{AB}(-\omega)
    \end{array}
    $
     \\
  \hline
 (iii) &   $\hat{A}^\dag$& $\hat{A}$&
   \ \  $\begin{array}{ccc}
   C_{AB}(\tau) &=& C_{AB}(-\tau)\in  \mathbbm{R} \\
   C_{AB}(\omega) &=& C_{AB}(-\omega) \in \mathbbm{R}
    \end{array}
    $
    \end{tabular}
    \caption{Symmetry properties of correlation functions in imaginary time/frequency domain for the three different cases discussed in the main text. These relations can be derived  starting from the definition in Eq.~(\ref{eq:def:corr}) and using the cyclic property of the trace and the fact that $C_{AB}(\tau)$ is a periodic function of period $\beta$.}
    \label{tab:corr:prop}
\end{table}
\renewcommand{\arraystretch}{1.}

In Table \ref{tab:corr:prop}, we list the properties of the correlation function in three different cases:
\begin{enumerate}
    \item[(i) ] $\hat{A}$ and $\hat{B}$ are two different Hermitian operators (mixed-correlator), \\
    \item[(ii) ] $\hat{A}$ and $\hat{B}$ are one the hermitian conjugate of the other (pair-like correlator),
    \item[(iii) ] $\hat{A}$ and $\hat{B}$ are identical and hermitian (auto-correlator). 
\end{enumerate}

In the paramagnetic case, mixed-correlators of different observables vanish and auto-correlators are bounded to be {\sl even} functions of the Matsubara frequency. This implies that in the limit of $\omega \to \infty$ they must decay at least as $1/\omega^2$. In the broken symmetry phase, instead, nonzero mixed correlators, which are not bound any longer to be even functions of the frequency, might appear. 
In that case, they may decay as $1/\omega$ in the limit of large frequencies.  
Indeed, this is the case of the mixed correlator $\chi^{xy}$ in the FM or $\chi^{x\bar{y}}$ in AF as discussed in the main text, which has important consequences on the high frequency behavior (asymptotics) of the self-energy as we show in Sec. \ref{sec:self:RPA-DGA}.
\section{Analysis of the two-particle reducible diagrams}

 In this appendix, we illustrate through an inspection of the relevant diagrammatics, how the simplified structure of the two-particle reducible contributions in Eqs.~(\ref{eqn::fdmft},\ref{eqn::fdmftAF}) of the main text arises. 
 
We consider explicitly the case  of the $ph$ channel in the AF phase, starting from the diagrams for $\Phi_{\rm ph}$,
 but the derivation can be repeated straightforwardly for the other channels (as well as for the PM/FM cases).
 
\begin{figure*}
\includegraphics[width=0.95\textwidth,angle=0]{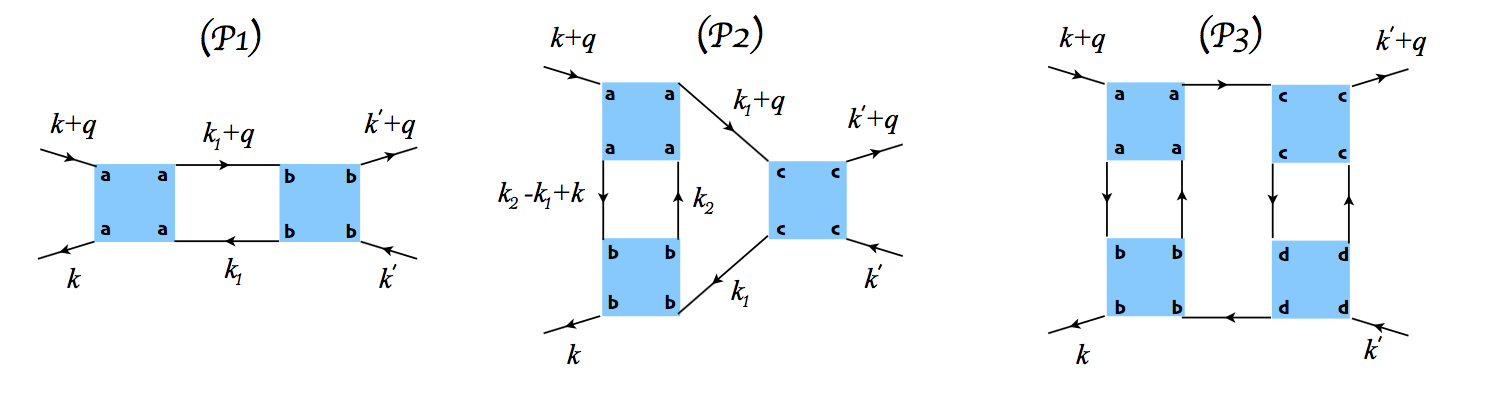}
\caption{Representative contributions to $\Phi_{ph}$, the two-particle {\sl reducible} vertex in the particle-hole channel in momentum space, see text. 
The labels $a,b,c,d$ run over the the lattice sites $A$,$B$.}
\label{fig:Phi}
\end{figure*}

 Once again, we follow the strategy of focusing on the first corrections to purely local diagrams.  Since non-locality is introduced by the reducible bubble terms, the representative three diagrams to be considered are those depicted in Fig.~\ref{fig:Phi}. Here all squared boxes correspond to purely local vertex contributions [e.g., ${\mathcal V}_a(\nu,\nu', \omega) \delta_{ab} \delta_{bc} \delta_{cd} $]. These three examples (${\mathcal P}_1$, ${\mathcal  P}_2$, ${\mathcal P}_3$)  also correspond to the three classes in which reducible diagrams can be subdivided, in terms of their dependence of the basis-lattice index ($2$,$3$,$4$) {\sl as well as} of high-frequency asymptotic  properties\cite{Rohringer2012,Tagliavini2018,Wentzell2020}. 

As for ${\mathcal P}_1^{ab}(k,k',q)$, one finds:
\begin{eqnarray} 
{\mathcal P}_1^{ab} & =  &\frac{1}{\beta V} \sum_{k_1}{\mathcal V}_a(\nu,\nu_1, \omega)  {\mathcal V}_b(\nu_1,\nu', \omega) \, G_{ab}(k_1)G_{ba}(k_1+ q) \nonumber \\
 & = & \frac{1}{\beta}\sum_{\nu_1} {\mathcal V}_a(\nu,\nu_1, \omega)  {\mathcal V}_b(\nu_1,\nu', \omega) \, {\mathcal B}^{ab}
 \left(\nu_1, \omega, X({\bf q})\right)  \nonumber \\\
\label{eqn::P1}
\end{eqnarray}
where ${\mathcal B}^{ab}(\nu_1, \omega, X({\bf q})) =  \frac{1}{V}\sum_{\bf k_1} G_{ab}(k_1)G_{ba}(k_1+ q)$,
where $G_{ab}$ is the Green's funcion calculated in DMFT, that reads:
\begin{equation}
    \bar{\bar{G}}_\sigma(k) = 
    \frac{1}{\zeta_{A\sigma}(\nu)\zeta_{B\sigma}(\nu)-\epsilon_\bk^2}\,\left(
    \begin{array}{cc}
    \zeta_{B\sigma}(\nu)&\epsilon_\bk \\
    \epsilon_\bk  &  \zeta_{A\sigma}(\nu)
    \end{array}
    \right),
\end{equation}
with $\zeta_{a\sigma(\nu)}  = i \nu+\mu -\Sigma_{\sigma a}(\nu)$.
\par For the calculation of $\mathcal{B}_{ab}$ we could extend the summation over the whole BZ because the integrand function is symmetric under translation of $\boldsymbol{\Pi}$
\footnote{The situation would be different if we had to calculate the following sum: $\sum_\bk^\prime G_{aa}(k)G_{bc}(k+q)$, with $ b \neq c$.
In this case, the integrand function would  acquire a minus sign when translated by $\boldsymbol{\Pi}$ and the summation over the whole BZ would yield zero. However,  by inspection of the corresponding diagrams, one sees that these kind of processes vanish in the limit of infinite dimensions.}.

\par Now we can express the bubble as a double integral with the integrand weighted by the two-particle density of states of the hyper-cubic lattice, i.e.
\begin{equation}
    \mathcal{B}^{ab}(\nu,\omega,X) = \iint d\epsilon_1 d\epsilon_2\, D^X_{\epsilon_1,\epsilon_2}G_{ab}(\nu,\epsilon_1)G_{ba}(\nu+\omega,\epsilon_2),
\end{equation}
where the explicit expression of $D^X_{\epsilon_1,\epsilon_2}$, which can be found in Ref.~[\onlinecite{Georges1996}], has the following property $D^{X = 0}_{\epsilon_1,\epsilon_2} =g(\epsilon_1)g(\epsilon_2)$, with $g(\epsilon)$ being the  DOS.
The latter property implies that $\mathcal{B}^{ab}(\nu,\omega ,0) = 0$ when $a\not = b$ , because the off-diagonal terms of the Green's function are odd in $\epsilon$ as opposed to the DOS that it is an even function.
Instead, when $a = b$, $\mathcal{B}^{ab}(\nu,\omega,0) = \mathcal{B}_{loc}^{a}(\nu,\omega)  = G^{AIM}_{a}(\nu)G_{a}^{AIM}(\nu+\omega)$. We can summarize these properties more concisely as follows:
\begin{equation}
    \mathcal{B}^{ab}(\nu,\omega,X) = \delta_{ab}\,\mathcal{B}_1(\nu,\omega,X) + \delta_{a\bar{b}}\,\mathcal{B}_2(\nu,\omega,X),
\end{equation}
where $\mathcal{B}_1(\nu,\omega,0) \! =\! G^{AIM}_{a}(\nu)G_{a}^{AIM}(\nu+\omega) $ and $\mathcal{B}_2(\nu,\omega,0) =  0$.

Similarly, for ${\mathcal P}_{2 \, c}^{\phantom{2} \,  ab}(k,k',q)$, we must  perform the internal momentum summation over ${\bf k_1}$ and ${\bf k_2}$ in the corresponding bubble terms:
\begin{eqnarray} 
& & \frac{1}{V^2}\sum_{{\bf k}_1, {\bf k}_2} G_{ab}(k_2-k_1+k)  G_{ba}(k_2)G_{ac}(k_1+ q) G_{cb}(k_1)  \nonumber  \\
& = & \frac{1}{V}\sum_{\bf k_1} {\mathcal B}^{ab}(\nu_2,\nu-\nu_1, X({\bf k} -{\bf k}_1)) \; G_{ac}(k_1+ q) G_{cb}(k_1) \nonumber \\
& =  & {\mathcal B}_1^{a}(\nu_2, \nu -\nu_1, X=0) \, \delta_{ab} \,  \frac{1}{V}\sum_{\bf k_1}  G_{ac}(k_1+ q) G_{ca}(k_1)  \nonumber \\
& = & {\mathcal B}^a_1(\nu_2, \nu -\nu_1, X=0) \, \delta_{ab} \,  {\mathcal B}^{ac}(\nu_1, \omega, X({\bf q}))\,, \nonumber \\ 
\label{eqn::P2}
\end{eqnarray}
where we used the fact that $\bk-\bk_1$ is a generic point in the BZ and that the special lines defined by $\bk-\bk_1$ have zero measure in the internal momentum summation. 
\par Finally, by exploiting the relations obtained in Eqs.(\ref{eqn::P1})-(\ref{eqn::P2}), one can apply the same procedure to the three internal momentum summations of ${\mathcal P}_{3 \, cd}^{\phantom{3} \,  ab}(k,k',q)$, obtaining
\begin{eqnarray}
\mathcal{B}_1^{a}(\nu_2,\nu-\nu_1,0)\mathcal{B}_1^{c}(\nu_3,\nu_1-\nu^\prime,0)\mathcal{B}^{ac}(\nu_1,\omega,X(\bq))\delta_{ab}\delta_{cd} \nonumber \\
\end{eqnarray}
obtaining  an analogous simplification of the non-local dependence, and thus, eventually,  of the final expression of the  $\Phi_{\rm ph}$  term anticipated in Eq.~(\ref{eqn::fdmftAF}). 
Evidently, since these results only depend on the non-local structure of  the bubble terms after performing the corresponding internal summations, the very same procedure can be applied to the other subsets of reducible diagrams $\Phi_{\rm \overline{ph}}$ and $\Phi_{\rm pp}$ appearing in Eq.~(\ref{eqn::fdmftAF}).
 
Finally, we note that analogous simplifications to that obtained by performing the internal summations over $\bk_1$ in Eq.~(\ref{eqn::P2}) are also responsible, in $d \rightarrow \infty$, for the disappearance of the momentum-dependence of any irreducible vertex in a given channel, once these vertices are inserted in the corresponding BSEs, consistent to Eq.~(\ref{eqn::gammaAFloc}) in the main text. 
 
 \vskip 1mm
\section{Vertices in the AF case}\label{app:matrices}
In this section we report explicitly the matrices appearing in Eq.(\ref{eq:BSE_long_1}), that read:
\begin{eqnarray}\label{matrices:AF:para1}
\bar{\bar{F}}_{\parallel}^{kk^\prime q}& = &
\left(
\begin{array}{cccc}
F^{AA}_{\uparrow\uparrow}&F^{AA}_{\uparrow\downarrow}&F^{AB}_{\uparrow\uparrow}&F^{AB}_{\uparrow\downarrow}\\ \\
F^{AA}_{\downarrow\uparrow}&F^{AA}_{\downarrow\downarrow}&F^{AB}_{\downarrow\uparrow}&F^{AB}_{\downarrow\downarrow}\\ \\
F^{AB}_{\downarrow\downarrow}&F^{AB}_{\downarrow\uparrow}&F^{AA}_{\downarrow\downarrow}&F^{AA}_{\downarrow\uparrow}\\  \\
F^{AB}_{\uparrow\downarrow}&F^{AB}_{\uparrow\uparrow}&F^{AA}_{\uparrow\downarrow}&F^{AA}_{\uparrow\uparrow}
\end{array}
\right), 
\end{eqnarray}
\begin{eqnarray}
\label{matrices:AF:para2}
\bar{\bar{\Gamma}}_{\parallel}^{kk^\prime q} &= &
\left(
\begin{array}{cccc}
\Gamma^{A}_{\uparrow\uparrow}&\Gamma^{A}_{\uparrow\downarrow}&0&0\\ \\
\Gamma^{A}_{\downarrow\uparrow}&\Gamma^{A}_{\downarrow\downarrow}&0&0\\ \\
0&0&\Gamma^{A}_{\downarrow\downarrow}&\Gamma^{A}_{\downarrow\uparrow}\\  \\
0&0&\Gamma^{A}_{\uparrow\downarrow}&\Gamma^{A}_{\uparrow\uparrow}
\end{array}
\right),
 \end{eqnarray}
 \begin{eqnarray} 
 \label{matrices:AF:para3}
\bar{\bar{\chi}}_{0,\parallel}^{kk^\prime q} &= &
\left(
\begin{array}{cccc}
\chi^{AA}_{0,\uparrow\uparrow}&0&\chi^{AB}_{0,\uparrow\uparrow}&0\\ \\
0&\chi^{AA}_{0,\downarrow\downarrow}&0&\chi^{AB}_{0,\downarrow\downarrow}\\ \\
\chi^{AB}_{0,\downarrow\downarrow}&0&\chi^{AA}_{0,\downarrow\downarrow}&0\\  \\
0&\chi^{AB}_{0,\uparrow\uparrow}&0&\chi^{AA}_{0,\uparrow\uparrow}
\end{array}
\right),\hspace{2.5cm},
\end{eqnarray}

\section{Intraband bubble contributions}\label{app:intra}

In this appendix, we discuss more in detail the 
contributions to the bubble terms arising from intraband processes in the AF case. 
\par The bubble terms expressed in the physical basis read:
\begin{eqnarray}
\chi_0^{xx}(q) &=&\frac{1}{N}\sum_\bk \left\{
u^+_\bq\left[ \,f_\omega\left(\omega^+_{\bk+\bq},\omega^+_{\bk}\right) +f_\omega\left(\omega^-_{\bk+\bq},\omega^-_{\bk}\right)\right]
\right. \nonumber \\
&+&
\left.
u^-_\bq \left[f_\omega\left(\omega^-_{\bk+\bq},\omega^+_{\bk}\right) + \,f_\omega\left(\omega^-_{\bk+\bq},\omega^+_{\bk}\right)\right]
\right\}, \nonumber\\
\end{eqnarray}
\begin{eqnarray}
\chi_0^{x\bar{y}}(q) &=&\frac{1}{N}\sum_\bk \left\{
w^+_\bq \left[f_\omega\left(\omega^-_{\bk+\bq},\omega^-_{\bk}\right)- \,f_\omega\left(\omega^+_{\bk+\bq},\omega^+_{\bk}\right)\right]
\right. \nonumber \\
&+&
\left.
w^{-}_\bq \left[f_\omega\left(\omega^+_{\bk+\bq},\omega^-_{\bk}\right) -  \,f_\omega\left(\omega^-_{\bk+\bq},\omega^+_{\bk}\right)\right]
\right\}, \nonumber\\
\end{eqnarray}
where $f_\omega(x,y) = \frac{n_F(x)-n_F(y)}{y-x+\omega}$,
$u_\bq^\pm =\frac{1}{4}
\left[1 \pm \frac{\epsilon_\bk \epsilon_{\bk +\bq}-\Delta^2}{E_\bk E_{\bk + \bq}}\right] $ and $w_\bq^\pm = i\,\Delta
\left(
\frac{\pm E_\bk-E_{\bk+\bq}}{4 E_\bk E_{\bk+\bq}}
\right)$, with $\Delta = m_S U /2$.
\par We shall now focus on the intraband terms and analyze their low-energy properties.  For $\bq \sim 0$, and in the case of particle doping, the intraband terms read:
\begin{eqnarray}
\chi_{0,\mbox{\footnotesize intra}}^{xx} &=& -\int \frac{d\bk}{4\pi^2}\,\frac{\epsilon_\bk^2}{2\,E_\bk^2}\delta(\omega^+_\bk)\frac{\nabla \omega_\bk^+\cdot \bq}{\omega -\nabla \omega_\bk^+\cdot \bq + i0^+}, \nonumber \\
\nonumber \\
\chi_{0,\mbox{\footnotesize intra}}^{\bar{x}\bar{x}} &=& \frac{\Delta^2}{8}\int \frac{d\bk}{4\pi^2}  \left(\frac{\nabla\epsilon_\bk\cdot\bq}{E_\bk^2}\right)^2\,\delta(\omega_\bk^+)\,\frac{\nabla \omega_\bk^+\cdot\bq}{\omega-\nabla\omega_\bk^+\cdot\bq + i0^+}, \nonumber \\
\nonumber \\
\chi_{0,\mbox{\footnotesize intra}}^{x\bar{y}} &=& -i \Delta \int \frac{d\bk}{4\pi^2}  \frac{\nabla E_\bk\cdot \bq}{4E_\bk^2}\,\delta(\omega_\bk^+)\,\frac{\nabla \omega_\bk^+\cdot\bq}{\omega-\nabla\omega_\bk^+\cdot\bq + i0^+},\nonumber \\
\end{eqnarray}
where $\chi_0^{\bar{x}\bar{x}}(q) = \chi_0^{xx}(q + \Pi)$.
As these contributions are given by line integrals along the FS, we can expand  the bands  around the point $\bk = (\pi,0)$, that is the center of one of  the four Fermi pockets in the BZ, and therefore  $\omega^+_\bk \sim -4\alpha\,t + \Delta+2\,\alpha\,t[(k_x-\pi)^2  + k_y^2] -\mu$ and $\nabla \omega_\bk^+ \sim 4\alpha\,t\,(k_x-\pi,k_y)$. Hence, the intraband bubble terms become:
\begin{eqnarray}\label{eq:intra:app}
    \chi_{0,\mbox{\footnotesize intra}}^{xx} &=& A\,q_F^4\, I^{xx}\left(\frac{\omega}{4\alpha t\,q_F\,|\bq|},\cos 4\theta\right), \nonumber \\
    \chi_{0,\mbox{\footnotesize intra}}^{\bar{x}\bar{x}} &=& 
    A\,|\bq|^2\,q_F^2\, I^{\bar{x}\bar{x}}\left(\frac{\omega}{4\alpha t\,q_F\,|\bq|},\cos 4\theta\right),
    \nonumber \\
    \chi_{0,\mbox{\footnotesize intra}}^{x\bar{y}} &=& 
    A\,|\bq|\,q_F^3\, I^{x\bar{y}}\left(\frac{\omega}{4\alpha t\,q_F\,|\bq|},\cos 4\theta\right),
\end{eqnarray}
where $q_F$ is the absolute value of the Fermi momentum, $A={1}/\left[{4\pi\,(\Delta/t)\,\sqrt{\alpha}}\right)]^2$ and:
\begin{eqnarray}\label{eq:integrals:app}
I^{xx}(\lambda,b) &=&  -2\int_0^\pi d\phi\,\frac{(1+b\,\cos 4\phi)\cos\phi}{\lambda - \cos\phi +i0^+},\nonumber \\
I^{\bar{x}\bar{x}}(\lambda,b) &=&  -\int_0^\pi d\phi\,\frac{(1+b\,\cos 2\phi)\cos\phi}{\lambda - \cos\phi +i0^+},
\nonumber \\
I^{x\bar{y}}(\lambda,b) &=&  -i\int_0^\pi d\phi\,\frac{(\cos\phi+b\,\cos 3\phi)\cos\phi}{\lambda - \cos\phi +i0^+},
\end{eqnarray}
with $-1<b<1$.

In presence of doping the system becomes anisotropic and this is formally encoded in the angle ($\theta$) dependence of the intra-band terms in Eq.(\ref{eq:intra:app}).  Further, 
when $-1<\lambda <1$, the integrals in Eq.(\ref{eq:integrals:app}) have a non-vanishing imaginary part. This introduces a damping of the Goldstone modes, which survives at low energy through the leading contribution $\chi^{xx}_{0,\mbox{\footnotesize intra}}$,  as the latter does not tend to zero for $|\bq| \to 0$. 

\newpage
\bibliography{DGA-AF}

\end{document}